\documentclass[twoside,12pt]{article}
\usepackage{epsf,graphicx}
\usepackage{times,latexsym,euscript,amstext,amssymb,amsbsy,amsmath,amsfonts}
\usepackage{epsfig}
\usepackage{slashbox}

\def\vDe{\varDelta}

\newcommand{\be}{\begin{equation}}
\newcommand{\ee}{\end{equation}}
\newcommand{\bea}{\begin{eqnarray}}
\newcommand{\eea}{\end{eqnarray}}

\def\figlab#1{\label{fig:#1}}

\def\barr{\left(\begin{array}{c}}
\def\earr{\end{array}\right)}
\def\bmat{\left(\begin{array}{cc}}
\def\emat{\end{array}\right)}
\def\Figref#1{Fig.~\ref{fig:#1}}

\begin{document}

\title{\vspace{1cm}Nucleon Electromagnetic Form Factors}
\author{C. F.\ Perdrisat,$^{1}$ V.\ Punjabi,$^2$ M.\ Vanderhaeghen
$^{1,3}$ \\
\\
$^1$ College of William and Mary, Williamsburg, VA 23187\\
$^2$ Norfolk State University, Norfolk, VA 23504\\
$^3$ Thomas Jefferson National Accelerator Facility, Newport News, VA 23606\\
}
%\altaffiliation{ {\it {Email address:perdrisa@jlab.org}} \\
\maketitle
\begin{abstract} There has been much activity in the measurement of the 
elastic electromagnetic proton and neutron form factors in the last decade, 
and the quality of the data has been greatly improved by performing 
double polarization experiments, in comparison with previous unpolarized data.
Here we review the experimental data base in view of the 
new results for the proton, and neutron, obtained at MIT-Bates, MAMI, and
JLab. The rapid evolution of phenomenological models triggered 
by these high-precision experiments will be discussed, 
including the recent progress in the determination of the valence 
quark generalized parton distributions of the nucleon, 
as well as the steady rate of improvements 
made in the lattice QCD calculations. \\

\end{abstract}

%\newpage

\tableofcontents

\section{Introduction}

The characterization of the structure of the nucleon is a defining problem
of hadronic physics, much like the hydrogen atom is to atomic physics. 
Elastic nucleon form factors (FFs) 
are key ingredients of this characterization. As such, a
full and detailed quantitative understanding of the internal structure
of the nucleon is a necessary precursor to extending our understanding
of hadronic physics.

The electromagnetic (e.m.) interaction provides a unique tool to investigate 
the internal structure of the nucleon. The measurements of e.m. 
FFs in elastic as well as inelastic scattering, 
and the measurements of structure functions in deep 
inelastic scattering of electrons, have been a rich source of 
information on the structure of the nucleon. 

The investigation of the spatial distributions of the charge and 
magnetism carried by nuclei
started in the early nineteen fifties; it was profoundly affected by 
the original work of one of its earliest pioneers, 
Hofstadter and his team of researchers~\cite{hofs53}, at the Stanford 
University High Energy Physics Laboratory. 
Quite early the interest turned to the nucleon; 
the first FF measurements of the proton were reported in 
1955~\cite{hofs55}, and the first 
measurement of the neutron magnetic FF was reported by Yearian and 
Hofstadter \cite{yearian} in 1958. Simultaneously 
much theoretical work was expanded to the development of 
models of the nucleus, as well 
as the interaction of the electromagnetic probe with 
nuclei and the nucleon. The prevailing
model of the proton at the time, was developed by 
Rosenbluth~\cite{rosenbluth}, and consisted of a neutral
baryonic core, surrounded by a positively charged pion cloud.

Following the early results obtained at the Stanford 
University High Energy Physics Laboratory, similar programs started at 
several new facilities, 
including the Laboratoire de l'Acc\'{e}lerateur Lin\'{e}aire in Orsay, 
(France), the Cambridge Electron Accelerator, 
the Electron-Synchrotron at Bonn, the Stanford Linear Accelerator Center (SLAC), 
Deutsches Elektronen-Synchrotron (DESY) in Hamburg, the 300 MeV linear accelerator at Mainz, 
the electron accelerators at CEA-Saclay, and at Nationaal Instituut voor Kernfysica en Hoge Energie Fysica (NIKHEF). 
The number of electron accelerators and laboratories, 
and the beam quality, grew steadily, reflecting the increasing interest 
of the physics problems investigated 
and results obtained using electron scattering. 
The most recent generation of electron accelerators, which combine 
high current with high polarization electron beams, at MIT-Bates, the Mainz 
Microtron (MAMI), and the Continuous Electron Beam Accelerator 
Facility (CEBAF) of the Jefferson Lab (JLab), 
have made it possible to investigate the internal structure 
of the nucleon with unprecedented precision. 
The CEBAF accelerator adds the unique feature of high energy which allows to 
perform measurements of nucleon e.m. FFs to large momentum transfers. 
Sizable parts of the programs at these facilities were and are oriented 
around efforts to characterize the spatial distribution of charge and 
magnetization in nuclei and in the nucleon. 

The recent and unexpected results from JLab of using the polarization transfer 
technique to measure the proton electric over magnetic FF ratio, 
$G_{Ep}/ G_{Mp}$~\cite{jones,gayou2,punjabi05}, 
has been the revelation that the FFs obtained using
the polarization and Rosenbluth cross section separation methods, 
were incompatible
with each other, starting around $Q^2 = 3$~GeV$^2$. 
The FFs obtained from cross section data had 
suggested that $G_{Ep}\sim G_{Mp}/\mu_p$, where $\mu_p$ is the proton magnetic 
moment; the results obtained from recoil polarization data 
clearly show that the ratio 
$G_{Ep}/G_{Mp}$ decreases linearly with increasing momentum transfer $Q^2$. 
The numerous attempts 
to explain the difference in terms of radiative corrections which affect 
the results of the Rosenbluth separation method very significantly, 
but polarization 
results only minimally, have led to the previously neglected calculation 
of two hard photon exchange with both photons sharing the momentum transfer.   

These striking results for the proton e.m. FF ratio as well as high precision 
measurements of the neutron electric FF, obtained through 
double polarization experiments, 
have put the field of nucleon elastic e.m. FFs into
the limelight, giving it a new life. Since the publication of the JLab ratio 
measurements, 
there have been two review papers on the subject of nucleon e.m.  
FFs \cite{gao,charleskees}, with a third one just recently completed 
\cite{arrreview}. The present review 
complements the previous ones by 
bringing the experimental situation up-to-date, and gives an overview of the 
latest theoretical developments to understand the nucleon e.m. FFs from the 
underlying theory of the strong interactions, Quantum Chromodynamics (QCD). 
We will focus in this review on the space-like nucleon e.m. FFs, as they have 
been studied in much more detail both experimentally 
and theoretically than their 
time-like counterparts \cite{Baldini:2005vn}. We will also not discuss the strangeness FFs of the
nucleon which have been addressed in recent years through dedicated parity
violating electron scattering experiments. For a recent review of the field of
parity violating electron scattering and strangeness FFs, see
e.g. Ref.~\cite{Beise:2004py}.   

This review is organized as follows. Section 2 is 
dedicated to a description of the beginning of the field of 
electron scattering on the nucleon, and the development of the theoretical 
tools and understanding required to obtain the fundamental FFs. 
Elastic differential cross section data lend themselves
to the separation of the two e.m. FFs of proton and neutron by the 
Rosenbluth, or LT-separation method. 
All experimental results obtained in this way are shown and discussed. 

Section 3 discusses
the development of another method, based on double polarization, either 
measuring the proton recoil 
polarization in $\vec{e}p\rightarrow{e}\vec{p}$,
or the asymmetry in $\vec{e}\vec{p}\rightarrow{ep}$. 
The now well documented and abundantly discussed
difference in the FF results obtained by Rosenbluth separation on 
the one hand, and double polarization experiments on the other hand, 
is examined in section \ref{DiscussFF}. 
The radiative corrections, including two-photon exchange corrections, 
essential to obtain the Born 
approximation FFs, are discussed in details in section 
\ref{RadCorr}.   

In Section 4, we present an overview of the theoretical understanding of the 
nucleon e.m. FFs. In Sect.~4.1, we firstly discuss 
vector meson dominance models and the latest dispersion relation fits. 
To arrive at an understanding of the nucleon e.m. FFs  
in terms of quark and gluon degrees of freedom, we next examine in Sect.~4.2 
constituent quark models. We discuss the role of relativity 
when trying to arrive at a microscopic description of nucleon FFs based on
quark degrees of freedom in the few GeV$^2$ region. The present limitations in
such models will also be addressed. In Sect.~4.3, we highlight the spatial 
information which can be extracted from the nucleon e.m. FFs, 
the role of the pion cloud, and the issue of shape of hadrons. 
Sect.~4.4 discusses the chiral effective 
field theory of QCD and their predictions for the nucleon e.m. FFs at low
momentum transfers. Sect.~4.5 examines the {\it ab initio} calculations of 
nucleon e.m. FFs using lattice QCD. We will compare the most recent results 
and the open issues in this field. We also explain how the chiral
effective field theory can be useful in extrapolating lattice QCD calculations 
for FFs, performed at larger than physical pion mass values, to the physical 
pion mass. In Sect.~4.6, we present the quark structure of the nucleon and
discuss how the nucleon e.m. FFs are obtained through sum rules from
underlying generalized (valence) quark distributions. We show the present 
information on GPDs, as obtained from fits of their first moments
to the recent precise FF data set. 
Finally, in Sect.~4.7, we outline the predictions made by perturbative QCD at
very large momentum transfers and confront them with the FF data at the
largest available $Q^2$ values. 

We end this review in Section~5 with our conclusions and spell out some open
issues and challenges in this field.

\section{Nucleon form factors from $eN$ cross sections}
\label{NFF_crossection}

In this section we outline the development of what was, in the early nineteen fifties, 
a new and exciting field of investigation of the structure of nuclei, 
using the elastic scattering of electrons with several hundreds of MeV energy. We also discuss the 
evolution of the Rosenbluth separation method to its present form, and show all FF results obtained 
using this method for both the proton and the neutron.   
 
In the late forties, several papers had pointed out the possibility of measuring the shape and
size of nuclei by observing deviations from Mott scattering by a point charge; most influential 
were the papers by Rose ~\cite{rose}, who argued that ``high energy'' 
electrons would be most suited for such studies, with 50 MeV a best value; and by 
Rosenbluth~\cite{rosenbluth}
for the proton, who provided explicit scattering formula taking into account both charge and the 
anomalous magnetic moment, with the use of ``effective'' charge and magnetic moment. 

An early report of work done at the Stanford University High Energy Physics Laboratory at 
energies larger 
than 100 MeV, was reported by Hofstadter, Fechter and McIntyre~\cite{hofstad53}, who detected deviations
from scattering by a point charge in carbon and gold.
The first review paper of the field, written by Hofstadter
in 1956~\cite{hofst56} included measurement of the proton FF, up to a momentum transfer 
squared of $q^2=13.3$ fm$^{-2}$, or 0.52 GeV$^2$.

\subsection{Early nucleon structure investigations}
\label{Early}

In the middle nineteen fifties, it had been known 
for more than 20 years that the proton could not be just a mathematical point
charge and point magnetic moment. Indeed the measurement of the proton's magnetic
moment by Stern \cite{stern} had revealed a value $\sim$2.8 times larger 
than expected for a spin-$\frac{1}{2}$ Dirac particle. 

Earliest definitions of a FF are usually credited to Rosenbluth 
\cite{rosenbluth}; in this early reference Rosenbluth discussed a model of the proton 
consisting of a neutron core and a positively charge meson cloud, known 
then as the 
weak meson coupling model. A high energy electron was expected to penetrate 
the mesonic cloud and to ``feel'' reduced charges and magnetic moments, $e'$
and $\kappa'e'$.
Expressions for such quantities as $\frac{e'}{e}$ and $\frac{\kappa'e'}
{\kappa_0{e}}$ had been derived by Schiff in 1949 \cite{schiff}. 

In his seminal review paper Hofstadter \cite{hofst56} was the first to
relate the results of McAllister and Hofstadter \cite{mcallis56} for the
$ep$ cross section in elastic scattering at given angle and energy, to 
the Mott cross section for the scattering of a spin $\frac{1}{2}$ electron 
by a spin-less proton, $\sigma_{Mott}$, with internal charge density 
distribution $\rho(r)$, as follows:
\begin{equation}
\sigma(\theta_e)=\sigma_{Mott}\left|\int_{volume}
\rho(\vec{r})e^{i\vec q.\vec r}d^{3}{\vec{r}} \right|^2=\sigma_{Mott}|F({\mathrm q})|^2,
\label{eq:Fsquared}
\end{equation}
\noindent
where:
\begin{equation}
\sigma_{Mott}=\left (\frac{e^2}{2E_{beam}}\right )^2\left (\frac{\cos^2\frac{\theta_e}{2}}
{\sin^4\frac{\theta_e}{2}}\right ),
\label{eq:Mott}
\end{equation}
\noindent
where $E_{beam}$ and $\theta_e$ are the electron incident energy and laboratory scattering 
angle, respectively, and the target mass is infinite.
\begin{figure}[h]
\begin{minipage}[b]{0.45\linewidth}
%\begin{center}
%\epsfxsize=\textwidth
\centerline{\epsfig{file=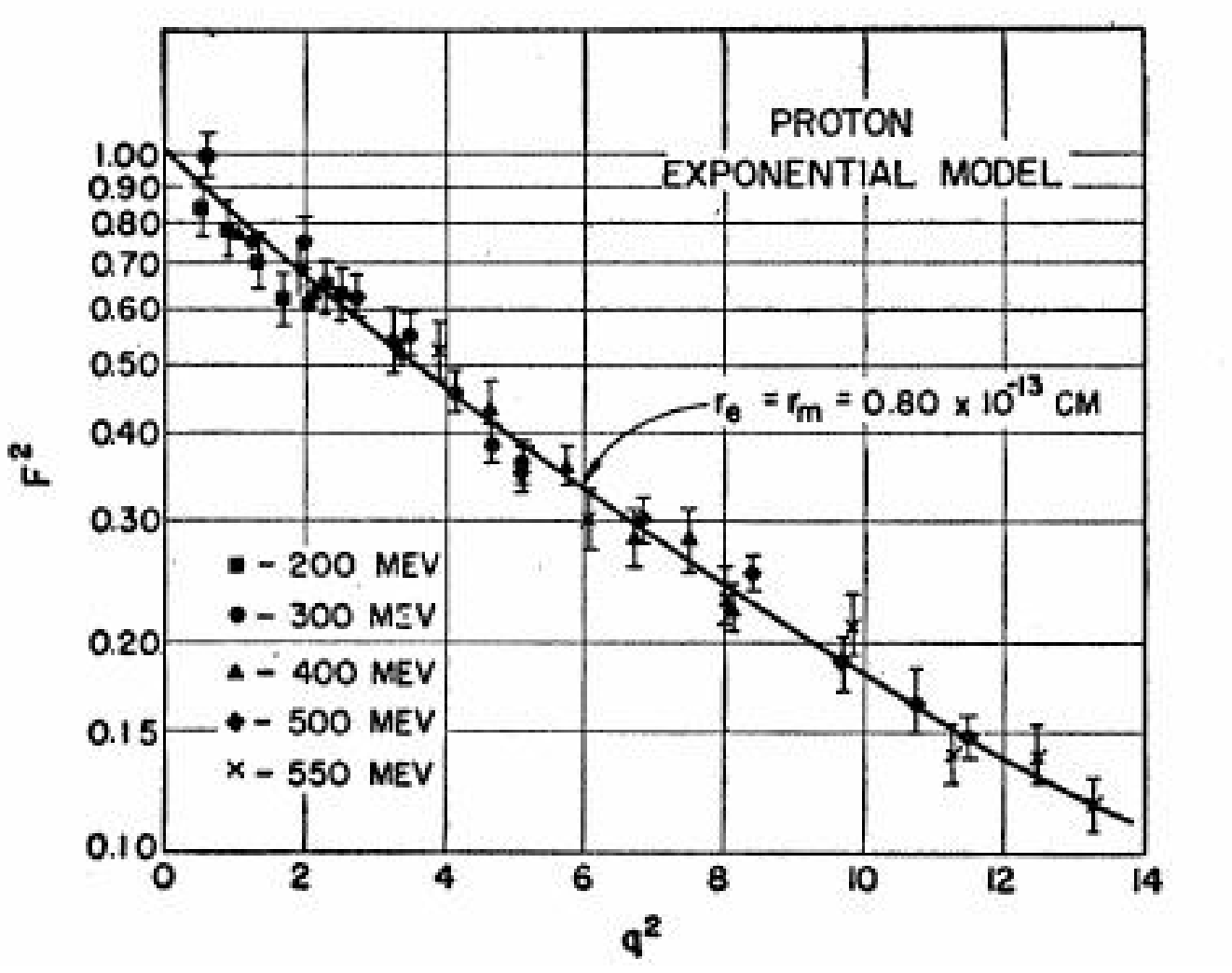,height=2.35in}}
\caption{\small {Fig.~27 in ~\cite{hofst56}, with figure caption ``The square of the
FF plotted against ${\mathrm q}^2$.~ ${\mathrm q}^2$ is given in units of $10^{26}cm^{-2}$.
The solid line is calculated for the exponential model with rms 
radii=$0.80\times10^{-13}$cm.''}}
\label{fig:hofs_first}
%\end{center}
\end{minipage}\hfill
\begin{minipage}[b]{0.45\linewidth}
\epsfxsize=\textwidth
\epsfig{file=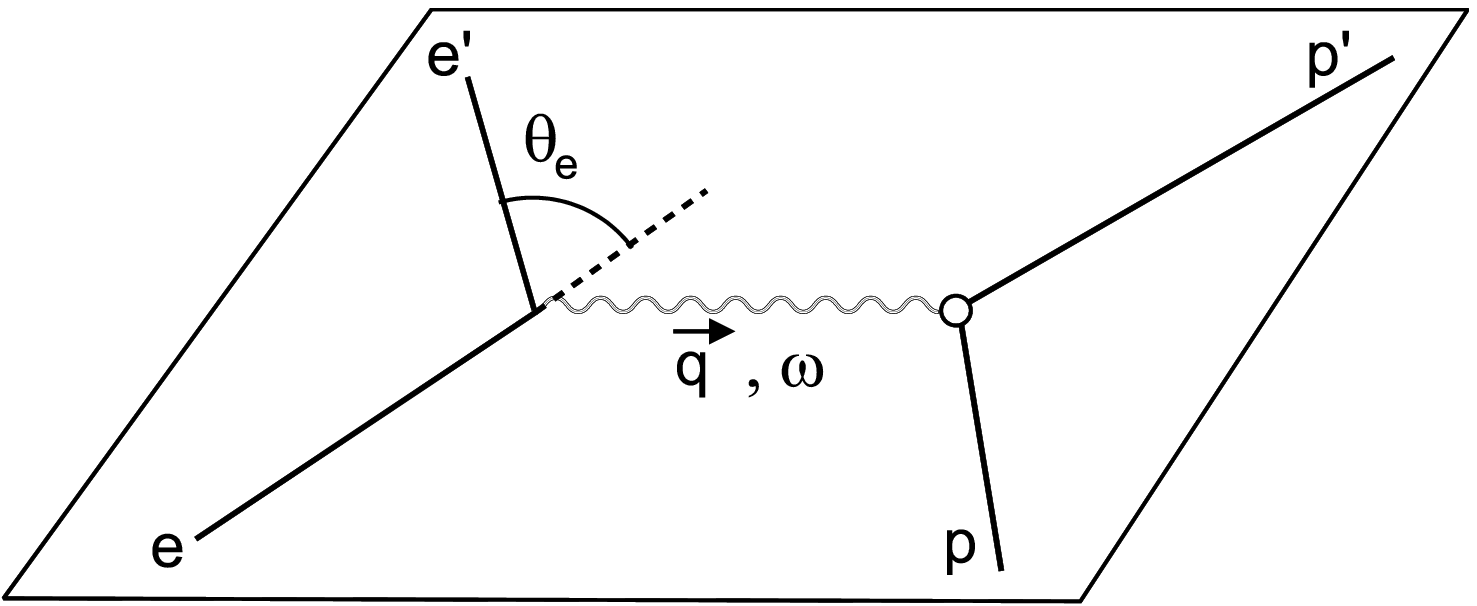,height=1.25in}
\vspace{2.75cm}
\caption{\small{Feynman diagram for the single-photon exchange, 
or Born term, for elastic $ep$ scattering.}}
\label{fig:born}
\end{minipage}
\end{figure}
In this early framework a phenomenological FF squared was 
obtained from 
absolute differential cross section measurements simply as:
\begin{equation}
[F({\mathrm q})]^2=\frac{\sigma({\mathrm q})}{\sigma_{Mott}({\mathrm q})}, ~~~\mbox{with } ~~~\vec {q}=\vec p_{beam}-
\vec p_e ~\mbox{ and }~ {\mathrm q}=|\vec{q}|,
\end{equation}
\noindent
where $\vec q$, $\vec p_{beam}$ and $\vec p_e$ are the center-of-mass (CM) momentum transfer, and 
incident and scattered electron momenta, respectively. The historically significant results of these
measurements of the proton FF are in Fig.~\ref{fig:hofs_first}.

\subsubsection{The Dirac and Pauli nucleon form factors}
\label{DiracPauli}
A direct connection between the reduced charge and magnetic moments discussed
in \cite{rosenbluth} and measurable observables was first proposed by Clementel and Villi 
\cite{clementel},
who defined FFs on the basis of Rosenbluth's discussion of effective
charge and magnetic moments, following \cite{schiff}, as 
$F_1(q)=\frac{e'}{e}$ and $F_2(q)=\frac{\kappa'e'}{\kappa_0{e}}$, with 
$q=2\sqrt{E_{beam}E_e}\sin\frac{\theta_e}{2}$. These FFs were 
then introduced in experimental 
papers by Hofstadter and coworkers~\cite{hofst56,mcallis56,hofst58}, who 
generalized the ``effective'' charge and 
magnetic moment concepts by associating the first with
the deviation from a point charge Dirac particle (Dirac FF, $F_1$), and the second
with the deviation from a point anomalous magnetic moment (Pauli FF, $F_2$).

In lowest order, elastic scattering of an electron by the proton is the result 
of the exchange of a single virtual photon of invariant mass squared  
${q}^2=\omega^2-\vec{q}~^{2}=-4E_{beam}E_e\sin^2\frac{\theta_e}{2}$, 
(the last step neglects the electron mass), where 
$\omega=E_{beam}-E_e$, the energy loss of 
the electron, and $\vec q=\vec p_{beam}-\vec p_e$, the vector momentum 
change of the electron; $\theta_e$ is the Lab electron scattering angle. 
For scattering in the space like region, 
${q}^{2}$ is negative.
\footnote{In this review we will use natural units, with energy and mass in GeV, 
momentum in GeV/c and invariant four-momentum transfer squared in (GeV/c)$^2$.
As is common practice in the literature we will put c=1 for convenience and
denote momentum transfer squared in GeV$^2$, although (GeV/c)$^2$ is understood.}

 The time-like region, where ${q}^{2}$ is 
positive, 
can be accessed for example in $e^-e^+\rightarrow{p}\bar{p}$ or 
${p}\bar{p}\rightarrow e^-e^+$; it will not be discussed in this review. 

Given the smallness of the fine structure constant 
$\alpha~\sim 1/137$, it has been common until recently, to neglect 
all higher order terms, except for the next order in $\alpha$ which is 
treated as a radiative correction, thus implicitly assuming
that the single photon diagram, corresponding to the Born approximation,
is determinant of the relation between cross section and FFs; 
we will revisit this point in section~\ref{RadCorr}. 
In the single photon-exchange process illustrated in Fig. \ref{fig:born}, 
and following the notation of \cite{rekalo}, the amplitude 
for elastic scattering can be written as
the product of the four-component leptonic and hadronic currents, 
$\ell_\nu$ and ${\mathcal J}_{\mu}$ , respectively:
\begin{eqnarray}
-i{\mathcal M}=\frac{-ig_{\mu\nu}}{{q}^2}\left[ie\bar{u}(p_e)\gamma^{\nu}u(p_{beam})\right]
\left[-ie\bar{N}(p')\Gamma^{\mu}(p',p)N(p)\right]=\frac{i}{{q}^2}\ell_{\mu}{\mathcal J}^{\mu},
\end{eqnarray}
\noindent
where $\Gamma^{\mu}$ contains all information of the nucleon structure, $u$
and $N$ are the electron- and nucleon spinors, respectively, $g_{\mu\nu}$ 
is the metric tensor and $k$, $k'$, $p$ and $p'$ are the four-momenta of the incident 
and scattered electron and proton, respectively. To ensure relativistic 
invariance of the amplitude 
$\mathcal M$, $\Gamma^{\mu}$ can only contain $p$, $p'$ and $\gamma^{\mu}$, 
besides scalars, masses and $Q^2$. 

As was shown by Foldy \cite{foldy52}, the most general form for the hadronic 
current for a spin $\frac{1}{2}$-nucleon with internal structure, satisfying 
relativistic invariance and current conservation is:\\
\begin{equation}
{\mathcal J}_{hadronic}^{\mu}=e\overline{N}(p')\left[\gamma^{\mu}{{F_1(Q^2)}}+
\frac{i\sigma^{\mu\nu}q_{\nu}}{2M}{{F_2(Q^2)}}\right]N(p),
\label{eq:Jhadron}
\end{equation}
\noindent
where $Q^2=-{q}^2$, is the negative of the square of the 
invariant mass of the virtual photon
in the one-photon-exchange approximation in $ep$ scattering, and $F_{1}$ 
and $F_2$ are the two only FFs allowed by relativistic invariance. Furthermore,
the anomalous part of the magnetic moment for the proton is $\kappa_p=\mu_p-1$, and for the neutron 
$\kappa_n=\mu_n$, in nuclear magneton-units, $\mu_N=\frac{e\hbar}{2M}$, with values 
$\kappa_p=$1.7928 and $\kappa_n=-1.9130$, respectively; ${M}$ is the nucleon mass. It follows that in 
the static limit, $Q^2=0$, $F_{1p}(0)=1, F_{2p}(0)=\kappa_p, F_{1n}(0)=0, F_{2n}(0)=\kappa_n$, for the proton and neutron, respectively. 

In the one-photon-exchange approximation $F_1(Q^2)$ and $F_2(Q^2)$ are real functions which depend upon 
$Q^2$ only, and are therefore relativistically invariant. When higher order terms with two photons 
exchange are included, there are in general 6 invariant amplitudes, 
which can be written in terms of 3 complex ones \cite{guichon}.

The Lab cross section is then:
\begin{equation}
\frac{d\sigma}{d\Omega_e}=\frac{\overline{|{\mathcal M}|}^2}{64\pi^2}
\left(\frac{E_2}{E_1}\right)^2\frac{1}{M}~~~\mbox{   with  }~~~\overline{|{\mathcal M}|}^2=\frac{1}{(Q^2)^2}\overline{|\ell\cdot{\mathcal J}|}^2, 
\end{equation}

Following the introduction above, we can now write the standard form for the Lab frame 
differential cross section for  $ep$ or $en$ elastic scattering as:
\begin{equation}
\frac{d\sigma}{d\Omega_e} = \left(\frac{d\sigma}{d\Omega}\right)_{Mott}\frac{E_e}{E_{beam}} 
\left\{F_1^2(Q^2) + \tau\left[F_2^2(Q^2)+2\left(F_1(Q^2)+
F_2(Q^2)\right)^2\tan^2\frac{\theta_e}{2}\right]\right\},
\label{eq:1cs}
\end{equation}
\noindent
where $\tau=Q^2/4M^2$, and $\frac{E_e}{E_{beam}}=(1+\frac{2E_{beam}}{M}\sin^2\frac{\theta_e}{2})^{-1}=(1+\tau)^{-1}$ is
the recoil factor.
Eq. (\ref{eq:1cs}) is the most general form for the cross section, as required by Lorentz invariance, 
symmetry under space reflection and charge conservation. Experimentally, the first
separate values for $F_1$ and $F_2$ were obtained by the intersecting ellipse method 
described by Hofstadter \cite{hofst60}. The early data of 
Bumiller {\it et al.} \cite{bumiller} showed that $F_2$
decreased with $q^2$ faster than $F_1$, even suggesting a diffractive behavior
for the proton cross section. Typically these results 
show $F_1/F_2$-ratio values  which are several times larger than modern values for 
the proton.

\subsubsection{The electric and magnetic form factors}
\label{GEGM}

Another set of nucleon FFs, $F_{ch}$ and $F_{mag}$, was first introduced by Yennie, 
Levy and Ravenhall \cite{yennie57}; Ernst, Sachs and Wali \cite{ernst} 
connected $F_{ch}$ and $F_{mag}$ to the charge and current 
distributions in the nucleon; the interpretation that $F_{ch}$ and $F_{mag}$ measure
the interaction with static charge and magnetic fields was given by Walecka \cite{walecka}. The 
following FFs, $F_{ch}$ and $F_{mag}$, were defined in \cite{ernst}:
$F_{ch}=F_1-\frac{Q^2}{2M}F_2,~\mbox{and}~F_m=\frac{1}{2M}F_1+F_2$.
A similar definition of FFs for charge and magnetization,  $G_E$ and $G_M$, which is the one in use today, 
was first discussed extensively by Hand, Miller and Wilson \cite{hand63}
who noted that with $G_{E}=F_{1}-\tau F_{2}~\mbox{and}~G_{M}=F_{1}+ F_{2}$,
the scattering cross section in Eq.~(\ref{eq:1cs}) 
can be written in a much simpler form, without interference term, leading to a 
simple separation method for $G_{Ep}^2$ and $G_{Mp}^2$:
\begin{equation}
\frac{d\sigma}{d\Omega} = \left(\frac{\alpha}{2E_{beam}\sin(\frac{\theta_e}{2})}\right)^2
\frac{E_e}{E_{beam}}\left(\frac{\cot^2(\frac{\theta_e}{2})}{1+\tau}\left[G_{E}^2+\tau{G_{M}}^2\right]
 + 2\tau{G_{M}}^2\right),
\label{eq:gepgmp}
\end{equation}
\noindent
$G_{Ep}$, $G_{Mp}$, $G_{En}$ and $G_{Mn}$ are now customarily called the electric- 
and magnetic Sachs FFs, for the proton and neutron, respectively; 
at $Q^2=0$ they have the static values of the charge and magnetic moments, of the proton 
and neutron, respectively: $G_{Ep}(0)=1, G_{Mp}(0)=\mu_p, G_{En}(0)=0~\mbox{and}~G_{Mn}(0)=\mu_n.$  

\subsubsection{Form factors in the Breit frame}
\label{Breitframe}
The physical meaning of the electric and magnetic FFs, $G_E$ and 
$G_M$, is best understood when the hadronic current is written in the Breit frame. 
In that frame the scattered electron transfers momentum $\vec{q}_B$ but no energy 
($\omega_B=0$). Therefore, the proton likewise undergoes only a change of 
momentum, not of energy, from $-\vec{q}_B/2$ to $+\vec{q}_B/2$; thus 
$Q^2=\vec{q_B}^2$. The four components of the hadronic current in the Breit frame 
are:
\begin{eqnarray}
J^0&=&e2M{\chi'^{\dagger}}\chi(F_1-{\tau}{F_2})=e2M{\chi'^{\dagger}}{\chi}G_E,\\
\vec{J}&=&ie\chi'^{\dagger}(\vec\sigma\times\vec{q}_B)\chi(F_1+{F_2})=ie
\chi'^{\dagger}(\vec\sigma\times\vec{q}_B)\chi{G_M}.
\label{eq:breit}
\end{eqnarray}
Only in the Breit frame can the electric and magnetic FFs 
$G_{E}$ and $G_{M}$ be associated with charge and magnetic current density 
distributions through a Fourier transformation. However the Breit frame is a mathematical 
concept without physical reality: there is a Breit frame for every $Q^2$ value; and above 
a few GeV$^2$, the Breit frame moves in the Lab with relativistic velocities, resulting 
in a non-trivial relation between Breit frame quantities and Lab frame quantities: the 
transformation affects both the kinematics and the structure. A model dependent 
procedure to transform these distributions 
from the Breit-- to the Lab frame has been recently developed by Kelly
\cite{kelly02}, with interesting results to be discussed later in section \ref{sec:rad}.

\subsection{Rosenbluth form factor separation method}
\label{Rosenbluth}

The Rosenbluth method has been the only technique available to obtain separated values for 
$G_E^2$ and $G_M^2$ for proton and neutron until the 1990s. The method requires measuring the 
cross section for 
$eN$ scattering at a number of electron scattering angles, for a given value of $Q^2$; this 
is obtained by varying both the beam energy and the electron scattering 
angle over as large a range as experimentally feasible.

The cross section for $ep$ scattering in Eq. (\ref{eq:1cs}), when written in terms of the electric- 
and magnetic FFs, $G_{E}$ and $G_{M}$, takes the following form:
\begin{equation}
\frac{d\sigma}{d\Omega} = \left(\frac{d\sigma}{d\Omega}\right)_{Mott}\times\left(G_{E}^2+
\tau\left[1+2(1+\tau)\tan^2\frac{\theta_e}{2}\right]G_{M}^2\right)/(1+\tau),
\label{eq:rosenb1}
\end{equation}
\noindent
and in the notation preferred today, this cross section can be re-written as:
\begin{equation}
\frac{d\sigma}{d\Omega} =\left(\frac{d\sigma}{d\Omega}\right)_{Mott}\times\left[G_{E}^2
+\frac{\tau}{\epsilon}G_{M}^2\right]/(1+\tau),
\label{eq:rosenb}
\end{equation}
\noindent
where $\epsilon=[{1+2(1+\tau)\tan^2 \frac{\theta_e}{2}}]^{-1}$ is the virtual photon polarization. 

In early versions of the Rosenbluth separation method for the proton, a correspondingly defined reduced 
cross section was plotted 
either as a function of $\cot^2\frac{\theta_e}{2}$ \cite{hand63,wilsonrr} or 
$\cos\theta_e$ \cite{berger}. For
example in \cite{hand63}, the function 
$R(Q^2,\theta_e)=\left[G_{E{p}}^2+\tau{G}_{M{p}}^2\right]\cot^2\frac{\theta_e}{2}
+\tau(1+\tau)G_{M{p}}^2$ was defined.
In 1973 Bartel {\it et al.} chose a form linear in ${\cos^2\frac{\theta_e}{2}}$, namely 
$\cos^2\frac{\theta_e}{2}
\times\left(\frac{d\sigma}{d\Omega}\right) / \left(\frac{d\sigma}{d\Omega}
\right)_{Mott}$ \cite{bartel}. Neither of these
linearization procedures fully disentangles $G_{E{p}}^2$ and $G_{M{p}}^2$.

\begin{figure}[h]
\begin{minipage}[b]{0.45\linewidth}\
\hspace{-0.5cm}
\begin{center} 
\epsfig{file=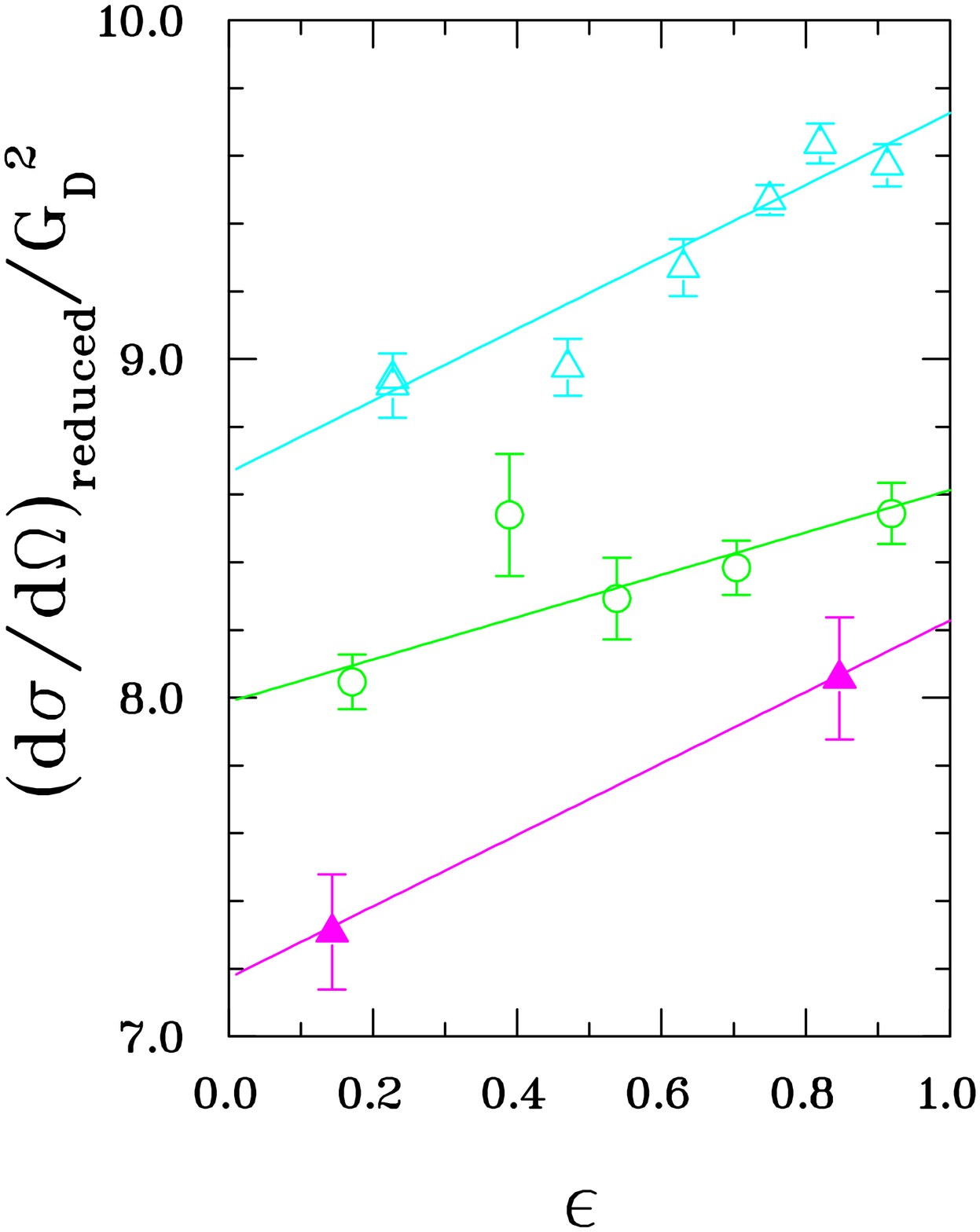,height= 3.15in}
\vspace{0.45cm}
\caption[]{\small{Demonstration of the Rosenbluth separation method based on the data from
~\cite{andivahis}. The $Q^2$ values shown are 2.5 (open triangle), 5.0 (circle) and 7.0 
(filled triangles) GeV$^2$. }}
\label{fig:redcsvseps}
\end{center}
\end{minipage}\hfill
\begin{minipage}[b]{0.45\linewidth}
%\end{figure}
%\begin{figure}[h]
\begin{center} 
\epsfxsize=\textwidth
\epsfig{file=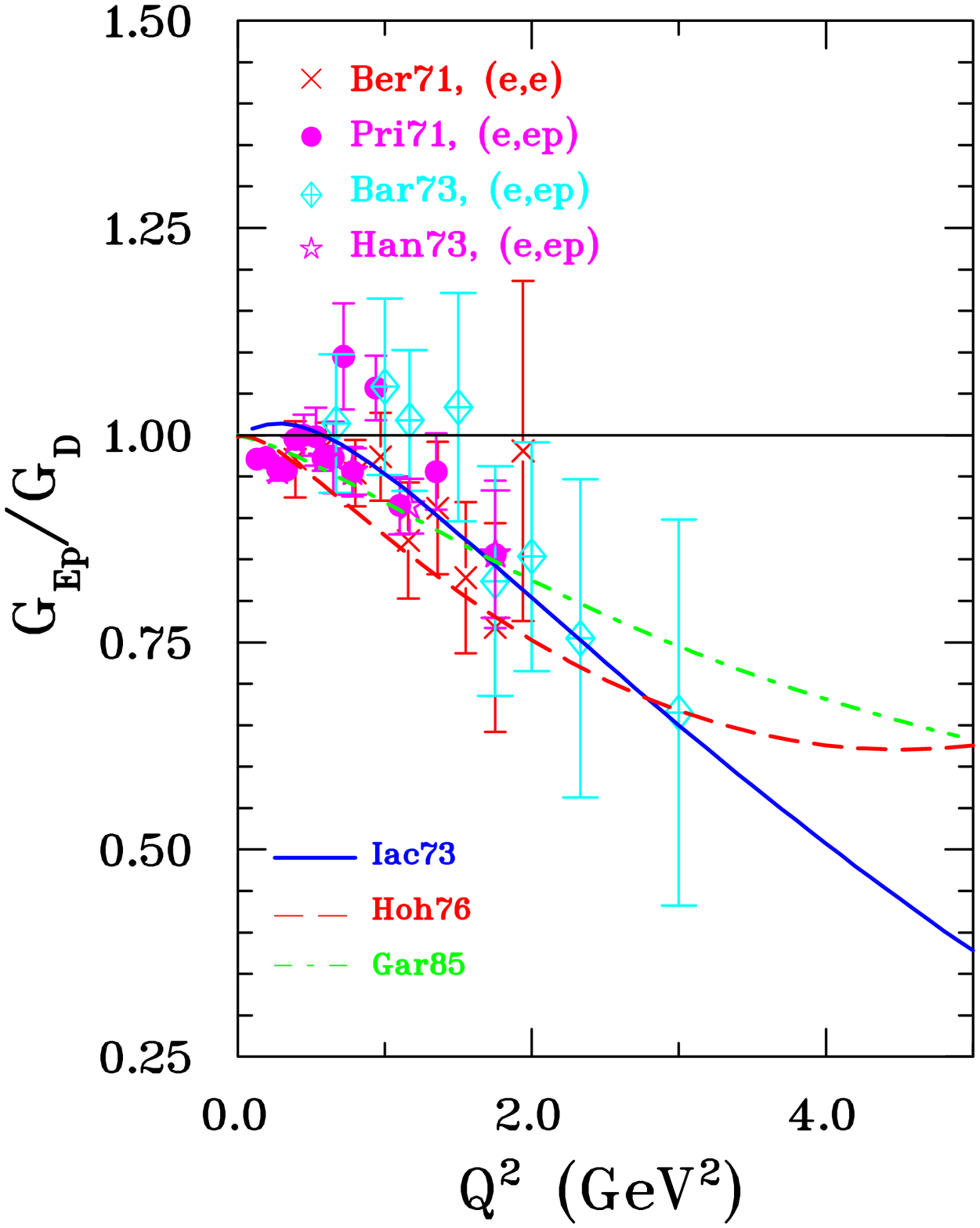,height=3.06in}
%\vspace{0.40cm}
\caption[]{\small{Early Rosenbluth separation data for $G_{Ep}$, up to 1973 
\cite{berger,price,bartel,hanson}, but not including the 1970 SLAC experiment of 
Litt {\it et al.} \cite{litt}.}}
\label{fig:earlysep}
\end{center}
\end{minipage}
\end{figure}

The modern version of the Rosenbluth separation technique takes advantage 
of the linear dependence in $\epsilon$ of the FFs in the reduced 
cross section based on Eq. (\ref{eq:rosenb}) and is defined as follows:
\begin{equation}
\left(\frac{d\sigma}{d\Omega}\right)_{reduced} = \frac{\epsilon(1+\tau)}{\tau}
\left(\frac{d\sigma}{d\Omega}\right)_{exp}/ \left(\frac{d\sigma}{d\Omega}
\right)_{Mott}=G_{M}^2+\frac{\epsilon}{\tau}G_{E}^2,
\label{eq:redcs}
\end{equation}
\noindent 
where $(d\sigma/d\Omega)_{exp}$ is a measured cross section. 
A fit to several measured reduced cross section values at the same $Q^2$, 
but for a range of $\epsilon$-value, gives independently $\frac{1}{\tau}G_{E{p}}^2$ 
as the slope and $G_{M{p}}^2$ as the intercept, as shown in Fig. \ref{fig:redcsvseps};
the data displayed in this figure are taken from \cite{andivahis}.

\subsubsection{Proton form factor measurements}
\label{ProtonFF}

Figure \ref{fig:earlysep} shows  
Rosenbluth separation results performed in the 1970's as the ratio 
$G_{E{p}}/G_{D}$, where $G_{D}$ is the dipole FF given below by Eq. \ref{eq:dipole}; it is noteworthy that these results strongly 
suggest a decrease of $G_{Ep}$ with increasing $Q^2$, a 
fact noted in all four references \cite{berger,price,bartel,hanson}. As will be 
seen in section \ref{DiscussFF}, the 
slope of this decrease is about half the one found in recent recoil 
polarization experiments.
%except for the data of ~\cite{hanson}, which is statistically 
%compatible with the new polarization data. 
Left out of this figure are the data of Litt {\it et al.} 
\cite{litt}, the first of a series of SLAC experiments which were going to 
lead to the 
concept of ``scaling'' based on Rosenbluth separation results, namely the 
empirical relation $\mu_pG_{E{p}}/G_{M{p}}\sim 1$. Predictions of the 
proton FF $G_{E{p}}$ 
made in the same period and shown in Fig. \ref{fig:earlysep} are from 
Refs.~\cite{iach,hohler,gari}, all three are based on a dispersion relation
description of the FFs, and related to the vector meson dominance 
model (VMD). 
\begin{figure}[h]
\begin{minipage}[b]{0.45\linewidth}
\begin{center}
\epsfig{file=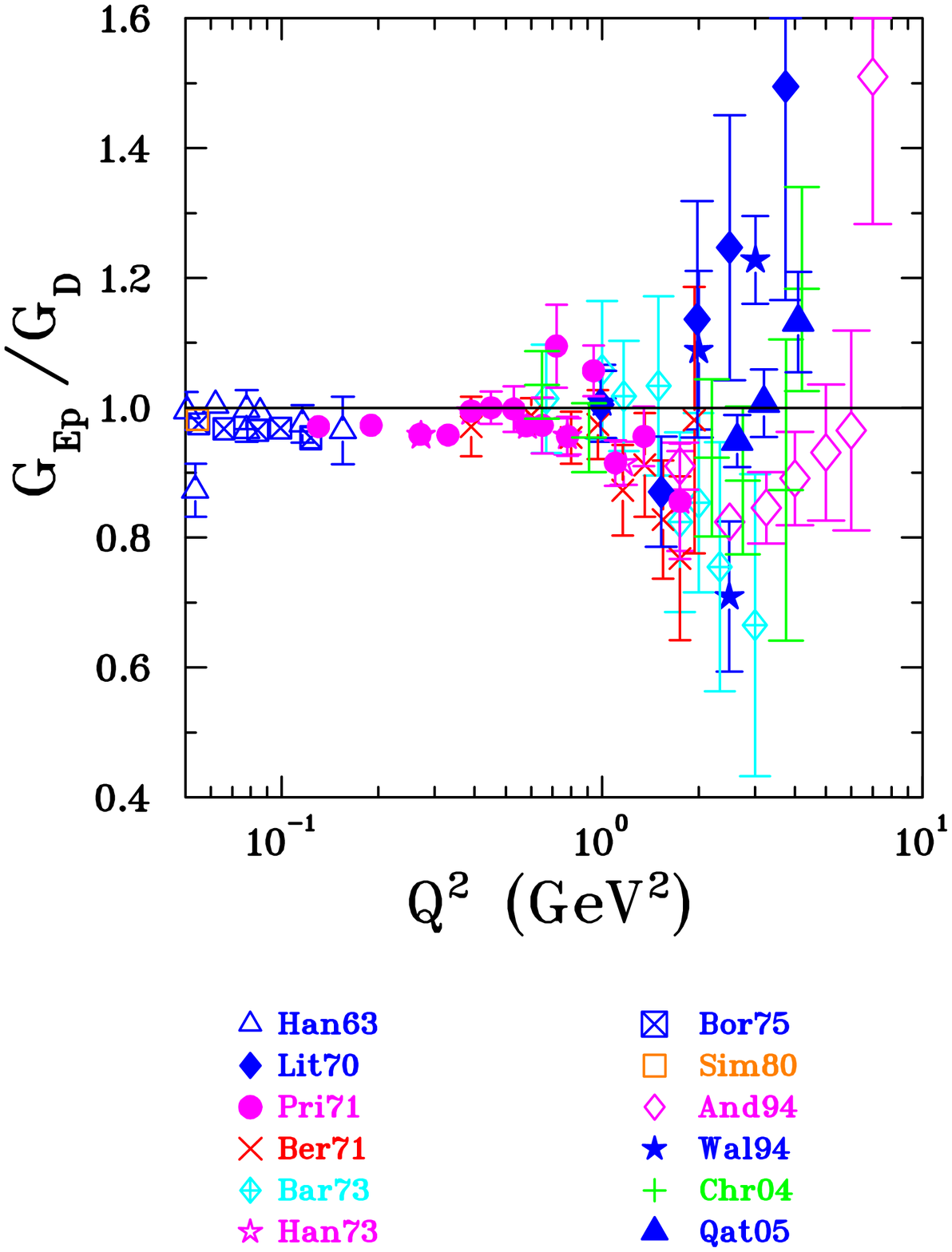,height=3.87in}
%\vspace{0.7cm}
\caption[]{\small{Data base for $G_{E{p}}$ obtained by the Rosenbluth method; 
the references are \cite{hand63,litt,price,berger,bartel,hanson,bork,simon,andivahis,walker,christy,qattan05}.}}
\label{fig:gepgd}
\end{center}
\end{minipage}\hfill
\begin{minipage}[b]{0.45\linewidth}
\begin{center}
\epsfxsize=\textwidth
%\hspace{-2.0cm}
\epsfig{file=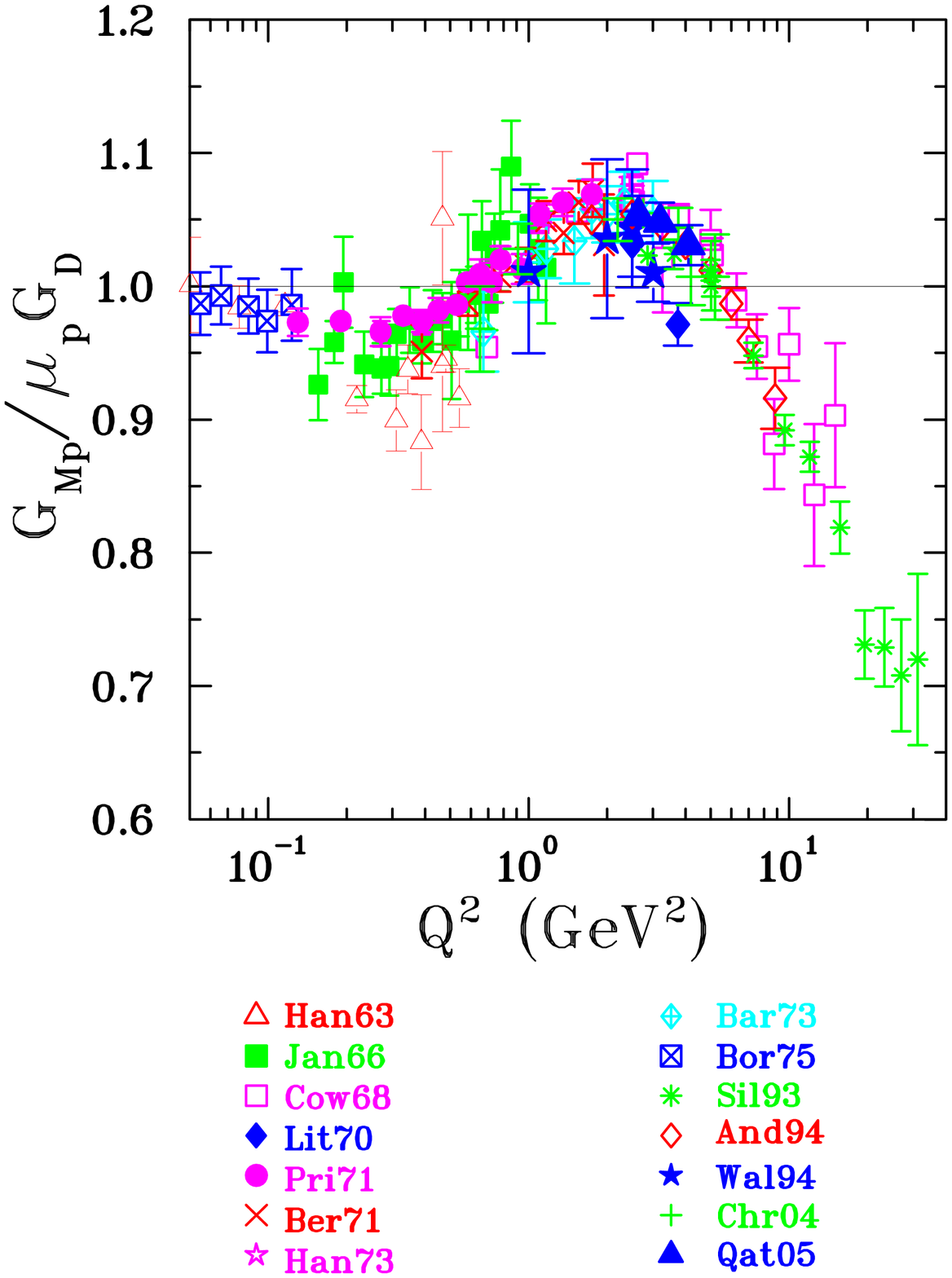,height=3.87in}
\caption[]{\small{Data base for $G_{M{p}}$ obtained by the Rosenbluth method; the references 
are \cite{hand63,janssens,coward,litt,price,berger,
hanson,bartel,bork,sill,andivahis,walker,christy,qattan05}. }}
\label{fig:gmpgd}
\end{center}
\end{minipage}\hfill
\end{figure}
%********************************************************************************
\begin{figure}[h]
\begin{minipage}[b]{0.45\linewidth}
\begin{center}
\epsfig{file=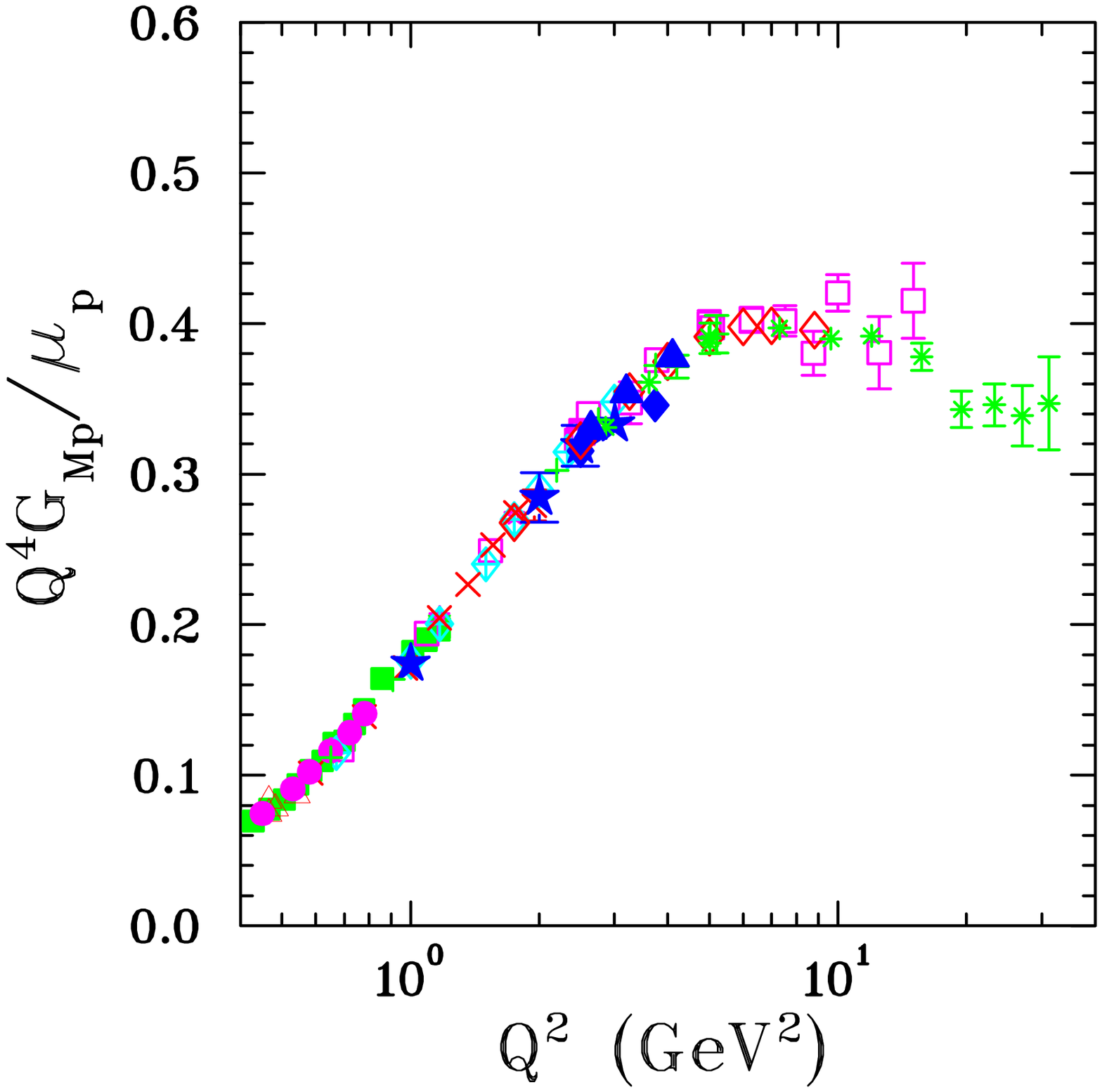,height=2.8in}
\caption[]{\small{The $G_{Mp}$ data follow the pQCD scaling law, 
$G_{Mp}\propto\frac{1}{Q^4}$, 
as was first demonstrated in \cite{arnold1}; data as in Fig. \ref{fig:gmpgd}.}} 
\label{fig:gmp_pQCD}
\end{center}
\end{minipage}\hfill
\begin{minipage}[b]{0.45\linewidth}
%\begin{center}
\epsfxsize=\textwidth
\hspace{-1.0cm}
\epsfig{file=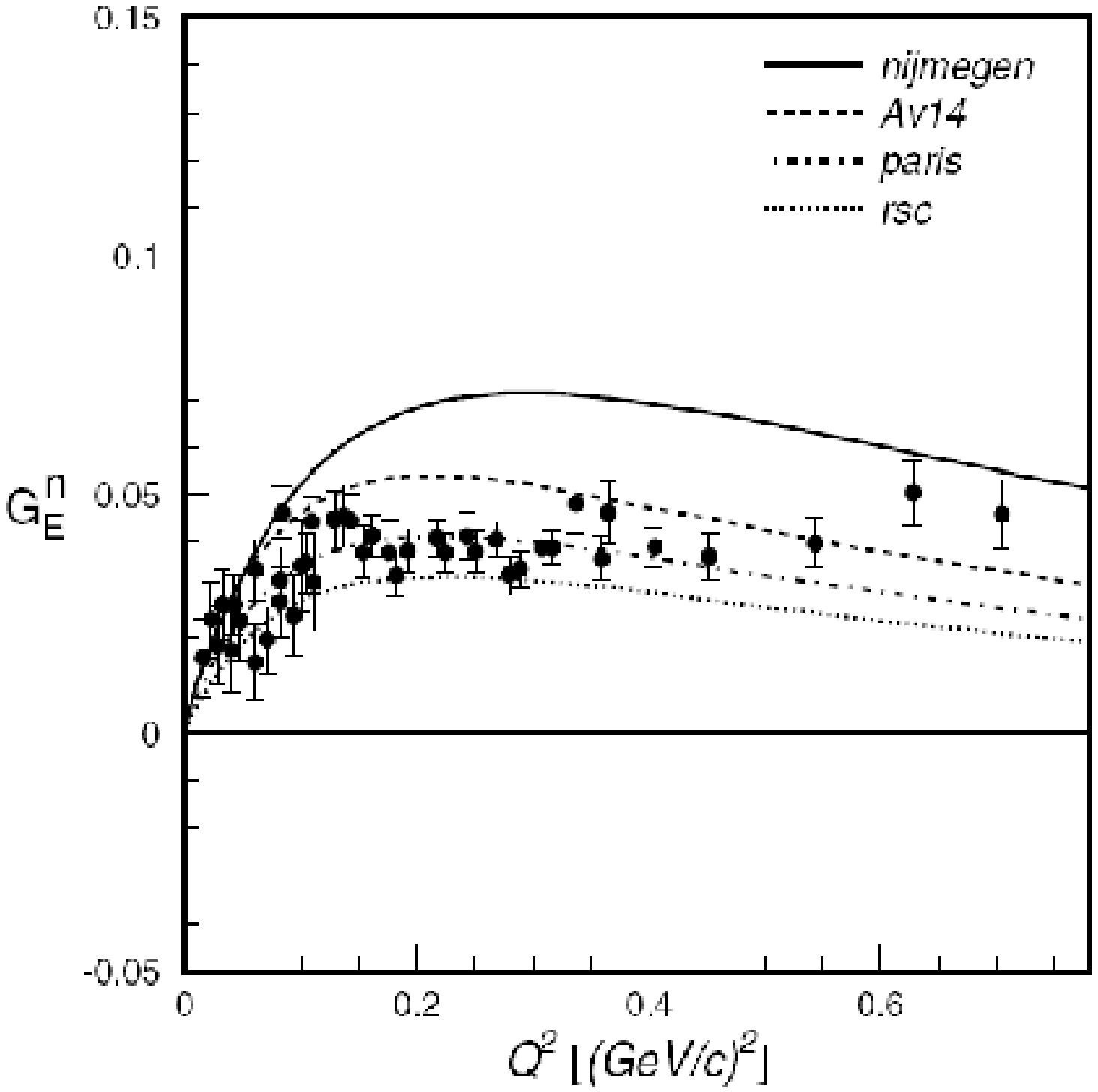,height=2.8in}
\vspace{0.25cm}
\caption[]{\small{Fig. 4 from Ref.~\cite{gao}. The 1990 Platchkov data \cite{platchkov} for $G_{En}$, with 
fits to several NN potentials determining the deuteron wave function.}}
\label{fig:genplatchkov}
%\end{center}
\end{minipage}\hfill
\end{figure}
%*********************************************************************************

A compilation of all $G_{Ep}$ and $G_{Mp}$ data obtained by the the Rosenbluth separation 
technique is shown in
Figs.~\ref{fig:gepgd} and \ref{fig:gmpgd}; in these two figures both $G_{Ep}$ and $G_{Mp}$ 
have been 
divided by the dipole FF $G_D$ given by:
\begin{equation}
G_D = \frac{1}{(1+Q^2/0.71{\mathrm GeV^2})^2}~~\mbox{with}~~ G_{Ep}=G_D,~G_{Mp}=\mu_pG_D, ~~\mbox{and}~~ G_{Mn}=\mu_nG_D.
\label{eq:dipole}
\end{equation}
It is apparent from Fig. \ref{fig:gepgd} that the cross section data have lost track of $G_{Ep}$ 
above $Q^2\sim 1$ GeV$^2$. It is difficult to obtain $G_{E}^2$ for large $Q^2$ values by Rosenbluth separation 
from $ep$ cross section data for several reasons; first, the factor $\frac{1}{\tau}$ multiplying $G_E^2$ 
in Eq. (\ref{eq:redcs}) automatically reduces 
the contribution of this term to the cross section as $Q^2$ increases; and second, even at small $Q^2$, 
$G_M^2\sim\mu_p^2G_E^2$, hence the contribution of $G_{E}^2$ to the cross section is reduced by 
a factor 7.80. 

In sharp contrast with the situation for $G_{Ep}$, the $G_{Mp}/\mu_{p}G_D$ ratios shown 
in Fig.~\ref{fig:gmpgd} display excellent internal consistency, up to $Q^2=30$ GeV$^2$, for the
$G_{Mp}$-values obtained from cross section data; note that the large $Q^2$- data in 
\cite{arnold1} were obtained without Rosenbluth separation, with the assumption that 
$G_{Ep}=G_{Mp}/\mu_p$; the ratio $G_{Mp}/\mu_{p}G_D$ becomes distinctly smaller than 1 above 
$\sim$ 5 GeV$^2$. 

It was first observed by Arnold {\it et al.} \cite{arnold1}
that the proton magnetic FF, $G_{Mp}$ follows the pQCD 
prediction of Brodsky and Farrar \cite{brodsky}, as illustrated in Fig.~\ref{fig:gmp_pQCD};
the pQCD prediction is based on quark counting rules. Indeed  $Q^4G_{Mp}$ becomes nearly constant
starting at $Q^2=8$ GeV$^2$. However, the $1/Q^4$-behavior of the proton magnetic 
FF was first mentioned by Coward {\it et al.} \cite{coward} based on their data extending 
to 20 GeV$^2$; these authors discussed the 1/$Q^4$ behavior in light of the vector meson 
exchange model prevailing at the time \cite{schwinger}.

\subsubsection{Neutron electric form factor measurements}
\label{NeutronEFF}
The ``neutrality'' of the neutron requires the electric FF to be zero at $Q^2=0$, 
and small at non-zero $Q^2$; historically, 
the fact that the electric FF is non-zero has been explained in terms of 
a negatively charged pion cloud in the neutron, which surrounds 
a small positive charge \cite{fermi}. 

Early attempts to determine the neutron FF were based on measurements 
of the elastic $ed$ cross section. The scattering by an electron from
the spin 1 deuteron requires 3 FFs in the hadronic current operator, 
for the charge, quadrupole and 
magnetic distributions, $G_C$, $G_Q$ and $G_M$, respectively. In the original
impulse approximation (IA) form of the cross section 
developed by Gourdin \cite{gourdin}, the elastic $ed$ cross section is: 
\begin{equation}
\frac{d\sigma}{d\Omega}=\frac{d\sigma}{d\Omega}_{Mott}\left(A(Q^2)+B(Q^2)\tan^2(\frac{\theta_e}{2})\right),
\end{equation}
\noindent
where $A(Q^2)=G_C^2(Q^2)+\frac{8}{9}\eta^{2}G_Q^{2}(Q^2)+
\frac{2}{3}\eta(1+\eta)G_M^{2}$ and $B(Q^2)=\frac{3}{4}
\eta(1+\eta)^2G_M^2(Q^2)$, with $\eta=Q^2/4M_D^2$.\\
The charge, quadrupole and magnetic FFs 
can be written in terms of the isoscalar electric and magnetic FFs 
~\footnote{The isoscalar ($F_i^S$) and isovector ($F_i^V$)
Dirac ($i = 1$) and Pauli ($i = 2$) FFs are usually defined from the
corresponding proton and neutron FFs as~:
$F_{i}^S = F_{i p} + F_{i n}$, and $F_{i}^V = F_{i p} - F_{i n}$.
Analogous relations hold for the Sachs FFs defining
$G_{E}^S = G_{E p} + G_{E n}$, and $G_{E}^V = G_{E p} - G_{E n}$.}
as follows:
\begin{center}
$G_C=G_{E}^{S}C_E,~~\mbox{  }~~G_Q=G_{E}^{S}C_Q ~~\mbox{  and } ~~G_M=\frac{M_D}{M_p}(
G_{M}^{S}C_S+\frac{1}{2}G_{E}^{S}C_L)$,
\end{center}
\noindent
where the coefficients $C_E$, $C_Q$, $C_L$ and $C_S$
are Fourier transforms of specific combinations of the S- and D-state 
deuteron wave functions, $u(r)$ and $w(r)$~\cite{gourdin}.

The 1971 DESY experiment of Galster {\it et al.} \cite{galster} measured elastic 
$ed$ cross sections up to 0.6 GeV$^2$ with good accuracy and 
provided a data base for the extraction of $G_{En}$; it had been 
preceded by a series of experiments started at the Stanford MARK III accelerator,
including McIntyre and Dhar~\cite{mcintyre}, Friedman, Kendall and Gram~\cite{friedman}, 
Drickey and Hand~\cite{drickey}, 
Benaksas, Drickey and  Fr\`{e}rejacque~\cite{benaksas}, and Grosset\^{e}te, Drickey and 
Lehmann~\cite{grossetetel}. On the basis of 
these data, and using Hamada-Johnston \cite{hamada} and Feshbach-Lomon 
\cite{feshbach} deuteron wave functions, the following fitting function was proposed in \cite{galster}:
\begin{equation}
G_{En}(Q^2)=-\frac{\mu_n \tau}{1+5.6\tau}G_{Ep}(Q^2).
\label{eq:galster}
\end{equation}
\noindent
The often quoted Galster fit uses Eq. (\ref{eq:galster}) with $G_{Ep}$ replaced 
by the dipole FF $G_D$ (see Eq. (\ref{eq:dipole})).

The next and last experiment to measure the elastic $ed$ cross section to 
determine $G_{En}$ is that of Platchkov {\it et al.} \cite{platchkov}. 
These data extend to $Q^2$ of 0.7 
GeV$^2$, with significantly smaller statistical uncertainties than all previous 
experiments. The data from Platchkov {\it et al.} \cite{platchkov} are shown in 
Fig. \ref{fig:genplatchkov}.
The FF $A(Q^2)$ is very sensitive to the deuteron
wave function, and therefore to the $NN$ interaction. Furthermore, the shape of 
$A(Q^2)$ cannot be explained by the IA alone. Corrections for  meson exchange
currents (MEC) and a small contribution from relativistic effects were found 
to significantly improve the 
agreement between calculations and the shape of $A(Q^2)$ observed. The 
authors included the constraint from the slope of the neutron electric FF as determined in $ne$
scattering, which at the time was $dG_{En}/dQ^2$=0.0199 fm$^2$ \cite{koester}; they 
proposed a modified form of the Galster fit using several $NN$ potentials, and including MEC 
as well as relativistic 
corrections, of the form $G_{En}(Q^2)=-a\frac{\mu_n \tau}{1+b\tau}G_{Ep}(Q^2)$,
corresponding to a slope at $Q^2=0$: $dG_{En}/dQ^2=-a\mu_n/4M^2$.
For the Paris $NN$ potential for example, $a$=1.25 and $b$=18.3; this fit 
will be compared with the double-polarization
data shown later in this review, in Fig. \ref{fig:gen_comp}.
Starting in 1994, all $G_{En}$ measurements have used either polarization 
transfer or beam-target asymmetry to take advantage of the interference 
nature of these observables: terms proportional to $G_{En}G_{Mn}$ are 
measured, instead
of the $G_{En}^2$ contribution to the cross section; these experiments 
will be reviewed in section \ref{DoubleExp}.

\subsubsection{Neutron magnetic form factor measurements}
\label{NeutronMFF}

In an early experiment Hughes {\it et al.} \cite{hughes} performed a 
Rosenbluth separation of quasi elastic $d(e,e')$  cross sections in the range 
$Q^2=0.04$ to 1.17 GeV$^2$; they observed non-zero values of $G_{En}$ only 
below 0.2 GeV$^2$ but measured $G_{Mn}$ up to 1.17 GeV$^2$; the technique 
consisted in comparing quasi-elastic $ed$- with elastic $ep$ cross 
sections. The several experiments following Hughes'  
can be subdivided into 3 groups: cross 
section measurements in quasi-elastic $ed$ scattering (single arm) 
\cite{hanson,bartel}, 
which requires large final state interaction (FSI) corrections at small $Q^2$; 
elastic $ed$ cross
section measurements \cite{benaksas,grossetete}; and cross 
section measurements in $d(e,e'p)n$ \cite{budnitz,dunning}, or ratio of 
cross sections
$d(e,e'n)p/d(e,e'p)n$ \cite{stein}, which is less sensitive 
to the deuteron wave function, and MEC. 
\begin{figure}[h]
\begin{minipage}[b]{0.45\linewidth}
\begin{center} 
%\hspace{-0.7cm}
\epsfig{file=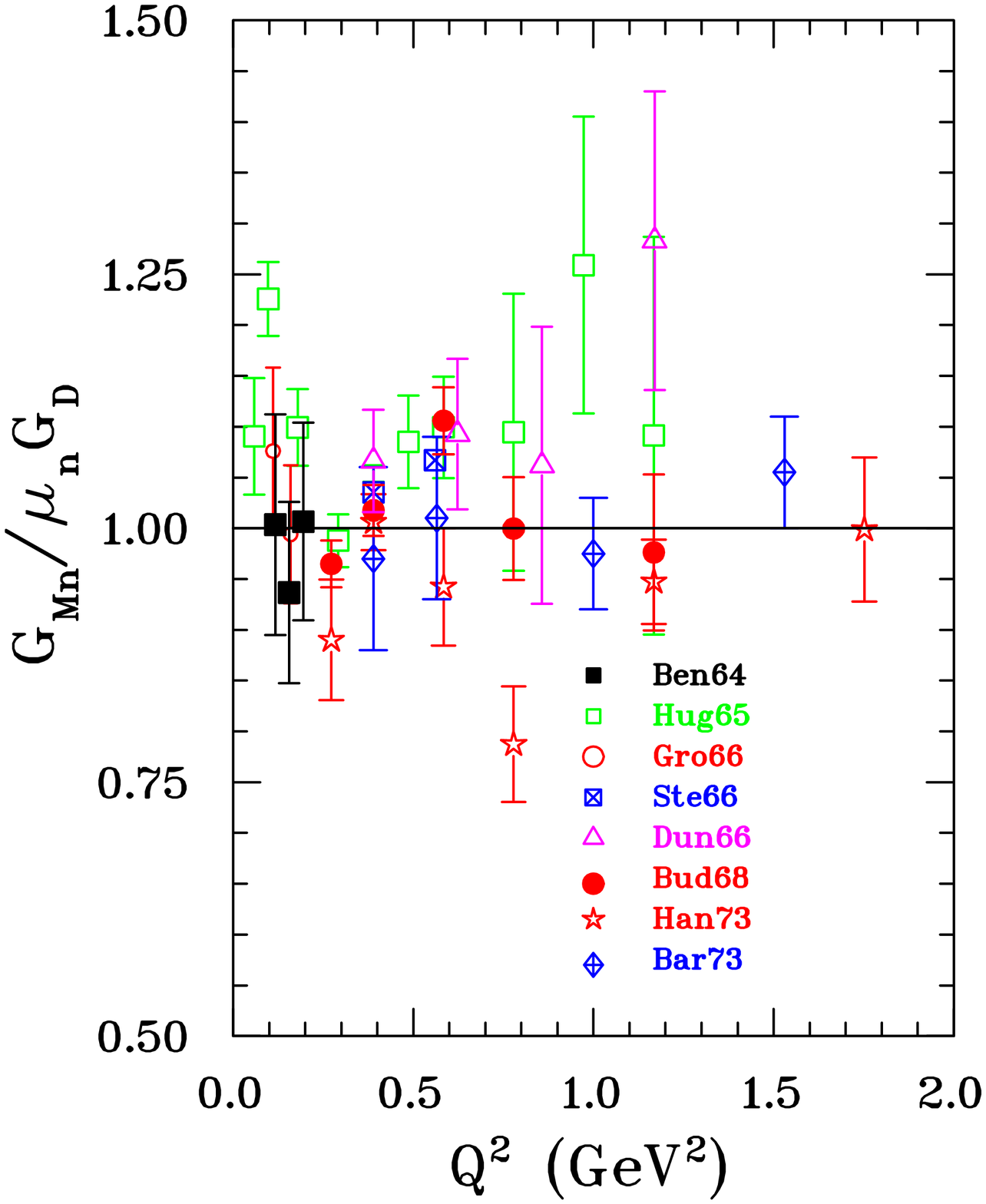,height=7.85cm}
\vspace{0.5cm}
\caption[]{\small{Early Rosenbluth separation data for $G_{Mn}$, up to 1973 \cite{
benaksas,hughes,grossetete,stein,dunning,budnitz,hanson,bartel}.}}
\label{fig:GMNEARLY}
\end{center}
\end{minipage}\hfill
\begin{minipage}[b]{0.45\linewidth}
\begin{center}
\epsfxsize=\textwidth
\epsfig{file=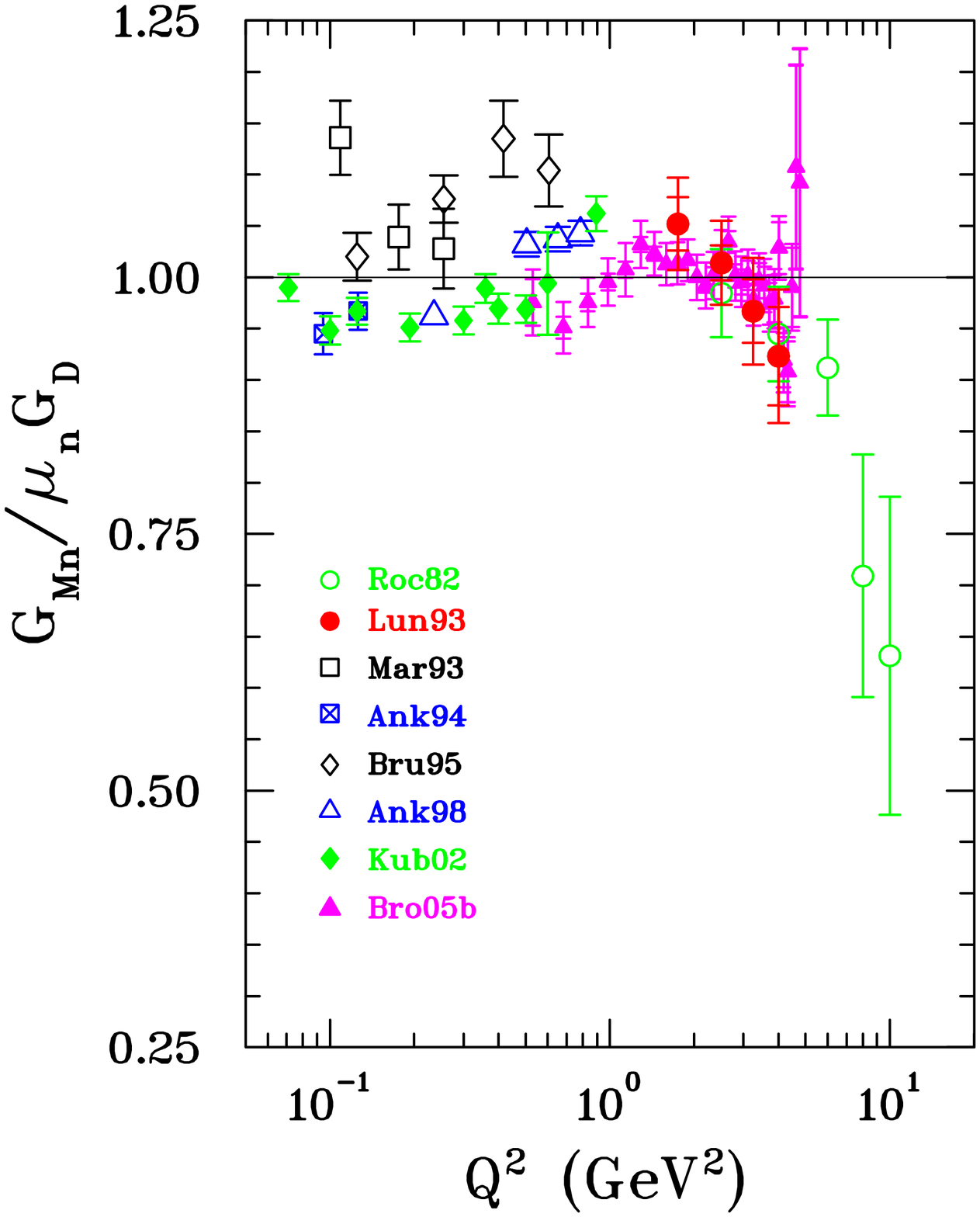,height=7.85cm}
\caption[]{\small{Recent $G_{Mn}$ data divided by $\mu_nG_D$, from cross section data only, 
starting in 1992 \cite{rock,lung,marko,anklin,bruins,anklin2,kubon,brooks}.}}
\label{fig:gmncsonly}
\end{center}
\end{minipage}
\end{figure}
All results published prior to 1973 are displayed in Fig. 
\ref{fig:GMNEARLY}, to be compared with the proton data from the same period in 
Fig.~\ref{fig:earlysep}. All more recent cross section results are in Fig. \ref{fig:gmncsonly},
allowing for a comparison of the progress made in this period
for the neutron. In Fig.~\ref{fig:gmncsonly}
all $G_{Mn}$ data obtained from cross section measurements are displayed,
including the SLAC experiments \cite{rock,lung}, which measured inclusive 
quasi-elastic $ed$ cross sections. The more recent
ELSA \cite{bruins} and MAMI \cite{anklin,anklin2,kubon} experiments 
are simultaneous measurements of the cross section for quasi elastic scattering 
on the neutron and proton in the deuteron,  
$d(e,e'n)p$ and $d(e,e'p)n$; the systematics is then dominated by
the uncertainty in the neutron detector efficiency; much attention was
given to that calibration in these experiments. In the ELSA experiment 
\cite{bruins} protons and neutrons were detected in the same scintillator, 
and the neutron efficiency was determined {\it in situ} with the neutrons from 
$^1H(\gamma,\pi)$. It has been argued in \cite{jourdan} that 3-body
electro-production contributes significantly and does not necessarily
lead to a neutron at the 2-body kinematic angle; these data points
are shown as $\diamondsuit$ in Fig.\ref{fig:gmncsonly}; a refutation
of these arguments is in \cite{bruins1}. In the Mainz experiments 
the dedicated neutron detector was calibrated in a neutron beam at SIN. The new data from Hall B at 
JLab, \cite{brooks}, are shown as filled triangles in Fig. \ref{fig:gmncsonly}; for these data from Hall B, in 
addition to the measurements of cross section ratio with the $^2H$ target, an in-line $^1H$ target was used 
for an {\it in-situ} determination of the neutron counter efficiency via $\pi^+$ electro-production.    

\subsubsection{Rosenbluth results and dipole form factor}
\label{DipoleFF}

\begin{figure}[t]
\begin{minipage}[b]{0.32\linewidth}
\begin{center}
\epsfig{file=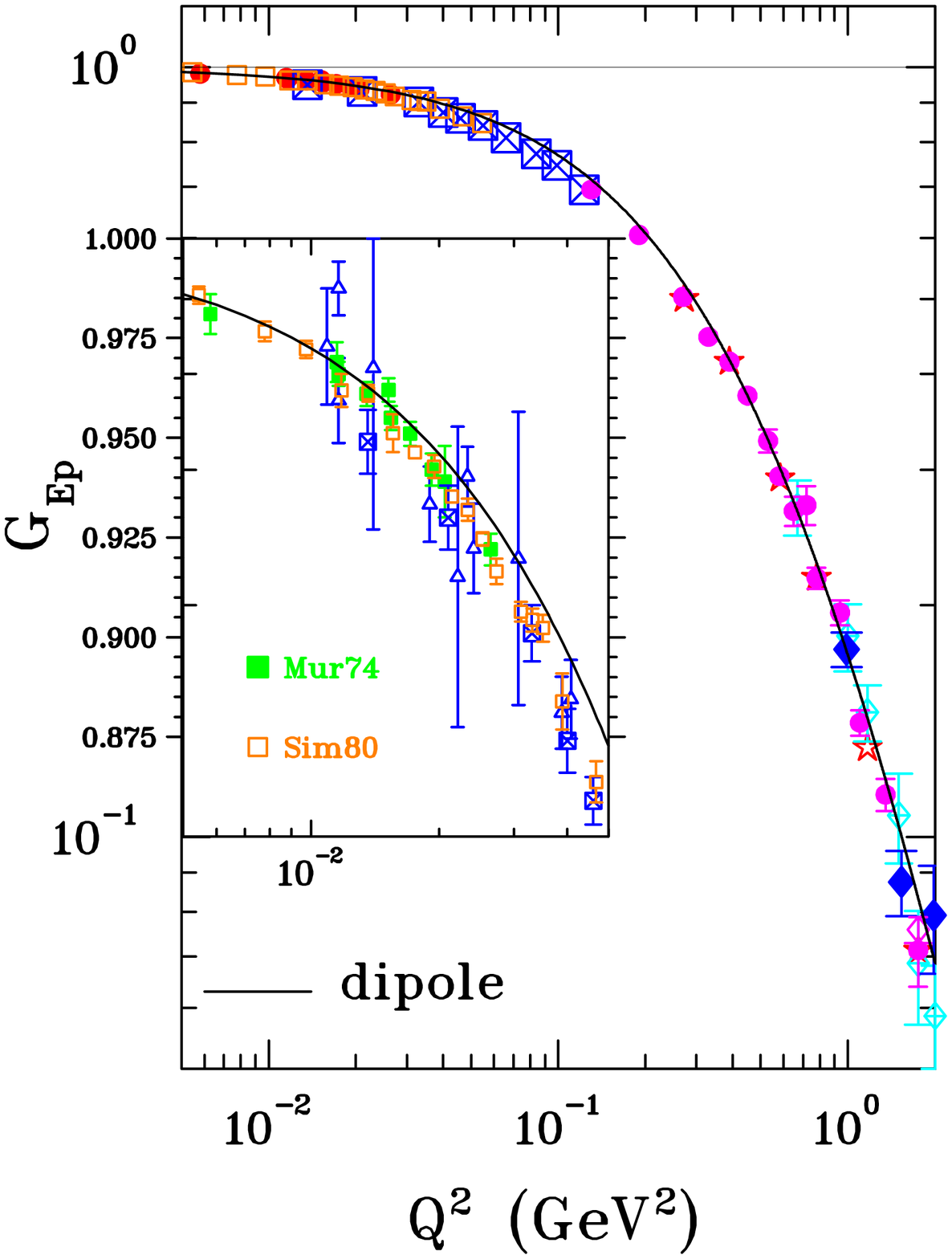,height=2.8in}
\caption{\small{Illustration of the quality of the dipole fit for $G_{E{p}}$; 
data from Refs. \cite{hand63,litt,price,berger,hanson,bartel,murphy,bork,
simon,walker,andivahis} in $Q^2$ range 0.005-2.0 GeV$^2$.}}
\label{fig:geplt2}
\end{center}
\end{minipage}\hfill
\begin{minipage}[b]{0.32\linewidth}
\begin{center}
\epsfxsize=\textwidth
%vspace{-1.0cm}
\epsfig{file=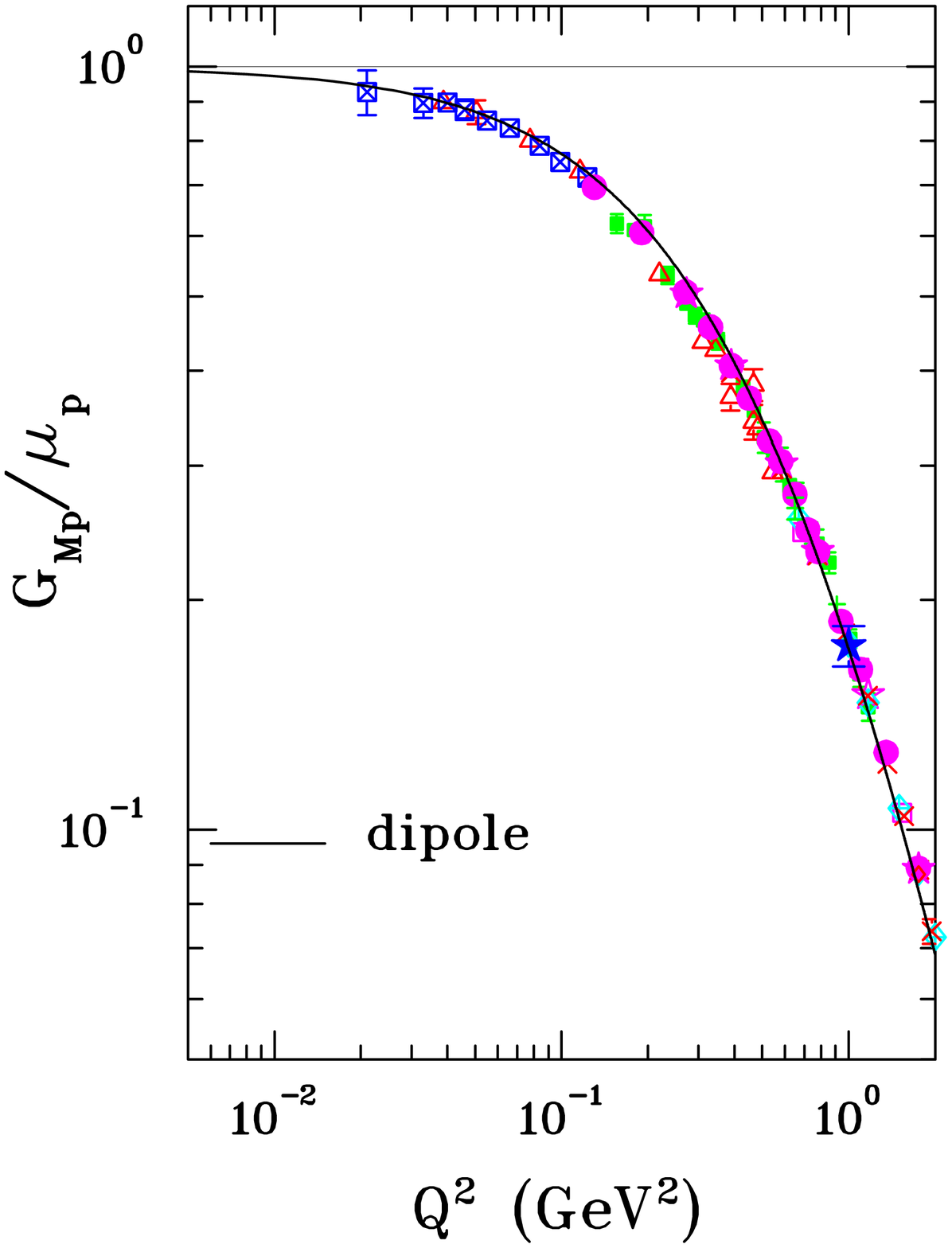,height=2.8in}
%\vspace{-0.3cm}
\caption{\small{Illustration of the quality of the dipole fit for $G_{Mp}$; the data of Refs.
\cite{hand63,janssens,litt,price,berger,hanson,bartel,bork,walker} in the 
range 0.005-2.0 GeV$^2$ are included here.}}
\label{fig:gmplt2}
\end{center}
\end{minipage}\hfill
\begin{minipage}[b]{0.32\linewidth}
\begin{center}
\epsfxsize=\textwidth
\epsfig{file=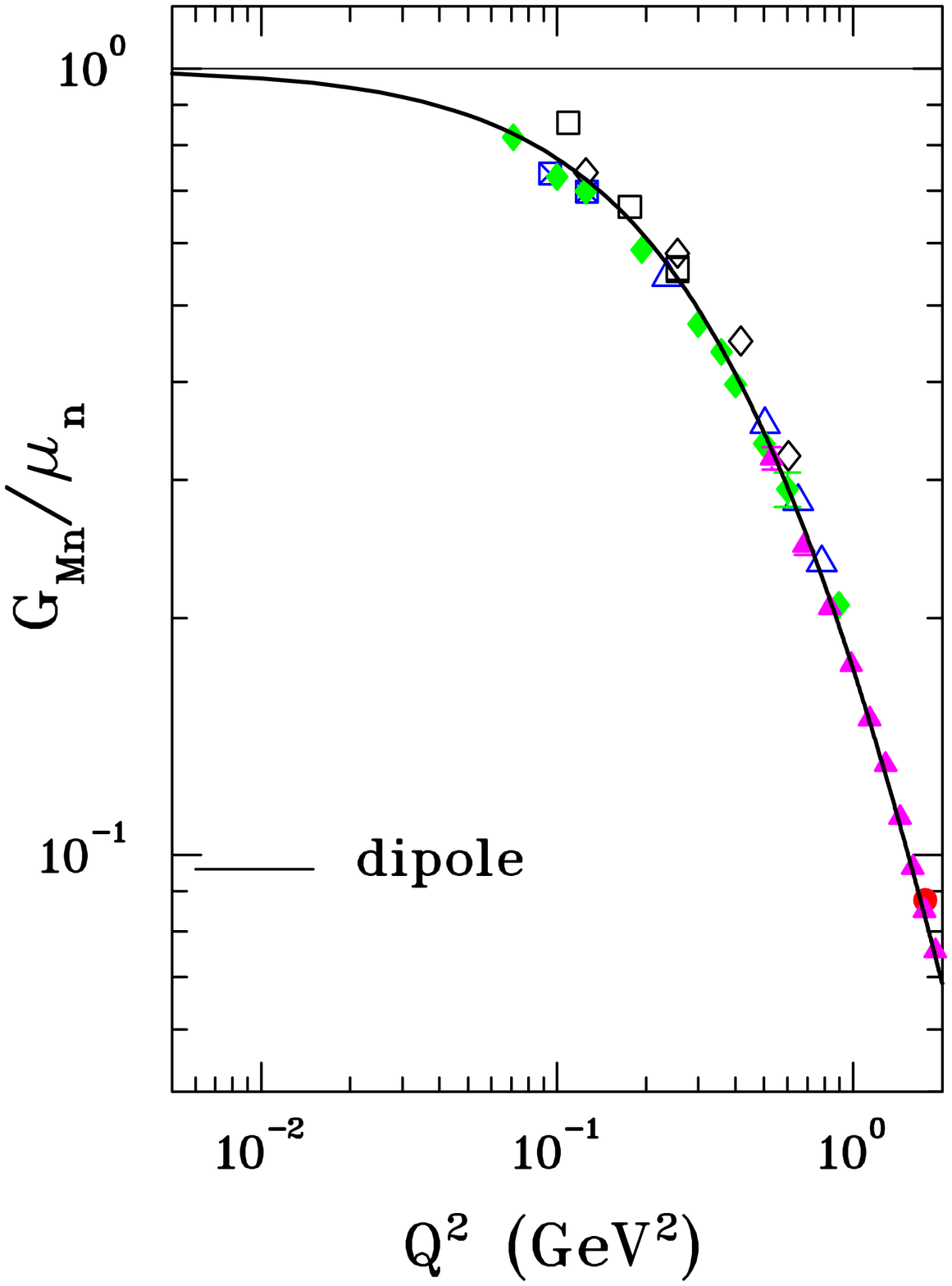,height=2.8in}
%\vspace{0.46cm}
\caption{\small{Illustration of the quality of the dipole fit for $G_{M{n}}$; 
the data included are the same as in Fig. \ref{fig:gmncsonly}, Refs. \cite{rock,lung,marko,anklin,bruins,anklin2,kubon,brooks} in 
the $Q^2$-range 0.005-2.0 GeV$^2$.}}
\label{fig:gmnlt2}
\end{center}
\end{minipage}
\end{figure}

In figures \ref{fig:geplt2}, \ref{fig:gmplt2} and \ref{fig:gmnlt2} the Rosenbluth
separation results  $G_{Ep}$, $G_{Mp}$ and $G_{Mn}$   
are shown in double logarithmic plots for $Q^2 < 2$ GeV$^2$, to emphasize 
the good agreement of these data with the dipole formula of Eq.~\ref{eq:dipole}. 

Noticeable is the lack of $G_{Mp}$ and $G_{Mn}$ data below $Q^2$ of 0.02 GeV$^2$, a 
consequence of the dominance of the electric FF at small $Q^2$ for the proton,
as seen in Eq. (\ref{eq:rosenb}).  

Although Hofstadter was the first to note that the proton FF data could be fitted by an 
``exponential model'', which corresponds to the ``dipole model'' for FFs in momentum space, it 
appears that the usage of dividing  data by $G_D$ was introduced first by Goitein {\it et al.} 
\cite{goitein}. 

The possible origin of the dipole FF has been discussed in a number of early papers. Within
the framework of dispersion theory the isovector and isoscalar parts of a FF is written as,  
$G_{E,M}^{V,S}=\Sigma_{i}\frac{\alpha_i^{V,S}}{1+Q^2/(M_i^{V,S})^2}$, where 
$G_{E,M}^V,\mbox{ }G_{E,M}^S$ 
are defined in footnote 2, and $M_i^{V,S}$ and $\alpha_i^{V,S}$ 
are the masses and residua of the isovector-, isoscalar vector mesons, respectively. A dipole
term occurs when the contribution of two vector mesons with opposite residua but similar 
masses are combined.

\section{Nucleon form factors from double polarization observables}
\label{DoublePol}
   
It was pointed out in 1968 by Akhiezer and Rekalo \cite{akh1} that ``for large 
momentum transfers the isolation of the charge FF of the proton is difficult'' using the elastic
$ep$ reaction with an unpolarized electron beam, for several reasons: 
one being $G_{Mp}^2/G_{Ep}^2 \geq \mu_{p}^2$ at any $Q^2$ value and the other is that at large 
$Q^2$ the contribution from the $\tau G_{Mp}^2$ term in $G_{Ep}^2+ (\tau/\epsilon) G_{Mp}^2$ increases  
(see Eq.(~\ref{eq:rosenb})) and makes the separation of the charge form 
factor practically impossible. In the same paper the authors also pointed out that the best way to 
obtain the proton charge FF is 
with polarization experiments, especially by measuring the polarization of the recoil proton. 
Further in a review paper in 1974 Akhiezer and Rekalo \cite{akh2} discussed specifically 
the interest of measuring an interference term of the form  $G_{E}G_{M}$ by measuring the transverse 
component of the recoiling proton polarization in the $\vec e p\rightarrow e \vec p$ reaction 
at large $Q^2$, to obtain $G_E$ in the presence of a dominating $G_M$.
In 1969, in a review paper Dombey \cite{dombey} also discussed the virtues of measuring 
polarization observables in elastic and inelastic lepton scattering; however 
his emphasis was to do these measurements with a polarized lepton on a polarized target. 
Furthermore in 1982 Arnold, Carlson and Gross \cite{arnold}emphasized that the best way 
to measure the electric FF
of the neutron would be to use the $^2H( \vec e,e' \vec n)p$ reaction. 
Both a polarized target, and a focal plane polarimeter (to measure recoil polarization), 
have been used to obtain nucleon FFs. We discuss below both methods to measure the elastic 
nucleon FFs, highlighting advantages and disadvantages of using polarized target and focal plane 
polarimeter.

\subsection{Polarization transfer}
\label{PolTransf}

\begin{figure}[h]
\begin{minipage}[b]{0.45\linewidth}
%\begin{center}
%\epsfig{file=epkin_recoil.eps,width=8.1cm}
\centerline{\epsfig{file=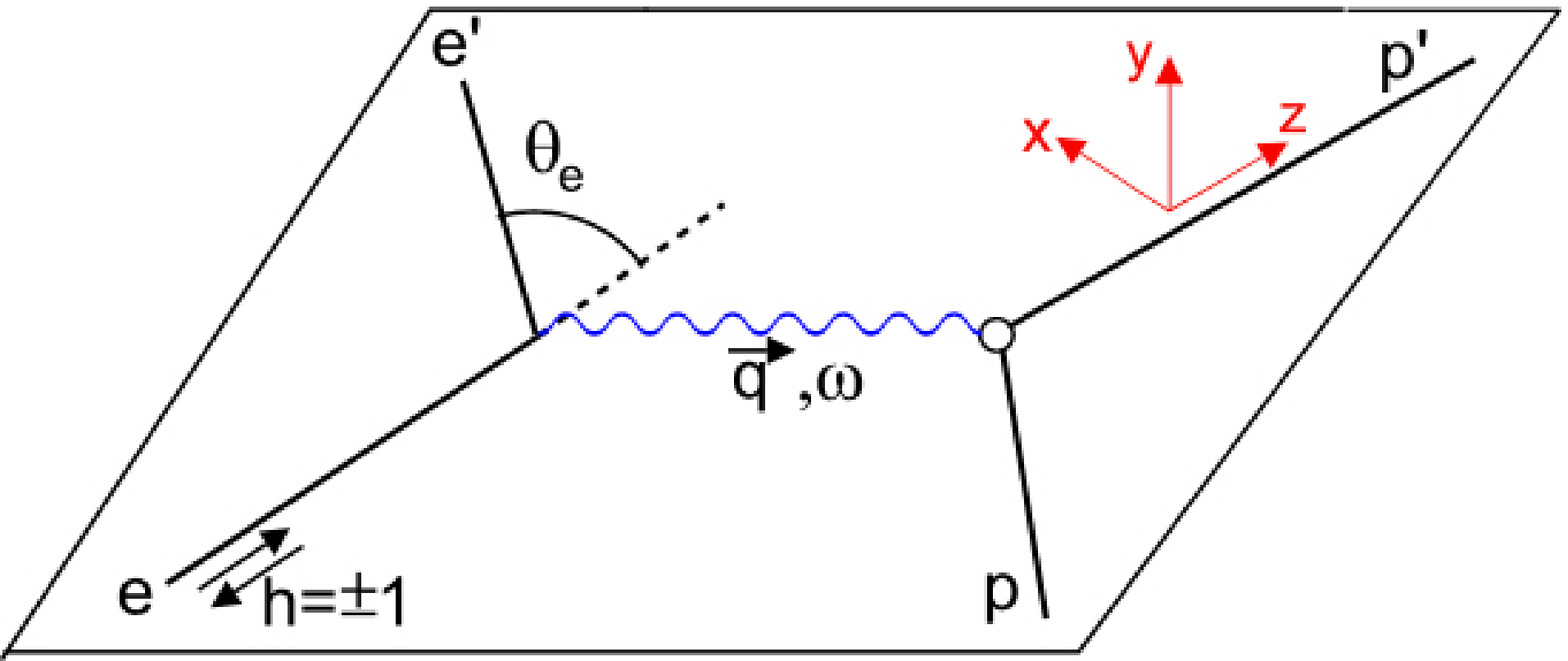,width=8cm}}
\vspace{0.2in}
\caption [] {\small{Kinematical variables for polarization transfer from a longitudinally 
polarized electron to a proton with exchange of a virtual photon.}}
\label{fig:nlt}
%\end{center} 
\end{minipage}\hfill
\begin{minipage}[b]{0.45\linewidth}
\epsfxsize=\textwidth
\centerline{\epsfig{file=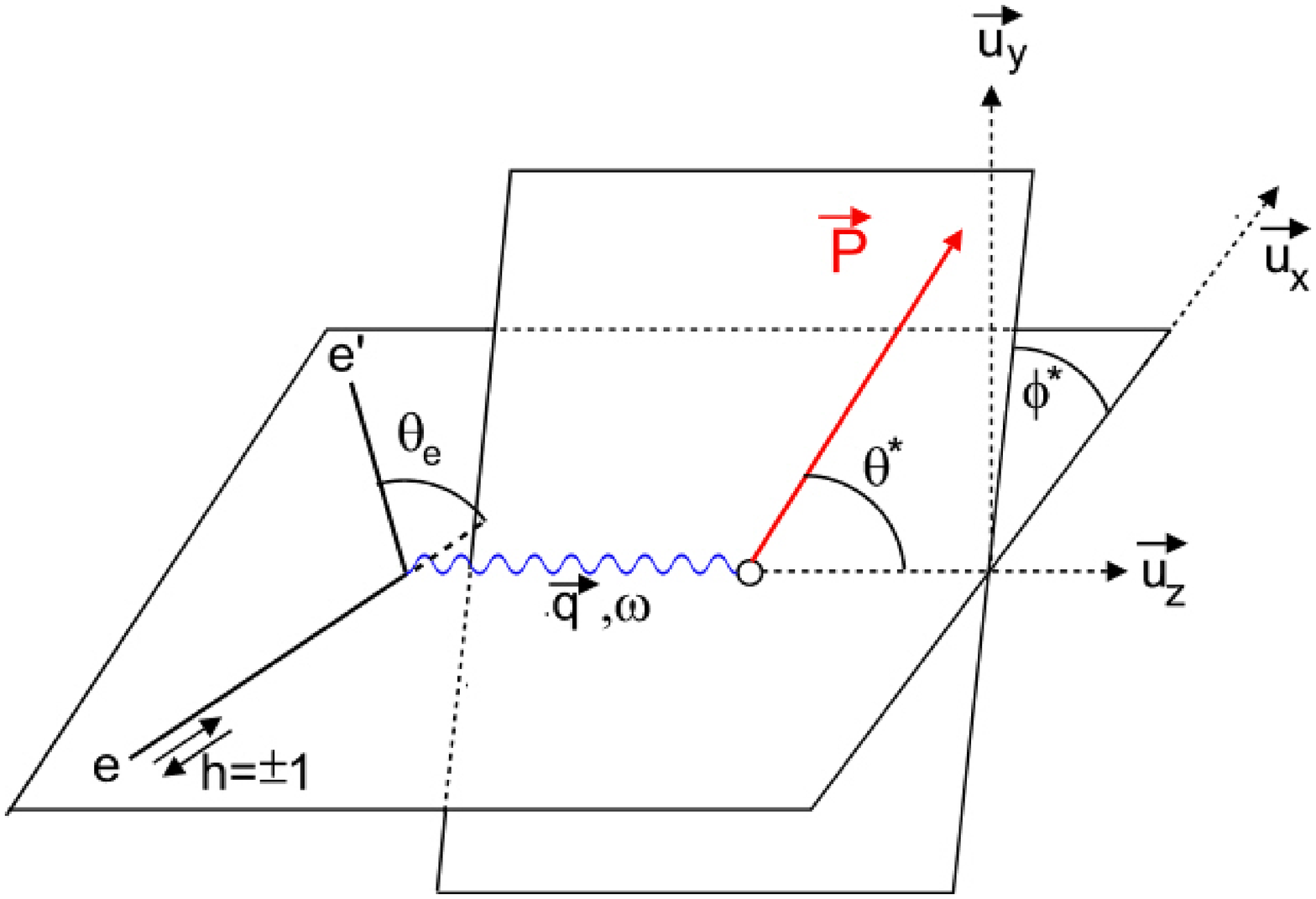,width=8.1cm}}
\vspace{0.5cm}
\caption{\small{Polarized electron scattering from a polarized target.}}
\label{fig:epkin_asym}
\end{minipage}
\end{figure}

Figure \ref{fig:nlt} shows the kinematical variables for the polarization transfer from a 
longitudinally polarized electron  to a struck proton in the one-photon exchange approximation. 

The electron vertex in Fig.~\ref{fig:nlt} can be described by basic Quantum Electrodynamics 
(QED) rules that involves the electron current, $\ell_{\mu}$, and the proton vertex can be 
described by QCD and hadron electrodynamics involving the hadronic current $J^{\mu}=\chi'^{\dagger}F_{\mu}\chi$. 

For elastic $ep$ scattering with longitudinally polarized electrons, the hadronic 
tensor, $W_{\mu \nu}$= $J_{\mu}J_{\nu}^*$, has four possible terms depending upon the 
polarization of the target and of the recoil proton:
\begin{equation}
W_{\mu \nu}~=~W_{\mu \nu}^{(o)}+W_{\mu \nu}({\bf P_1})+W_{\mu \nu}({\bf P_2}) +W_{\mu \nu}({\bf P_1,P_2}),
\end{equation}
\noindent
where the first term in the equation corresponds to unpolarized protons, 
the second and the third term correspond to the vector polarization of the initial 
and the final proton, respectively, and the last term describes the reaction when both, 
the initial and the final protons are polarized. 

Considering the case where only the polarization of the final proton is measured, $W_{\mu \nu}({\bf P})$ is:
\begin{equation}
W_{\mu \nu}({\bf P})~=~ \frac{1}{2}TrF_{\mu} F_{\nu}^{\dagger}\vec \sigma\cdot {\bf P}.
\end{equation}
\noindent
For the scattering of longitudinally polarized electrons off an unpolarized target, in the one-photon-exchange 
approximation, there are only two non-zero polarization components, transverse, $P_x$, 
and longitudinal, $P_{z}$; and these components are obtained by calculating the tensors 
$W_{\mu \nu}(P_x) = \frac{1}{2}TrF_{\mu} F_{\nu}^{\dagger} \sigma_x \sim 2m G_E G_M$ and 
$W_{\mu \nu}(P_z) = \frac{1}{2}TrF_{\mu} F_{\nu}^{\dagger} \sigma_z \sim G_M^2$.

The transformation from Breit to laboratory frame gives following expressions for the polarization components 
$P_x$ and $P_{z}$ in terms of the electric $G_{E}$, and magnetic, $G_{M}$ FF \cite{akh2,arnold}; 

\begin{eqnarray}
I_{0}P_{x}&=&-2\sqrt{\tau (1+\tau )}G_{E}G_{M}\tan \frac{\theta_{e}}{2}
\label{eq:px} \\
I_{0}P_{z}&=&\frac{1}{M}(E_{beam}+E_{e})\sqrt{\tau (1+\tau )}
G_{M}^{2}\tan ^{2}\frac{\theta_{e}}{2},  
\label{eq:pz}
\end{eqnarray}
\noindent
where $E_{beam}$ and $E_e$ are the energy of the incident and scattered electrons, 
respectively, $\theta_e$ is the scattered electron angle in the laboratory frame,  and $I_0$ is:
\begin{equation}
I_{0}=G_{E}^{2}(Q^{2})+\frac{\tau}{\epsilon} G_{M}^{2}(Q^{2}).
\end{equation}
\noindent
Eqs.~(\ref{eq:px}) and ~(\ref{eq:pz}) show that the transverse and longitudinal
polarization components are proportional to $G_{E}G_{M}$ and 
$G_{M}^2$, respectively. The ratio $G_{E}/G_{M}$ then can be obtained directly from the 
ratio of the two polarization components $P_x$ and $P_z$ as follows: 
\begin{equation}
\frac{G_{E}}{G_{M}}=-\frac{P_{x}}{P_{z}}\frac{(E_{beam}+E_{e})} 
{2M}\tan (\frac{\theta_{e}}{2}).
\label{eq:ratio}
\end{equation}
\noindent
Equation~(\ref{eq:ratio}) makes clear that this method offers several experimental advantages
over the Rosenbluth separation: (1) for a given $Q^2$, only a single measurement
is necessary, if the polarimeter can measure both components at the same time.
This greatly reduces the systematic errors associated with angle and beam energy change, and 
(2) the knowledge of the beam polarization and of the analyzing power of the 
polarimeter is not needed to extract the ratio, $G_{E}/G_{M}$.

\subsection{Asymmetry with polarized targets}
\label{Asymmetry}

It was pointed out by Dombey \cite{dombey} that the nucleon FFs 
can be extracted from the scattering of longitudinally polarized electrons off
a polarized nucleon target. In the one-photon-exchange approximation, following the 
approach of Donnelly and Raskin \cite{donnelly,raskin}, the
elastic $eN$ ($N=p$ or $n$) cross section can be written as a sum of an unpolarized part and a 
polarized part; the latter is non-zero only if the electron beam is 
longitudinally polarized:
\begin{equation} 
\sigma^{pol} = \Sigma + h \Delta, \\
\label{eq:ed_cs_pol}
\end{equation}
\noindent
where $h$ is the electron beam helicity, $\Sigma$ is the elastic un-polarized 
cross section given by Eq. (\ref{eq:rosenb}), and $\Delta$ is the polarized part
of the cross section with two terms related to the directions of the target polarization.
The expression for $\Delta$ can be written as \cite{donnelly,raskin}: 
\begin{equation}
\Delta = -2 \sigma_{Mott} \tan(\theta_e/2) \sqrt{\frac{\tau}{1+\tau}}
{ \Biggl \{\sqrt{\tau[1+(1+\tau)\tan^2(\theta_e/2)]}\cos\theta^{\ast}G_M^2+ 
\sin\theta^{\ast}\cos\phi^{\ast}G_E G_M \Biggr \}}
\label{eq:delta}
\end{equation}
\noindent
where $\theta^{\ast}$ and $\phi^{\ast}$ are the polar and 
azimuthal laboratory angles of the target polarization vector with $\vec q$ in the $\vec u_z$
direction and $\vec u_y$ normal to the electron scattering plane, as shown 
in Figure \ref{fig:epkin_asym}.

The physical asymmetry $A$ is then defined as 
\begin{equation}
A=\frac{\sigma_{+} -\sigma_{-}}{\sigma_{+} + \sigma_{-}}=\frac{\Delta}{\Sigma},
\label{eq:asy}
\end{equation}
\noindent
where $\sigma_+$ and $\sigma_-$ are the cross sections for the two beam helicities. 

For a polarized target, the measured asymmetry, $A_{meas}$, is 
related to the physical asymmetry, $A$, by 
\begin{equation}
A_{meas}=P_{beam}P_{target}A,
\label{eq:asy1}
\end{equation}
\noindent
where $P_{beam}$ and $P_{target}$ are electron beam- and target polarization, respectively, and,   
\begin{eqnarray}
A&=&
-\frac{2\sqrt{\tau(1+\tau)}\tan(\theta_e/2)}{G_E^2+\frac{\tau}{\epsilon}G_M^2}
\Big[ \sin\theta^{\ast}\cos\phi^{\ast}G_E G_M 
%\nonumber \\
+ \sqrt{\tau[1+(1+\tau)\tan^2(\theta_e/2)]} \cos\theta^{\ast}G_M^2 \Big].
\label{eq:asy2}
\end{eqnarray}
\noindent
It is evident from Eq. (\ref{eq:asy2}) that to extract $G_{E}$, the target polarization 
in the laboratory frame must be perpendicular with respect to the momentum transfer vector ${\vec q}$
and within the reaction plane, with $\theta^{\ast}= \pi/2$ and $\phi^{\ast}= 0^o$ or $180^o$. 
For these conditions, the asymmetry $A$ in Eq. (\ref{eq:asy2}) simplifies to:
\begin{equation}
A_{perp}=\frac{-2\sqrt{\tau(1+\tau)}\tan(\theta_e/2) \frac{G_E}{G_M}}{(\frac{G_E}{G_M})^2+\frac{\tau}{\epsilon}}.
\label{eq:asy3}
\end{equation}
\noindent
As $(G_E/G_M)^2$ is quite small, $A_{perp}$ is approximately proportional to $G_E/G_M$. In practice, the second term in 
Eq. (\ref{eq:asy2}) is not strictly zero due to the finite acceptance of the detectors, but these effects are small and depend
on kinematics only in first order and can be corrected for, so the ratio $G_E/G_M$ is not affected directly.

The discussion described above is only applicable to a free nucleon; corrections are required 
if nuclear targets, like $^2$H or $^3$He, are used instead in quasi-elastic scattering  
to obtain the FFs.

\subsection{Double polarization experiments}
\label{DoubleExp}

The polarization method, using polarized targets and focal plane polarimeter 
with longitudinally polarized electron beam, has been used to measure both the proton and the neutron 
e.m. FFs. Below we first describe the polarization experiments 
that measured the proton FFs and next those that measured the neutron FFs. 

\subsubsection{Proton form factor measurements with polarization experiments}
\label{ProtonFFpol}

The first experiment to measure the proton polarization observable in $ep$ elastic scattering was 
done at the Stanford Linear Accelerator Center (SLAC) by Alguard {\it et al.} \cite{alguard}. 
They measured the anti-parallel-parallel asymmetry in the differential cross sections by 
scattering longitudinally polarized electrons on polarized protons. From their 
result they concluded that the signs of $G_{Ep}$ and $G_{Mp}$ are the same; they also noted
that the usefulness of using polarized beam on polarized
target is severely limited by low counting rates. 

Next, the recoil polarization method to measure the proton e.m. FF was 
used at MIT-Bates laboratory \cite{milbrath,barkhuff}. In this experiment the proton 
FF ratio $G_{Ep}/G_{Mp}$ was obtained for a free proton and a bound proton in a deuterium 
target at two $Q^2$ values, 0.38 and 0.5 GeV$^2$ using polarization transfer from longitudinally
polarized electron to the proton in the target, and measuring the polarization of 
the recoiling proton with a focal plane polarimeter (FPP). The conclusion from these 
measurements was that 
the polarization transfer technique showed great promise for future measurements of $G_{Ep}$ and 
$G_{En}$ at higher $Q^2$ values. 

The ratio $G_{Ep}/G_{Mp}$ in elastic $^1H(\vec e,e' \vec p)$ 
was also measured at the MAMI in a dedicated experiment 
\cite{pospischil}, and as a calibration measurement \cite{dietrich} at 
a $Q^2$ of $\approx$0.4 Gev$^2$. The ratio values found were in agreement with other polarization 
measurements as well as Rosenbluth measurements. Most recently the BLAST group at MIT-Bates \cite{crawford} 
measured the ratio $G_{Ep}/G_{Mp}$ at $Q^2$ values of 0.2 to 0.6 GeV$^2$ with high precision.

Starting in late 1990's, the proton FF ratios $G_{Ep}/G_{Mp}$ were measured in two successive dedicated experiments 
in Hall A at JLab for $Q^{2}$ from 0.5 to 5.6~GeV$^{2}$~\cite{jones,gayou2,punjabi05}. 
Other measurements were also conducted in Hall A \cite{gayou1,strauch,hu} at lower $Q^2$ values, as 
calibration measurements for other polarization experiments, and one measurement in Hall C \cite{mac}. 

In the first JLab experiment the ratio, $G_{Ep}/G_{Mp}$, was measured  
up to $Q^{2}$ of 3.5~GeV$^{2}$~\cite{jones,punjabi05}. Protons
and electrons were detected in coincidence in the two high-resolution
spectrometers (HRS) of Hall A \cite{nimhallA}. The polarization of the recoiling 
proton was obtained from the asymmetry of the azimuthal distribution after
the proton re-scattered in a graphite analyzer.

The ratio, $G_{Ep}/G_{Mp}$, was measured at $Q^{2}$ = 4.0, 4.8 and 5.6~GeV$^{2}$ with an overlap 
point at $Q^{2}$ = 3.5~GeV$^{2}$ ~\cite{gayou2}, in the second JLab experiment. 
To extend the measurement to
higher $Q^2$, two changes were made from the first experiment. First, to increase the 
figure-of-merit (FOM) of the FPP, a CH$_{2}$ analyzer was used instead of graphite; hydrogen has 
much higher analyzing power \cite{spinka,dmiller} than carbon \cite{cheung}.  
Second,  the
thickness of the analyzer was increased from 50 cm of graphite to 100~cm of CH$_{2}$ to 
increase the fraction of events with a second scattering in the analyzer.
Third, the electrons were detected in a lead-glass calorimeter with
a large frontal area, to achieve complete solid angle matching with the 
HRS detecting the proton. At the largest $Q^{2}$ of 5.6 GeV$^2$ the solid
angle of the calorimeter was 6 times that of the HRS. 

Proton polarimeters are based on nuclear scattering
from an analyzer material like graphite or CH$_{2}$; 
the proton-nucleus spin-orbit interaction and proton-proton spin dependent interaction
results in an azimuthal asymmetry in the scattering distribution which can
be analyzed to obtain the proton polarization. 
The detection probability for a proton scattered by the analyzer
with polar angle $\vartheta$ and azimuthal angle $\varphi$ is given by:
\begin{equation}
f^\pm(\vartheta,\varphi) = \frac{\epsilon(\vartheta,\varphi)}{2\pi} 
\left (1 \pm A_y(P_x^{fpp}\sin{\varphi} - P_y^{fpp}\cos{\varphi}) \right ), 
\label{eq:azimuth}
\end{equation}
\noindent
where $\pm$ refers to the sign of the beam helicity, $P_{x}^{fpp}$ and $P_{y}^{fpp}$ are 
transverse and normal polarization components in the reaction plane at the analyzer, 
respectively, and $\epsilon(\vartheta,\varphi)$ is an instrumental asymmetry that 
describes a non-uniform detector response. 
Physical asymmetries are obtained from the difference distribution of $f^{\pm}$, 
\begin{eqnarray}
D_i &=& (f^{+}_i - f^{-}_i)/2 ~= ~\frac{1}{2 \pi}\left(A_y P_x^{fpp}\sin{\varphi_i} - A_y P_y^{fpp}\cos{\varphi_i} 
\right ),
\end{eqnarray}
\noindent 
and the sum distribution of $f^{\pm}$ separates the instrumental asymmetries $\epsilon_i$,
\begin{eqnarray}
E_i &=& (f^{+}_i + f^{-}_i)/2 ~= ~\frac{\epsilon_i}{2\pi}.  
\end{eqnarray}
The values of the two asymmetries at the FPP, $A_{y}P_{x}^{fpp}$ and
$A_{y}P_{y}^{fpp}$, can be obtained by Fourier analysis of the difference distribution $D_i$;
however to calculate the ratio $G_{Ep}/G_{Mp}$, the proton polarization components 
$P_x$ and $P_{z}$ are needed at the target. 

As the proton travels from the target to the focal plane
through the magnetic elements of the HRS, its spin precesses, 
and therefore the polarization components at the FPP and at the target are different. 
The hadron HRS in Hall A consists of three quadrupoles and one dipole with
shaped entrance and exit edges, as well as a radial field gradient.
The polarization vectors at the polarimeter, $\vec P^{fpp}$, are related to 
those at the target, $h\vec P$, through a 3-dimensional spin
rotation matrix, $(\bf S)$, as follows:

\begin{center}

$\left( 
\begin{array}{c}
P_{y}^{fpp} \\ 
P_{x}^{fpp} \\ 
P_{z}^{fpp}
\end{array}
\right)
=
\left( 
\begin{array}{ccc}
S_{yy} & S_{yx} & S_{yz}  \\ 
S_{xy} & S_{xx} & S_{xz} \\ 
S_{zy} & S_{zx} & S_{zz}
\end{array}
\right) 
\left( 
\begin{array}{c}
P_{y} \\ 
P_{x} \\ 
P_{z}
\end{array}
\right)$.
\end{center}
\noindent
The spin transport matrix elements $S_{ij}$ can be calculated using a model of the HRS with quadrupoles, 
fringe fields, and radial field gradient in the dipole, for each tuning of the spectrometer setting, and 
event by event with 
the differential-algebra-based transport code COSY \cite{bertz}. The spin transport method to obtain the 
two asymmetries at the target, $hA_{y}P_{x}$ and $hA_{y}P_{z}$, was developed by Pentchev  and described in 
detail in \cite{pentchev}, and discussed in \cite{punjabi05}. 
The ratio $G_{Ep}/G_{Mp}$ was calculated from the two asymmetries at the target from Eq. (\ref{eq:ratio}).
The fact that both beam polarization, and polarimeter analyzing power cancel out of this equation contributes 
to the reduction of the systematic uncertainties, however their values do influence the statistical errors.

The most recent acquisition of the FF ratio $G_{Ep}/G_{Mp}$ has been made at a $Q^2$ of 1.51 GeV$^2$ 
by measuring the beam-target asymmetry in an experiment in Hall C at JLab in elastic $ep$ scattering \cite{Jones06}.
This is the highest $Q^2$ at which the $G_{Ep}/G_{Mp}$ ratio has been obtained from a beam-target asymmetry measurement. 

\begin{figure}[hbt]
\begin{minipage}[b]{0.45\linewidth}
\centerline{\epsfig{file=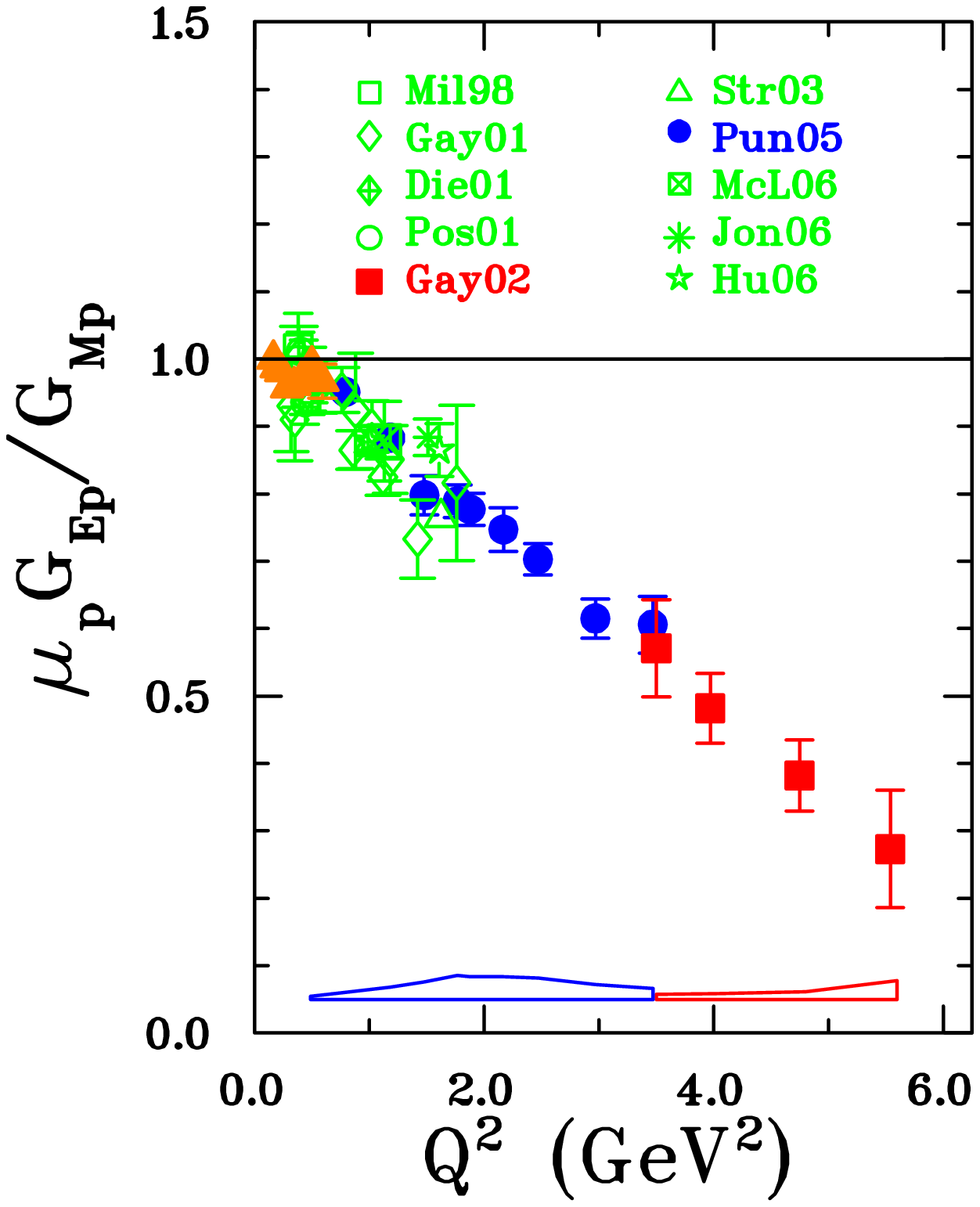,width=6.35cm}}
\vspace{0.5cm}
\caption{\small{The ratio $\mu_p G_{Ep}/G_{Mp}$ from the two JLab experiments 
\cite{gayou2,punjabi05} (filled circle and square), 
together with all other polarization 
transfer experiments \cite{milbrath,gayou1,pospischil,dietrich,strauch,hu,mac,Jones06}. The systematic 
uncertainties apply to the JLab data only.}}
\label{fig:gepgmp_pol}
\end{minipage}\hfill
\begin{minipage}[b]{0.45\linewidth}
\epsfxsize=\textwidth
\centerline{\epsfig{file=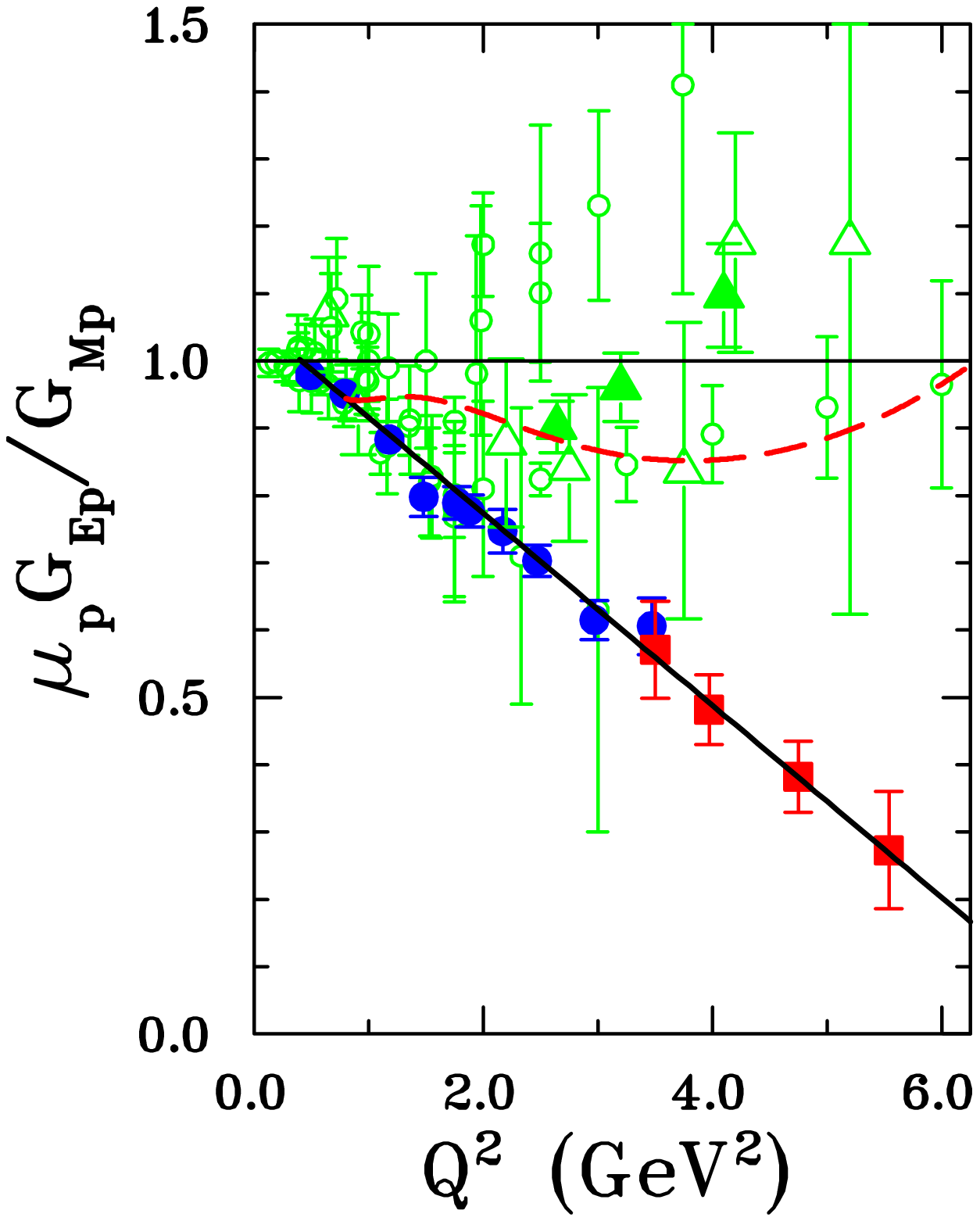,width=6.35cm}}
\caption{\small{Comparison of $\mu_p G_{Ep}/G_{Mp}$ from the two JLab polarization data 
\cite{gayou2,punjabi05}, and 
Rosenbluth separation; JLab Rosenbluth 
results  from \cite{christy,qattan05} shown as open, 
filled triangles, respectively. The rest of cross section data as 
in Figs.~\ref{fig:gepgd} and \ref{fig:gmpgd} (open circles). Dashed curve is a re-fit 
of Rosenbluth data~\cite{arring03}; solid curve is the fit of Eq. (\ref{eq:jlabfit}).}}
\label{fig:gepgmp_comp}
\end{minipage}
\end{figure}

The results from the two JLab experiments \cite{gayou2,punjabi05}, and other polarization  
measurements \cite{milbrath,gayou1,pospischil,dietrich,strauch,hu,mac,Jones06}, are plotted 
in Fig.~\ref{fig:gepgmp_pol} as the ratio $\mu_{p}G_{Ep}/G_{Mp}$ versus $Q^2$. All
data show only the statistical uncertainty; the systematic uncertainty for the data of \cite{gayou2,
punjabi05} are shown separately as a polygon; they are typical for all polarization data obtained
in Hall A at JLab. The new asymmetry 
data from BATES 
\cite{crawford} are not in this figure as they are in the range of $Q^2$-values smaller than 0.6 GeV$^2$; 
they appear in Fig.~\ref{fig:gepgdpol}. 
As can be seen from figure \ref{fig:gepgmp_pol}, data from different experiments are in excellent 
agreement and the statistical
uncertainty is small for all data points; this is unlike $G_{Ep}$ obtained from cross section data and shown 
in Fig. \ref{fig:gepgd}, where we see a large scatter in results  from different experiments as well 
as large statistical uncertainty at higher $Q^2$ values, underlining
the difficulties in obtaining $G_{Ep}$ by the Rosenbluth separation method.   

The results from the two JLab experiments \cite{jones,punjabi05,gayou2} showed 
conclusively for the first time a clear deviation of
the proton FF ratio from unity, starting at $Q^2\simeq 1$~GeV$^2$; older data from \cite{berger,price,bartel,hanson}
showed such a decreasing ratio, but with much larger statistical and systematic uncertainties, as seen in Fig. 
\ref{fig:earlysep}. 
The most important feature of the JLab data, is the sharp decrease of the ratio
$\mu_p G_{Ep}/G_{Mp}$ from 1 starting at $Q^2$ $\approx$ 1 GeV$^2$ to a value of
$\sim 0.28$ at $Q^2$= 5.6 GeV$^2$, which indicates that $G_{Ep}$ falls 
faster with increasing $Q^2$ than $G_{Mp}$. This was the first definite experimental indication
that the $Q^2$ dependence of $G_{Ep}$ and $G_{Mp}$ is different. 
If the $\mu _{p}G_{Ep}/G_{Mp}$-ratio continues the observed linear decrease with the 
same slope, it will cross zero at $Q^{2}\approx 7.5$~GeV$^{2}$ and become negative.

In Fig. \ref{fig:gepgmp_comp} all the ratio data 
obtained from the Rosenbluth separation method are plotted together with the results of the two JLab
polarization experiments. There are recent proton FF results obtained with the Rosenbluth 
separation method 
from two JLab experiments \cite{christy,qattan05}; these results agree with previous Rosenbluth results 
\cite {litt,berger,price,walker,andivahis} and confirm 
the discrepancy between the ratios obtained with the Rosenbluth separation method and 
the recoil polarization method. The two methods give definitively 
different results; the difference cannot be bridged by either simple re-normalization 
of the Rosenbluth data \cite{arring03}, or by variation of the polarization data within the 
quoted statistical and systematic uncertainties. This discrepancy has been known
for the past several years and is currently the subject of
intense discussion. A possible explanation is the hard two-photon exchange process, which affects 
both cross section and polarization transfer components at the level of only a few percents; 
however, in some calculations \cite{afanasev,blunden} the contribution of the two-photon 
process has drastic effect on the  Rosenbluth separation results, whereas in others 
it does not \cite{bystrit}; in either case it modifies the 
ratio obtained with the polarization method by a few percent
only (this will be discussed in section \ref{RadCorr}). There are several experiments planned 
at JLab \cite{lubomir,arring05} to investigate the two-photon effects in the near future.      

\subsubsection{Neutron electric form factor measurements with polarization experiments}
\label{NeutronFFpol}

Measurements of the FFs of the neutron are far more difficult than for the proton, 
mainly because there are no free neutron targets. Neutron FF measurements were started 
at about 
the same time as for the proton, but the data are generally not of the same quality as for the
proton, especially in the case of the electric 
FF of the neutron; the $Q^2$ range is limited also.
The early measurements of the FFs of the neutron are discussed in sections 
\ref{NeutronEFF} and \ref{NeutronMFF}; 
in this section we discuss only measurements made with longitudinally polarized electron beams on 
polarized $^2$H- or $^3$He-targets, and polarization transfer in the $^2$$H(\vec e,e'\vec n)p$ 
reaction. We start with the measurements of the charge FF, $G_{En}$, and proceed then to the
relatively few measurements of the magnetic FF, $G_{Mn}$.

The first measurement of the charge FF of the neutron, $G_{En}$,
by the polarization method was made at MIT-Bates using the exclusive $^2$$H(\vec{e},e'\vec{n})p$ 
reaction \cite{eden}. The advantage of polarization measurements on the deuteron in the quasi free 
kinematics is that the extracted neutron FF is quite insensitive to the choice of deuteron 
wave functions, and also to higher order effects like final state interaction (FSI), 
meson exchange currents (MEC) and isobar configurations (IC), when the momentum 
of the knocked out neutron is in the direction of three-momentum transfer $\vec q$ 
\cite{arenhovel,rekalo2,laget}.     

For a free neutron the polarization transfer coefficient $P_{x}$ is given by Eq. (\ref{eq:px}).
The relation between polarization transfer coefficient $P_{x}$,  
the beam polarization, $P_e$, and the measured neutron polarization component, $P'_x$, is 
$P'_x=P_e P_{x}$. The FF $G_{En}$ was extracted at a $Q^2$ of 0.255 GeV$^2$ 
in this experiment from the measured transverse polarization component $P'_x$ of the recoiling neutron, 
and known beam polarization, $P_e$. This 
early experiment  demonstrated the feasibility of extracting 
$G_{En}$ from the quasi-elastic $^2$$H(\vec{e},e'\vec{n})p$ reaction with the recoil polarization
technique, with the possibility of extension to larger $Q^2$ values.

Next, this same reaction $^2$$H(\vec{e},e'\vec{n})p$ was used to determine $G_{En}$ at  
MAMI \cite{herberg,ostrick} by measuring the neutron recoil polarization ratio $P_x/P_z$, at a $Q^2$
of 0.15 and 0.34 GeV$^2$. The ratio $P_x/P_z$ is related to $G_E/G_M$ as shown in Eq. (\ref{eq:ratio}).
The measurement of the ratio, $P_x/P_z$, has some  
advantage, as discussed earlier for the proton, over the measurement of $P_x$ only, because in the ratio the 
electron beam polarization and the polarimeter analyzing power cancel; as a result the systematic
uncertainty is small. In yet another experiment at MAMI  the ratio of polarization transfer components,  
$P_x/P_z$, 
was measured  using the same reaction $^2$$H(\vec{e},e'\vec{n})p$ and the electric FF $G_{En}$ was
obtained at $Q^2$ = 0.3, 0.6 and 0.8 GeV$^2$  \cite{glazier}; Glazier {\it et al.} concluded 
that their results were in good agreement with all other $G_{En}$ double-polarization measurements.

The experiment at JLab by Madey {\it et al.} \cite{madey,plaster} obtained the neutron FF ratios
$G_E/G_M$ at $Q^2$ values of 0.45, 1.13 and 1.45 GeV$^2$ using the same method of measuring the
recoil neutron polarization components $P_x$ and $P_z$ simultaneously, using a dipole with vertical B-field to 
precess the neutron polarization in the reaction plane, hence obtaining directly the ratio $G_{En}/G_{Mn}$. 
The best-fit values of $G_{Mn}$ were used to calculate values of $G_{En}$ from the ratio measurements. 
This is the first experiment that determined the value of $G_{En}$ with small statistical and systematic 
uncertainty and at the relatively high $Q^2$ values up to 1.45 GeV$^2$.  

Passchier {\it et al.} \cite{passchier} reported the first measurement of spin-correlation 
parameters $A^V_{ed}$ at a $Q^2$ of 0.21 GeV$^2$ in $^2$$\vec H(\vec{e},e'n)p$ reaction at NIKHEF; 
this experiment used a stored polarized electron beam and an internal vector polarized 
deuterium gas target; they extracted the value of $G_{En}$ from the measured sideways 
spin-correlation parameter in quasi-free scattering. 
 
Experiment E93-026 at JLab extracted the neutron electric FF at $Q^2$ = 0.5 and 
1.0 GeV$^2$ \cite{zhu,warren} from measurements of the beam-target asymmetry using the
$^2$$\vec{H}(\vec{e},e'n)p$ reaction in 
quasi elastic kinematics;  in this experiment the polarized electrons were scattered off a  
polarized deuterated ammonia (${ND_3}$) target. This experiment was the first 
to obtain $G_{En}$ at a relatively large $Q^2$ using a polarized target. 

Blankleider and Woloshyn, in a paper in 1984 \cite{blankleider}, proposed that a polarized $^3$He target  
could be used 
to measure $G_{En}$ or $G_{Mn}$. They argued that the $^3$He ground state is dominated by the 
spatially symmetric S-state in which the  two proton spins point in opposite directions, hence the spin of 
the nucleus is largely carried by the neutron. Therefore, the $^3$He target effectively serves as a 
polarized neutron target; and in the quasi-elastic scattering region the spin-dependent properties 
are dominated by the neutron in the $^3$He target. 

There were experiments in the early 1990's at MIT-Bates Laboratory that used a polarized $^3$He target and
measured the asymmetry with polarized electrons in spin-dependent quasi-elastic scattering 
\cite{cejones,thompson}, and extracted the value of $G_{En}$ using the prescription of 
Blankleider and Woloshyn \cite{blankleider}, at a $Q^2$=0.16 and 0.2 GeV$^2$. However,
Thompson {\it et al.} \cite{thompson} pointed out  
that significant corrections are necessary at $Q^2$=0.2 GeV$^2$ for spin-dependent 
quasi elastic scattering on polarized $^3$He according to the calculation of Laget \cite{laget};
hence no useful information on $G_{En}$ could be extracted from these measurements; but Thompson {\it et al.}
concluded that at higher $Q^2$ values the relative contribution of the 
polarized protons becomes significantly less and a precise measurements of $G_{En}$ using polarized 
$^3$He targets will become possible. 

\begin{figure}[h]
\begin{minipage}[b]{0.45\linewidth}
\centerline{\epsfig{file=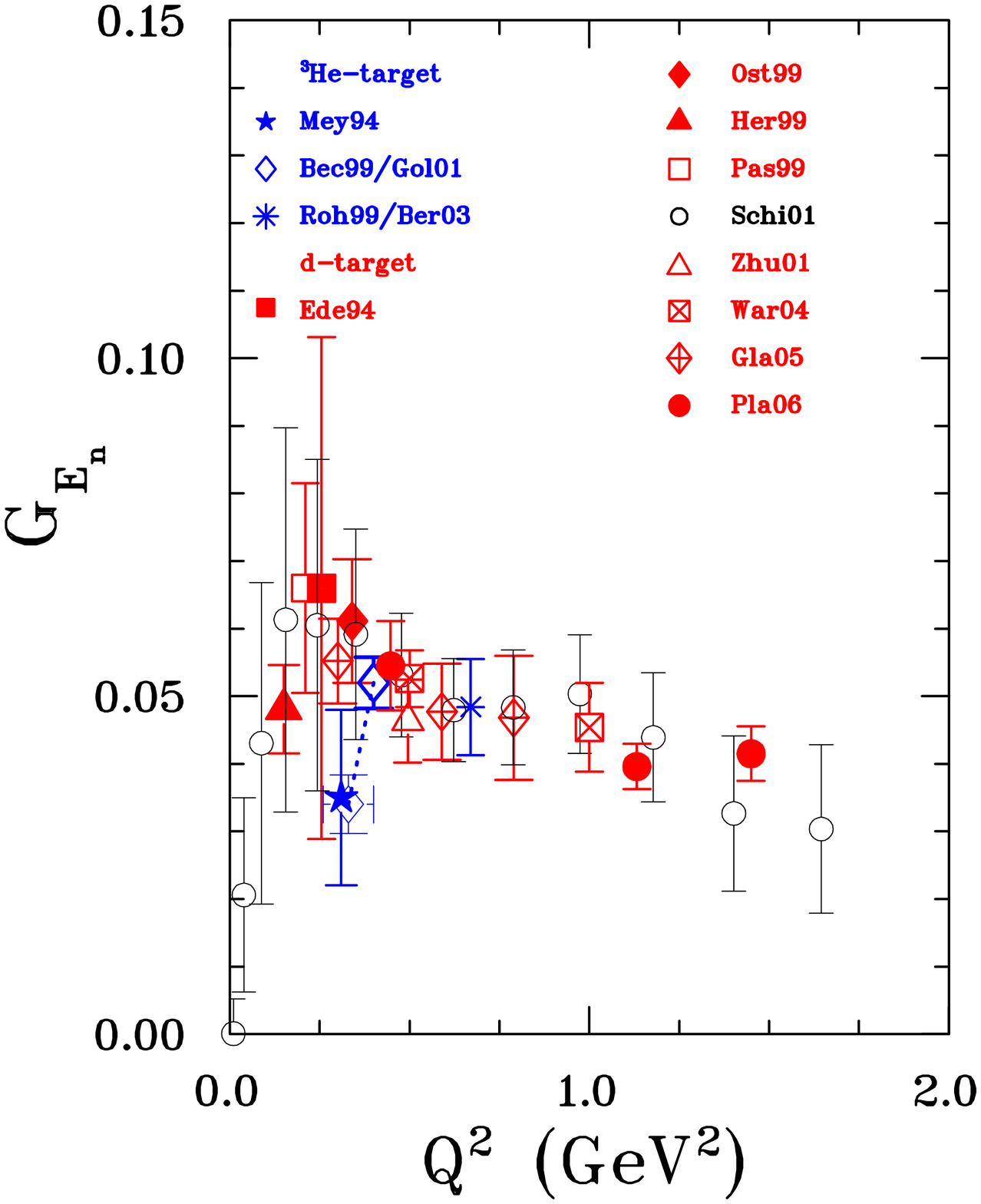,width=6cm}}
\caption{\small{Data for $G_{En}$ from beam asymmetry with polarized $D_2$ \cite{passchier,zhu,warren}, 
and $^3$He \cite{meyerhoff,becker,rohe,bermuth}, and recoil 
polarization with $D_2$ \cite{eden,herberg,ostrick,glazier,plaster}. Included
values obtained from $T_{20}$ \cite{abbott} in elastic $ed$, in \cite{schiavil}.}}
\label{fig:gen_pol}
\end{minipage}\hfill
\begin{minipage}[b]{0.45\linewidth}
\epsfxsize=\textwidth
\centerline{\epsfig{file=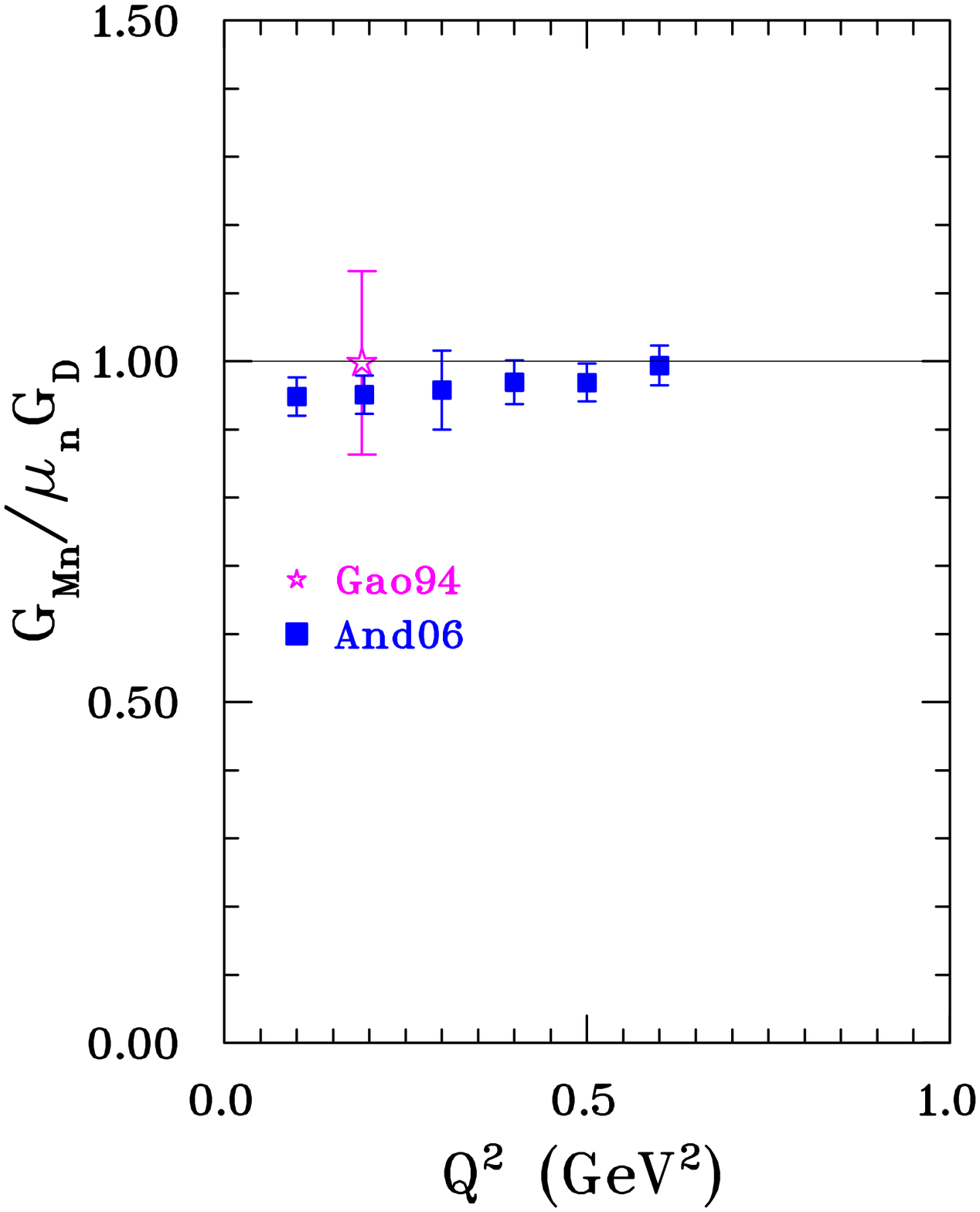,width=6cm}}
\vspace{1.24cm}
\caption{\small{The data for $G_{Mn}$ from the two polarization experiments which have been 
performed so far,
at Bates \cite{gao1} and JLab \cite{anderson}, respectively.}}
\label{fig:gmn_pol}
\end{minipage}
\end{figure}

Starting in the early 1990's, the neutron electric FF $G_{En}$ has been 
obtained in several experiments at MAMI, by measuring the beam-target asymmetry 
in the exclusive quasi-elastic scattering of electrons from polarized $^3He$ in the 
$^3$$\vec{He}(\vec{e},e'n)pp$ reaction \cite{meyerhoff,becker,rohe,bermuth}. All the $G_{En}$ data 
from polarization experiments are shown in Fig. \ref{fig:gen_pol}. 
In the first of these experiments, at MAMI, \cite{meyerhoff}, $G_{En}$ was obtained 
at $Q^2$ = 0.31 GeV$^2$. In following experiments at MAMI, $G_{En}$ was
extracted at $Q^2$ of 0.35 GeV$^2$ \cite{becker} and 0.67 GeV$^2 $ \cite{rohe,bermuth} using the 
same reaction. The 0.35 GeV$^2$ point of \cite{becker} was later corrected in \cite{golak}, based 
on Faddeev solutions and with some MEC corrections. The large effect of these corrections is 
illustrated in Fig. \ref{fig:gen_pol} 
with the dashed line connecting the open diamonds. The size of these corrections is expected to 
decrease with $Q^2$, although the corrections become increasingly difficult to calculate with increasing $Q^2$s.

\subsubsection{Neutron magnetic form factor measurements with polarization experiments}

Only two experiments have obtained the magnetic FF of the neutron, $G_{Mn}$, from 
polarization observables; both experiments used a polarized $^3$$\vec{He}$ target. 
The first experiment at the MIT-Bates laboratory, extracted 
$G_{Mn}$ from the measured beam-target asymmetry in inclusive quasi-elastic scattering 
of polarized electrons from polarized  $^3$$\vec{He}$ target  
at $Q^2$ of 0.19 GeV$^2$ \cite{gao1}; the uncertainty on $G_{Mn}$ was dominated by the 
statistics, with a
relatively small contribution from model dependence of the analysis. 
The second JLab experiment obtained $G_{Mn}$ for $Q^2$ values between 0.1 and 0.6 GeV$^2$,
by measuring the transverse asymmetry in the $^3$$\vec{He}(\vec{e},e')$ reaction 
in quasi-free kinematics \cite{xu00,xu02,anderson}. The values of $G_{Mn}$ were extracted 
in the plane wave impulse 
approximation (PWIA) at $Q^2$ of 0.3 to 0.6 GeV$^2$, and from a full Faddeev calculation 
at $Q^2$ of 
0.1 and 0.2 GeV$^2$. The authors of this paper asserted that the PWIA extraction of $G_{Mn}$ is 
reasonably reliable in  the $Q^2$ range of 0.3 to 
0.6 GeV$^2$; however, a more precise extraction of $G_{Mn}$ requires fully relativistic three-body 
calculations. The $G_{Mn}$ values from both experiments are shown in Fig. \ref{fig:gmn_pol}.
\begin{figure}[h]
\begin{minipage}[b]{0.45\linewidth}
\centerline{\epsfig{file=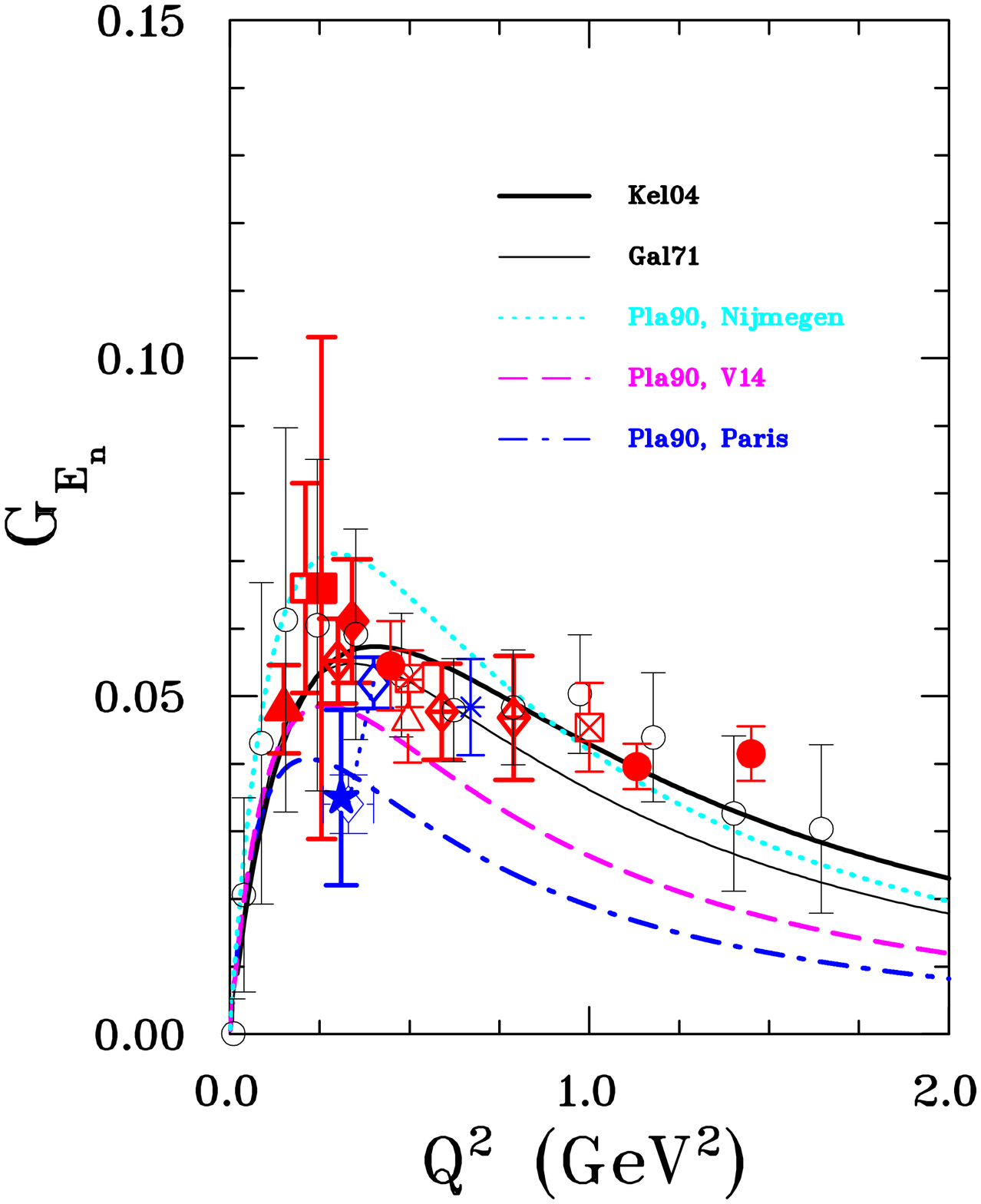,width=6cm}}
\caption{\small{$G_{En}$ data as in Fig. \ref{fig:gen_pol}, compared to the fits 
\cite{kelly04} (thick line) and \cite{galster} (thin solid line). Platchkov's fits \cite{platchkov} 
with 3 different
$NN$ potentials shown as dotted~\cite{reid}, dot-dashed \cite{lacombe} and long dashes~
\cite{wiringa} lines, respectively.}}
\label{fig:gen_comp}
\end{minipage}\hfill
\begin{minipage}[b]{0.45\linewidth}
\epsfxsize=\textwidth
\centerline{\epsfig{file=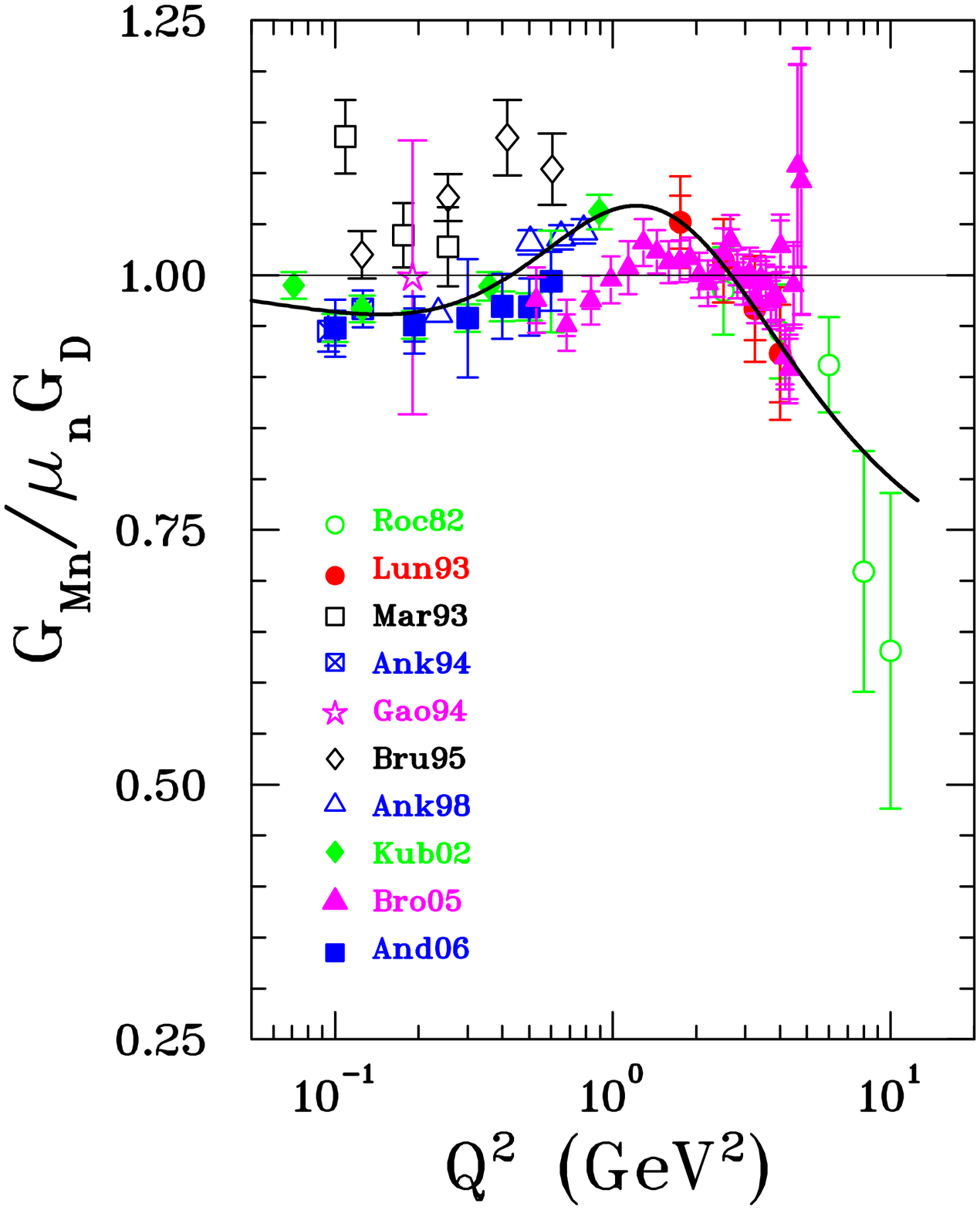,width=6cm}}
\vspace{0.15cm}
\caption{\small{The complete data base for $G_{Mn}$, from cross section and polarization measurements.
Shown as a solid curve is the polynomial fit by Kelly \cite{kelly04}; note that the recent data of 
~\cite{brooks} are not included in this fit.}}
\label{fig:gmn_comp}
\end{minipage}
\end{figure}

\subsection{Discussion of the form factor data}
\label{DiscussFF} 

\begin{figure}[h]
\begin{minipage}[b]{0.45\linewidth}
\begin{center}
\epsfig{file=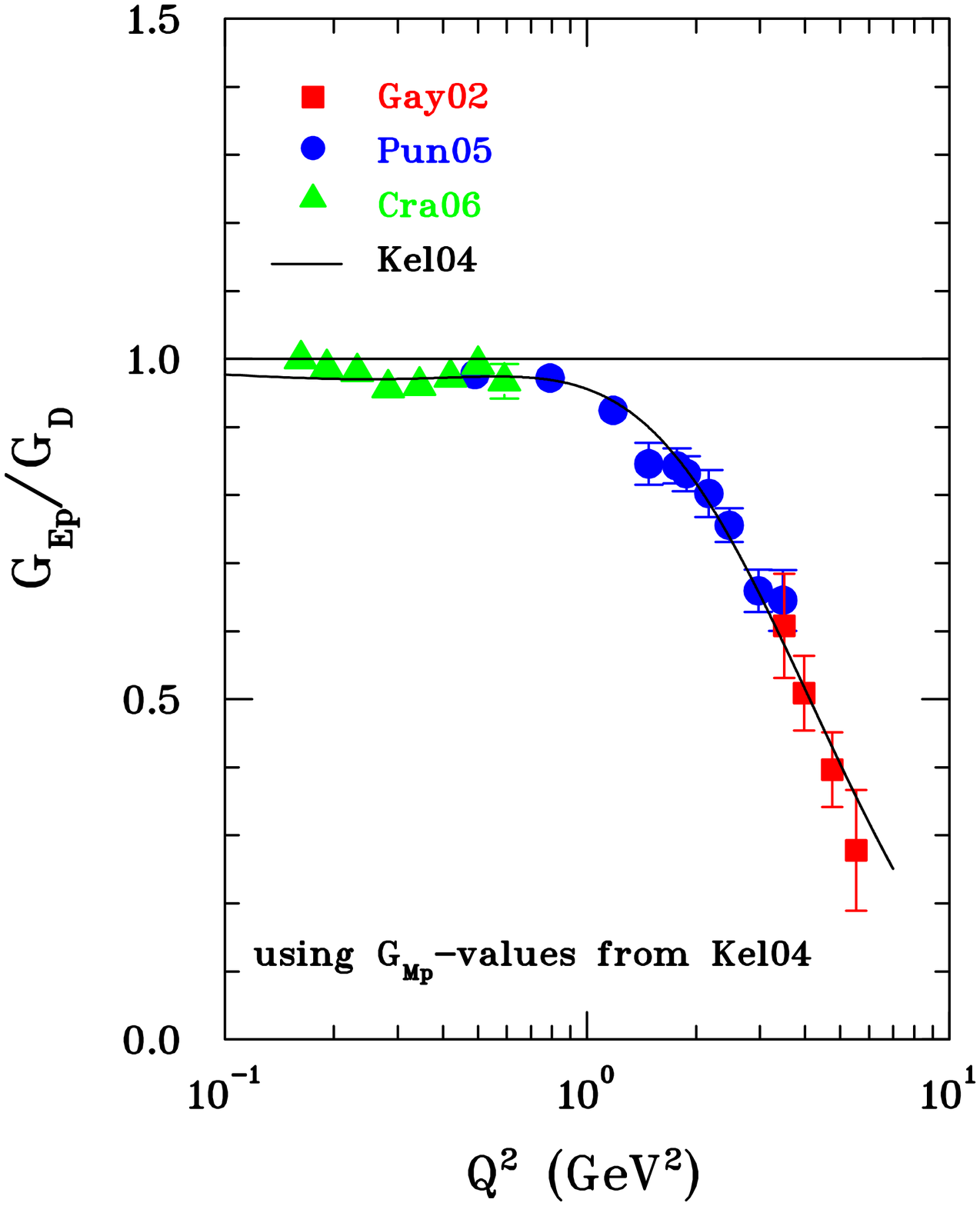,width=6.0cm}
\caption{\small{Polarization data presented as $G_{Ep}/G_D$, where $G_{Ep}$ is obtained from 
the ratio $G_{Ep}/G_{Mp}$ obtained from polarization data in~ \cite{punjabi05,gayou2,crawford}, 
multiplied by $G_{Mp}$ from the Kelly fit~\cite{kelly04}.}}
\label{fig:gepgdpol}
\end{center}
\end{minipage}\hfill
\begin{minipage}[b]{0.45\linewidth}
\hspace{-1.5cm}
\begin{center}
\epsfxsize=\textwidth
\epsfxsize=\textwidth\epsfig{file=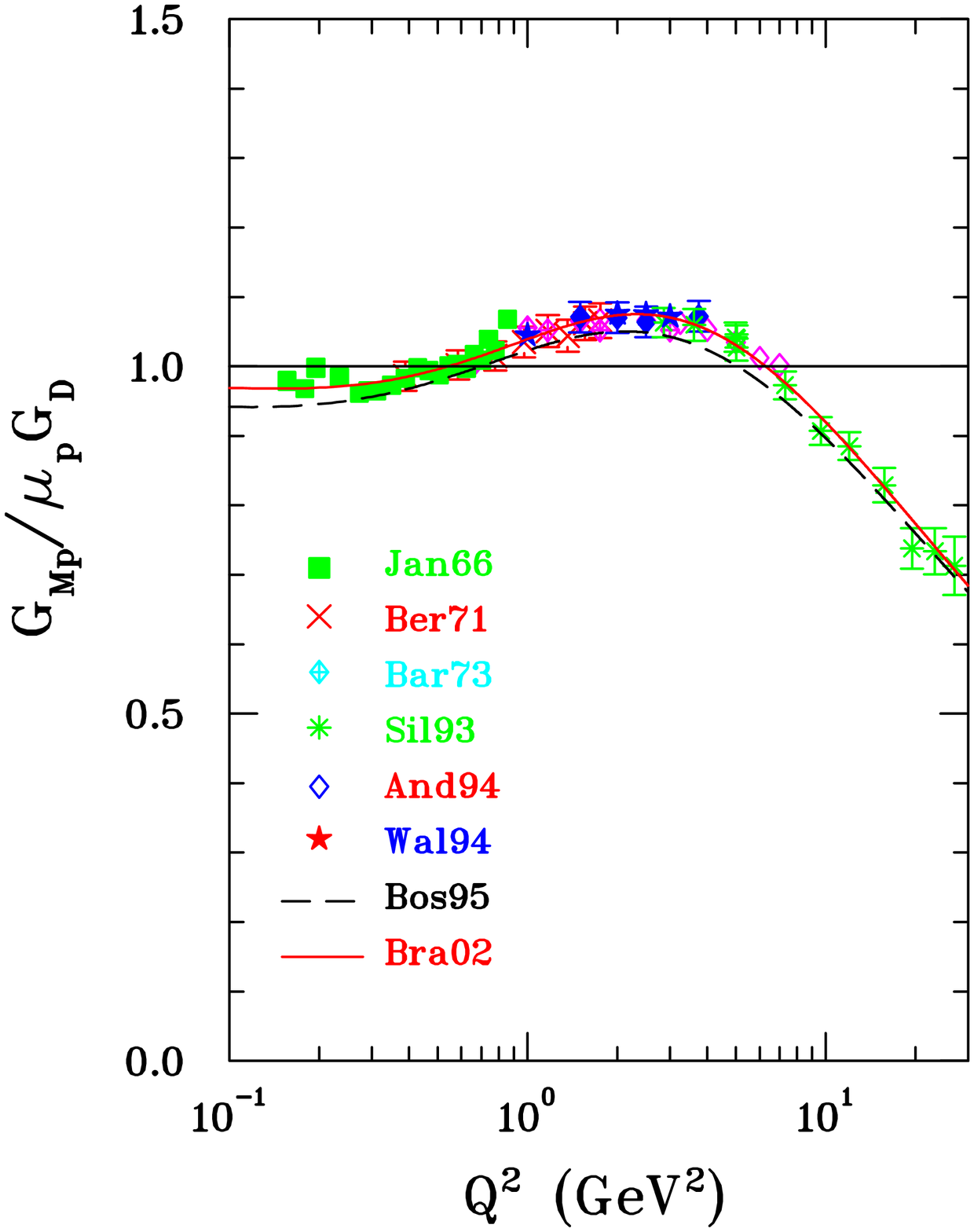,width=6.0cm}
\caption{\small{The $G_{Mp}$ data were refitted in \cite{brash} imposing the value of the 
$G_{Ep}/G_{Mp}$ from the recoil polarization data of Refs. \cite{punjabi05,gayou2}, leaving out 
Rosenbluth separation data above 1 GeV$^2$.}}
\label{fig:gmpbrash}
\end{center}
\end{minipage}
\end{figure}

Probably the most important advance in the characterization of the FFs of the nucleon
made in the last 10 years has been the realization that the so-called ``scaling''-behavior of 
the proton FFs:
\begin{equation}
 G_{Ep}\sim G_{Mp}/\mu_p ~\sim~ G_D,
\end{equation}
was limited to values of $Q^2$ smaller than 2 GeV$^2$. The recoil polarization data obtained at JLab
in 1998 and 2000 proved beyond any doubt that for $Q^2$-values larger than 2 GeV$^2$, $G_{Ep}$ 
decreases faster than $G_{Mp}/\mu_p$ with a slope of -0.14 per GeV$^2$. What we now have are distinctly 
different 
$Q^2$-dependences for $G_{Ep}$ and $G_{Mp}$; in the $Q^2$ region investigated so far, the 
scaling-behavior is violated by a factor of 3.66 $^{+1.71}_{-0.88}$ at 5.54 GeV$^2$.
The deviation of $G_{Ep}$ from the dipole FF is illustrated in Fig.~\ref{fig:gepgdpol}, where
only polarization results are shown. Of course it was well known that the dipole FFs, 
when Fourier transformed, produce unphysical 
distributions of charge or magnetization, with a discontinuity at zero radius. Nevertheless there 
were valid reasons, to believe that the dipole FF discussed in \ref{DipoleFF} 
may actually describe the FFs $G_{Ep}$, $G_{Mp}$ and $G_{Mn}$ of 
the nucleon. The data no longer 
support such expectations, as can be concluded by comparing the results in Figs. \ref{fig:gepgd} and 
\ref{fig:gepgdpol}.  

The discrepancy is related to the techniques used: all Rosenbluth separation of 
cross section data including the 2 new measurements from JLab \cite{christy,qattan05} 
give $\mu_p G_{Ep}/G_{Mp}$ ratios close to the scaling behavior, except the early data shown in Fig. 3; 
all recoil polarization results for the same ratio are 
clustered along an approximately straight line versus $Q^2$, with a best fit valid above $Q^2\sim0.4$ GeV$^2$ given by:
\begin{equation}
\mu_p G_{Ep}/G_{Mp}=1.0587 - 0.14265~Q^2.
\label{eq:jlabfit}
\end{equation}
A number of observations relative to this difference in results follows. 
First, there is one well established difference between the two techniques, cross section 
versus recoil polarization, and it is the relative importance of the 
radiative corrections required for them, as discussed in detail in section~\ref{RadCorr}. The total 
radiative 
corrections as routinely calculated in cross section measurements is typically 
10 to 30\%, and the corrections are strongly $\epsilon$ dependent; this $\epsilon$ dependence affects 
primarily the results for $G_{Ep}^2$, and for increasing $Q^2$ the 
accuracy requirement for the correction becomes very demanding.
Second,  polarization observables, in recoil polarization or target asymmetry
measurements, being ratios of cross sections, are 
only minimally affected by radiative corrections, and the ratio $G_{E}/G_{M}$ even less being a 
ratio of ratios. Nevertheless polarization data ultimately will require radiative corrections, 
particularly as experiments continue into the domain of yet larger $Q^2$.  
So is the discrepancy between Rosenbluth and polarization data entirely due to inaccuracy or 
incompleteness in the radiative correction? An immediate consequence of the previous 
statements is that radiative corrections for elastic $ep$ scattering in general
have to be reexamined, as in their presently practiced form they are unable to 
reconcile the cross section results with polarization results. 

Encouraging progress has been made including the one process certainly neglected in 
all previous radiative corrections, the exchange of two photons, 
neither one of them ``soft'' (this will be further discussed in section~\ref{RadCorr}). Several 
calculations \cite{guichon,afanbrod,blunden} suggest that this 
one  diagram may contribute significantly to the $\epsilon$-dependence of 
the cross section; other considerations lead to the conclusion that the 
contribution from the two-photon term is too small at the $Q^2$-values of interest 
\cite{bystrit}, and/or leads to a definite non-linearity in the Rosenbluth plot which has 
not been seen in the data so far \cite{tomasi1}.

Following the publication of the JLab recoil polarization 
$G_{Ep}/G_{Mp}$ ratios up to 5.54 GeV$^2$, the entire cross section data base for the proton has 
been reanalyzed by Brash {\it et al.} \cite{brash},
leaving all data above $Q^2=1$ GeV$^2$ out, using the data from 
\cite{jones,gayou2} 
above this value of $Q^2$, and allowing for relative renormalization of all cross section data so as to 
minimize the $\chi^2$ of a global fit for $G_{Mp}$. 
The fitting function is the inverse of a polynomial of order 5. The renormalized values of $G_{Mp}$ 
show less scatter than the original data base, and the net effect of imposing the recoil 
polarization results is to re-normalize all $G_{Mp}$ data upward by 1.5-3\% when compared with the
older Bosted parametrization \cite{bosted95}, as shown in Fig. \ref{fig:gmpbrash}.

Another useful fit to the nucleon FFs which gives a good representation of the data is the one by 
Kelly \cite{kelly04}. This fit uses ratios of polynomials with maximum powers chosen such that 
$G_{Ep}$, $G_{Mp}$ and 
$G_{Mn}$ have the asymptotic $1/Q^4$ behavior required by pQCD; in \cite{kelly04} $G_{En}$ was also 
re-fitted with a Galster FF, as shown in Fig.~\ref{fig:gen_comp}.

In Figs. \ref{fig:gen_comp} and \ref{fig:gmn_comp} we compare all the data available for 
$G_{En}$ and $G_{Mn}$, obtained from cross section and polarization observables.  
The $G_{En}$  data obtained in double polarization show reasonable consistency above 0.5 GeV$^2$; 
they are systematically higher than the older cross section results shown in Fig. 
\ref{fig:gen_comp} by the 3 Platchkov fits \cite{platchkov}. The revision by Kelly \cite{kelly04} 
of the Galster fit \cite{galster} gives an excellent representation of the data available today. 
%Much is expected from the recent experiment in Hall A at JLab that measured $G_{En}$ up to 3.4 GeV$^2$ \cite{cates}. 

Recently, more and better data have been obtained for $G_{En}$, exclusively by the polarization method, 
either recoil polarization transfer or target asymmetry, with deuterium and $^3$He targets and up
to $Q^2$=1.5 GeV$^2$. No drastic change of the general behavior 
of $G_{En}$ has been observed to this point in time. There is a new measurement of $G_{En}$ at JLab up to
a $Q^2$ of 3.4 GeV$^2$ \cite{cates}, but the data have yet to be analyzed. In general 
all polarization data for $G_{En}$ have given results larger than those obtained from elastic scattering;
these earlier data required considerable nuclear structure corrections, as illustrated in 
Fig. \ref{fig:gen_pol}; the sensitivity to the deuteron wave function, therefore to the $NN$ 
potential used, was extensively discussed at the time in ~\cite{platchkov}.

The data for $G_{Mn}$ come mostly from cross section measurements, except two polarization 
measurements, using polarized $^3$He target, one at MIT-Bates for low $Q^2$ with large 
uncertainty \cite{gao1} and the other a recent measurement at JLab \cite{xu00,xu02,anderson}. The most 
recent Hall B results
\cite{brooks}, which extend to $Q^2$ of nearly 5 GeV$^2$, and used quasi-elastic scattering on 
deuterium, reveal some internal inconsistency in the data base near 1 GeV$^2$ as shown in 
Fig. \ref{fig:gmn_comp}; as can be seen in Fig.~\ref{fig:gmn_comp} there is some disagreement 
between the results of different experiments in the $Q^2$ range of 0.3 to 1.5 GeV$^2$. . These measurements 
will be extended to 14 GeV$^2$ after the JLab upgrade to 12 GeV; similarly, the measurement
of $G_{Ep}/G_{Mp}$ will be continued to 13 GeV$^2$ after the upgrade.

Several experiments are planned at JLab to resolve the dichotomy in the $G_{Ep}/G_{Mp}$ ratio.  
One experiment will measure the ratio of the $e^-p$ and $e^+p$ cross sections, which determines 
directly the real part of the two-photon amplitude \cite{afanasev4}. Another experiment 
will measure the ratio $G_{Ep}/G_{Mp}$ at fixed Q$^2$=2.5 GeV$^2$ 
\cite{lubomir}, as a function of $\epsilon$, to detect the two hard photon contribution 
as a variation of this ratio; non constancy would 
be related to the real part of the two-photon amplitude. A third experiment will be a high statistics search
of non-linearity in the Rosenbluth plot in $ep$-scattering, which should also reveal the contribution of the
two-photon process (\cite{arring05}). Measurements of the induced 
polarization 
in ${e}p\rightarrow{e}\vec{p}$ (a byproduct of the experiment from \cite{lubomir}) , and of the single 
spin target asymmetry in quasi elastic scattering on the 
neutron in $^3{He}^{\uparrow}({e},{e}')$ for target polarization normal to the reaction plane \cite{todd}, will 
measure the imaginary part of the two-gamma contribution. The transverse beam spin asymmetry in $ep$ has
been measured at Bates \cite{wells} and MAMI \cite{maas}; it too originates from the imaginary part of 
the two-photon contribution.
%It is possible, in principle with an appropriate model, to connect the imaginary part to the real part 
%of the amplitude which contribute to the cross sections. 

\subsection{Rosenbluth results and radiative corrections}
\label{RadCorr}

All cross section measurements have been single arm experiments, $(e,e')$, except 
three early experiments at Cambridge \cite{price,hanson} and DESY  \cite{bartel}
in which both proton and electron were detected, $(e,e'p)$, and the most recent one of 
Qattan {\it et al.} \cite{qattan05}, in which only the proton was detected, $(e,p)$. In all cases, measured raw cross 
sections need to be corrected for QED processes to first order in 
$\alpha\sim1/137$, before accessing the cross section corresponding to one-photon-exchange, or 
Born term. Only to the extent that these corrections remain relatively
small, can one hope to obtain the Born term FFs $G_{E}^2$ and 
$G_{M}^2$, which are functions of $Q^2$ only, using the Rosenbluth method.

The effect of the radiative correction on the cross section is typically in 
the range 10-30\%; what is important however, is the fact that overall, the 
radiative corrections are $\epsilon$-dependent; i.e. they affect the slope of 
the Rosenbluth plot. Although 
radiative corrections have been applied to all data taken after 1966 using the ``recipe'' of
Tsai \cite{tsai1},  Mo and Tsai \cite{motsai} and \cite{tsai2}, not 
all corrections were applied in all data sets. This point was recently reviewed by 
Arrington \cite{arring03}, who reanalyzed some of the cross section data; the fit to the re-analyzed 
data is included in Fig. \ref{fig:gepgmp_comp}. Furthermore, in the  
references \cite{tsai1,motsai,tsai2} the effect of the structure of the nucleon was ignored, and a number of 
approximations were made. In more recent work on radiative corrections, Maximon and Tjon 
\cite{maxtjon} have included 
the structure of the proton by introducing the proton FF, and they also eliminated some 
of the soft-photon 
approximations made by  \cite{tsai1,motsai,tsai2}. In the current energy 
range of JLab, the difference for $\delta$, the radiative correction, used in 
$\frac{d\sigma}{d\Omega}=(1+\delta)\frac{d\sigma^{Born}}
{d\Omega}$ up to corrections of order $\alpha^3$, between the older and the new calculation is at the level of 
several~\%.

\begin{figure}[h]
\begin{minipage}[b]{0.45\linewidth}
\begin{center} 
\hspace{-0.7cm}
\epsfig{file=rad_el.eps,height=3.15in}
\vspace{-0.25cm}
\caption[]{Born term and lowest order radiative correction graphs for the electron in elastic $ep$.}
\label{fig:radel}
\end{center}
\end{minipage}\hfill
\begin{minipage}[b]{0.45\linewidth}
\begin{center}
\epsfxsize=\textwidth
\epsfig{file=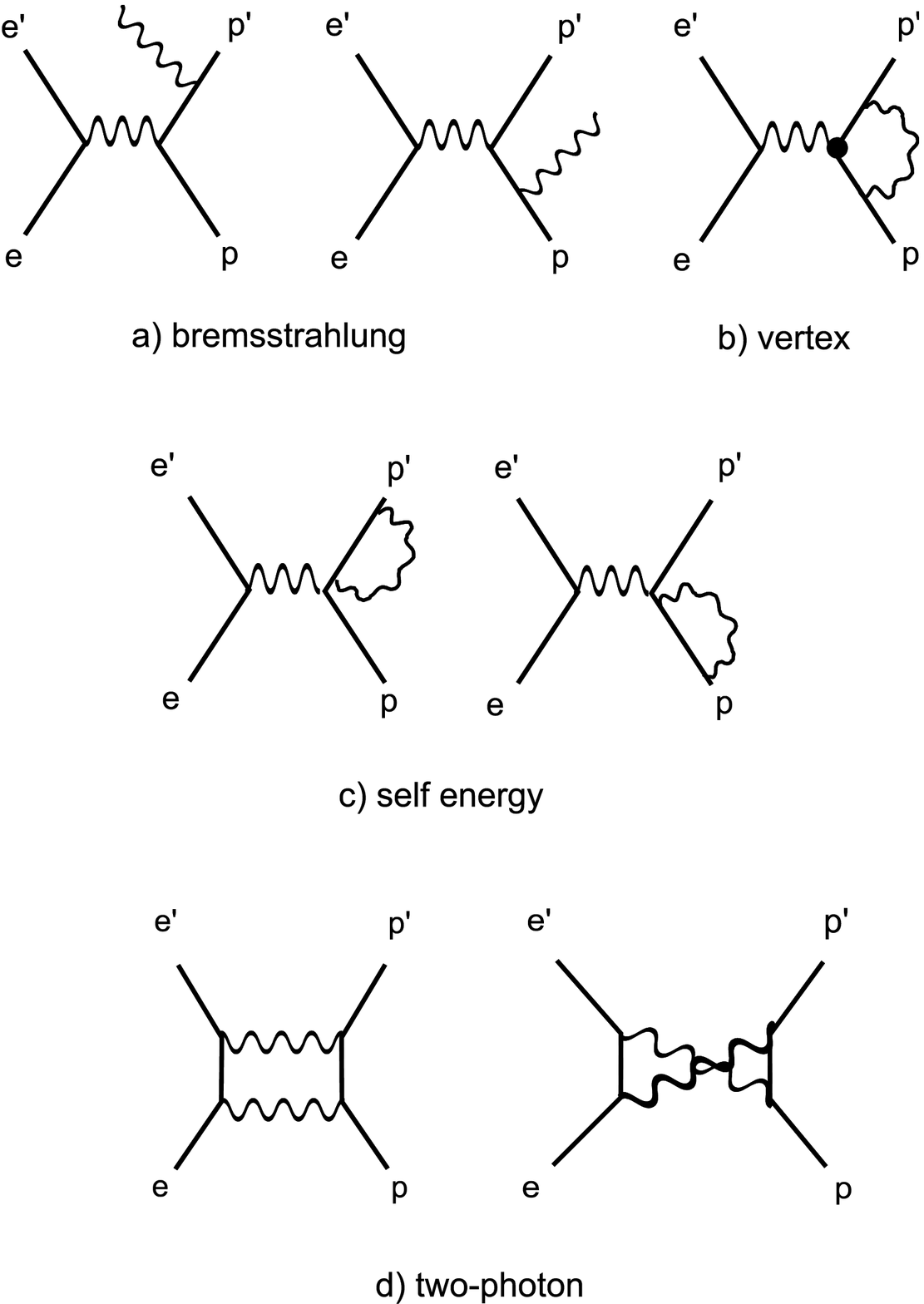,height=3.0in}
\caption[]{Lowest order radiative correction for the proton side in elastic $ep$ scattering.}
\label{fig:radp}
\end{center}
\end{minipage}
\end{figure}

The various internal radiative correction diagrams involving the electron are shown in Fig. \ref{fig:radel}.
The first order virtual radiative processes are the vertex diagram b), the photon self-energy diagram c) and the 
two self-energy diagrams for the electron d); 
the first order real radiative processes include emission of a real photon by either 
the initial or the final electron diagram e). Similarly 
diagrams for the proton include bremsstrahlung a), vertex b) and proton self energy c), shown in 
Fig. \ref{fig:radp}. Two-photon exchange is shown as diagram e). In addition 
there are external radiative
corrections due to the emission of real photons 
by the incoming and scattered electrons in the material of the target, as well as energy loss by ionization.

The virtual part of the internal radiative corrections depend exclusively upon $Q^2$, thus it 
generates no $\epsilon$-dependence, hence does not modify the value of 
$G_{Ep}^2$, but modifies the value of $G_{Mp}^2$ directly.

The radiative correction for real photon emission (bremsstrahlung) is energy, and therefore 
$\epsilon$ dependent, and it also results in a changed value of $Q^2$. 
In general the scattered electron energy spectrum is integrated up to a maximum energy loss which 
is kept below the pion threshold. The correction is different for different experiments; it depends on 
the procedure used to integrate over the scattered electron energy, or missing mass squared spectrum.

The contributions due to real photon emission by the initial
and final proton, as well as the proton vertex and two-photon exchange with 
one soft- and one hard photon are relatively small, but strongly $\epsilon$-dependent.

\begin{figure}[h]
\begin{minipage}[b]{0.32\linewidth}
%\begin{center}
\centerline{\epsfxsize=\textwidth\epsfig{file=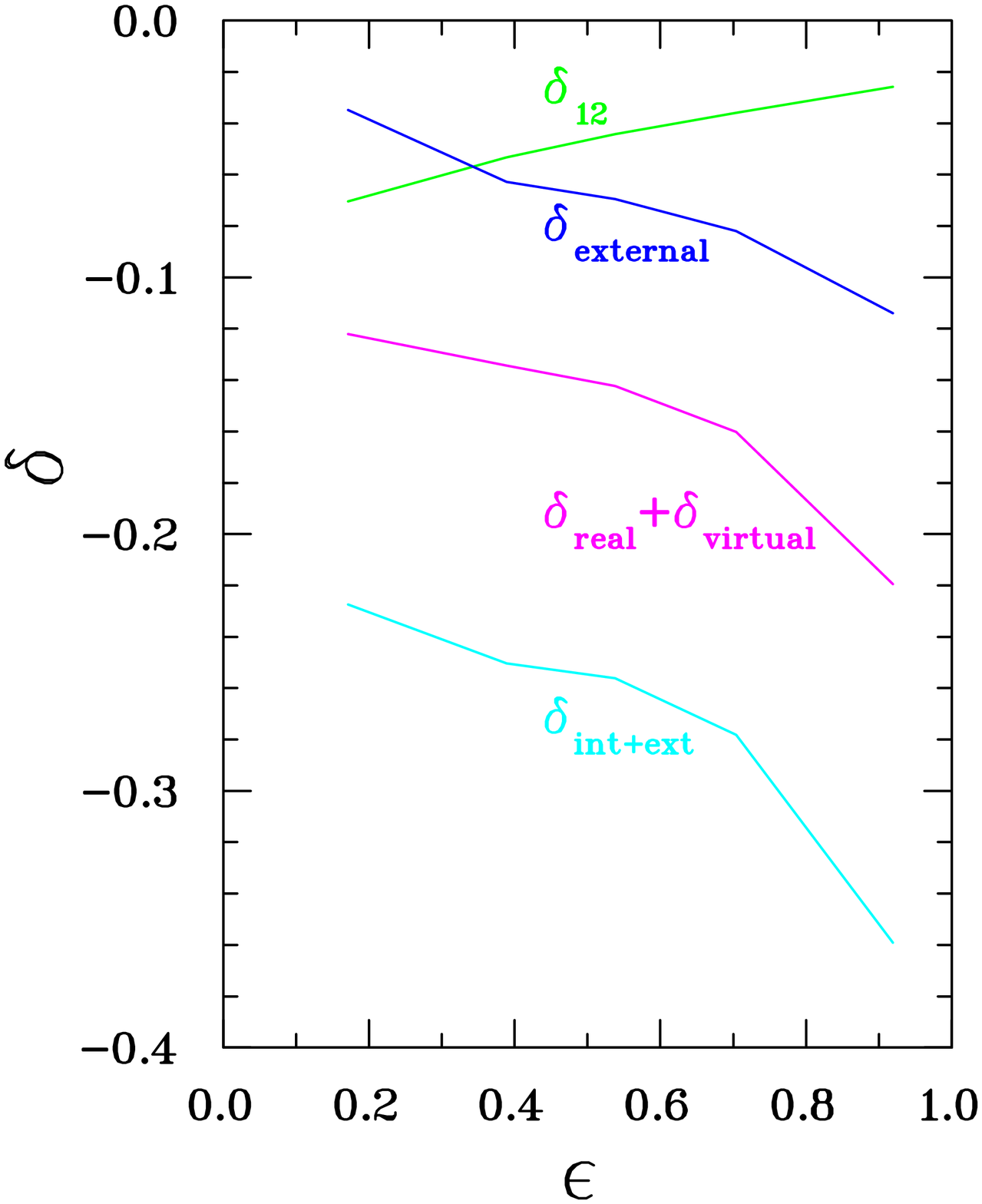,height=7.0cm}}
%\vspace{0.1in}
\caption[]{\small{The various contributions to the correction factor $\delta$ at $Q^2$=5 GeV$^2$,
calculated with the code from \cite{vdh00} based on the work of Maximon and Tjon \cite{maxtjon},
including the external correction for a 15 cm
target, taken from \cite{andivahis}.}} 
\label{fig:vdhrad} 
%\end{center}
\end{minipage}\hfill
\begin{minipage}[b]{0.32\linewidth}
\epsfxsize=\textwidth
%\begin{center}
\centerline{\epsfig{file=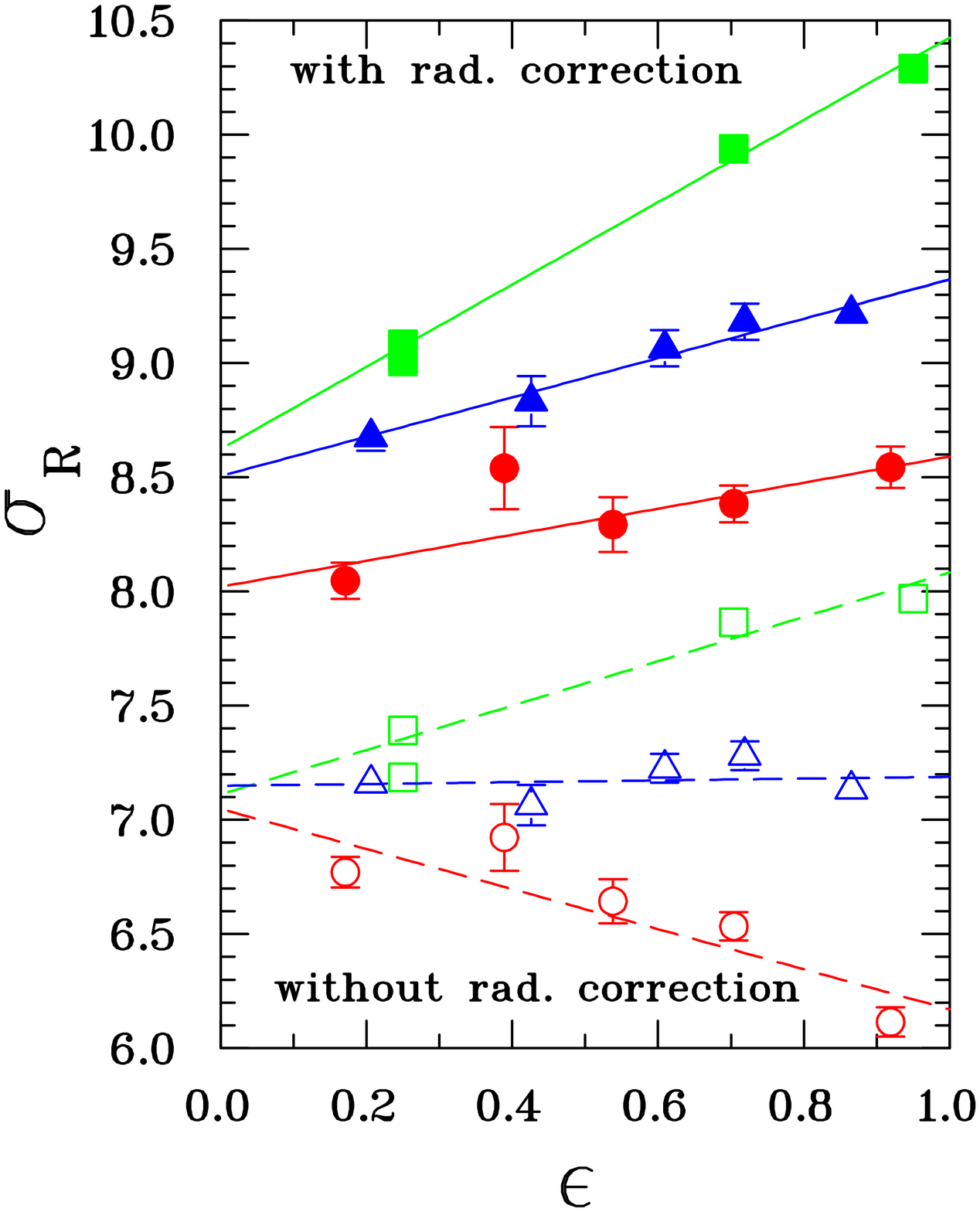,height=7.0cm}}
%\vspace{0.7cm}
\caption{\small{Rosenbluth plot for data of {\it et al.} \cite{andivahis}.
At bottom before radiative correction, at
top after radiative correction. Filled squares, triangles and circles for 
1.75, 3.25 and 5 GeV$^2$, respectively; empty symbols for uncorrected data.}}
\label{fig:rosenandi}
%\end{center}
\end{minipage}\hfill
\begin{minipage}[b]{0.32\linewidth}
%\begin{center}
\epsfxsize=\textwidth
\centerline{\epsfig{file=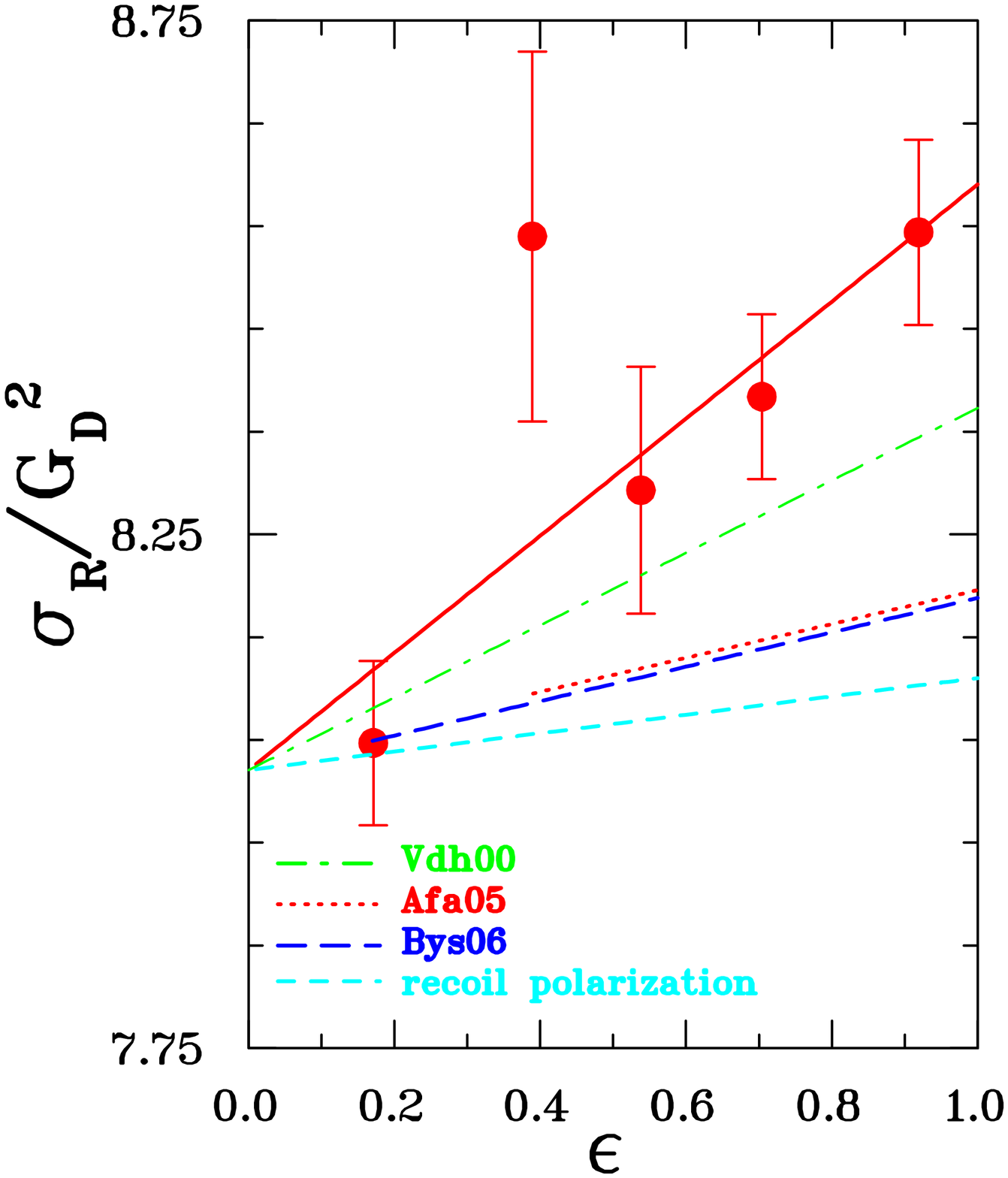,height=7.0cm}}
%\vspace{1.415cm}
\caption{\small{The $Q^2$= 5 GeV$^2$ data of \cite{andivahis}, 
with their best fit (thick solid line), compared with the results of the 
various calculations of the radiative correction, taken from 
Refs.~\cite{vdh00,afanbrod,bystrit}. Also shown is the slope from the JLab polarization data.}}
\label{fig:simplest}
%\end{center}
\end{minipage}
\end{figure}

The external part of the radiative corrections includes only real photon emission by the incident and 
scattered electron, and is not coherent with the $ep$ interaction. 
Although the correction for the incoming electron in the target is energy independent, and it can be 
averaged to a value at 
the center of the active area of the target for all kinematics of a given experiment, the correction 
for the scattered electron in the target depends directly 
upon the target length and diameter which determines the amount of 
target material traversed, and therefore the scattering angle. As the desired range of $\epsilon$ values is obtained by changing the electron 
scattering angle, this correction has $\epsilon$-dependence. 
For the data of  Andivahis {\it et al.} \cite{andivahis} the external corrections are 
one fourth to one half as large as the internal corrections 
from the smallest to the largest $\epsilon$-values as shown in Fig. \ref{fig:vdhrad}. The 
calculation of the 
external correction requires information on the spectrometer acceptance and on the target geometry, 
and is an integral 
part of the analysis of the data; it cannot be repeated on the basis of published data. However, it is
potentially a significant source of uncertainty in the $\epsilon$-dependence of the total 
radiative correction. 

To gain some appreciation of what term might be most strongly affecting the final result of the 
radiative correction, we show the values $\delta_{real}+\delta_{virtual}$, 
graphs b), c), d) and e) in Fig. \ref{fig:radel}, $\delta_{12}$ from graphs 
a), b),c) and d) in Fig. \ref{fig:radp}, and $\delta_{external}$ for the condition of the Andivahis
experiment~\cite{andivahis} separately in Fig~\ref{fig:vdhrad}. The curve in Fig.~\ref{fig:vdhrad}
labeled $\delta_{int+ext}$ determines the overall correction. Its slope versus $\epsilon$ 
is due to the combined effect of the real and external-contributions, with the proton 
contribution reducing it somewhat; $\delta_{virtual}$ has no $\epsilon$-dependence.

The importance of calculating the contributions to the radiative correction 
which are $\epsilon$-dependent accurately is illustrated in 
Fig. \ref{fig:rosenandi}. Shown in this figure are reduced cross sections defined in terms of the  
$(\frac{d\sigma}{d\Omega})_{reduced}$ from Eq. (\ref{eq:redcs}) as:
\begin{equation}
\sigma_R=\left(\frac{d\sigma}{d\Omega}\right)_{reduced}/G_D^2=\frac{\epsilon}{\tau}\frac{G_{E{p}}^2}{G_D^2}+
\frac{G_{M{p}}^2}{G_D^2},
\end{equation}
\noindent
as a function of $\epsilon$ for the data of Andivahis {\it et al.} \cite{andivahis}. If both 
FFs are functions of $Q^2$ only, the 
intercept of a straight line fit is $G_{M{p}}^2/G_D^2\sim\mu_p^2$, and the slope 
is $\frac{1}{\tau}\frac{G_{E{p}}^2}{G_D^2}$. 
Most noticeable in this figure is the negative slope of the uncorrected data, above $Q^2$=3 GeV$^2$. This 
figure dramatically illustrates the importance of the radiative correction and gives a measure of the 
accuracy that is required to obtain the FFs with the desired accuracy. 
The final value of $G_{Ep}^2$ obtained from cross section data depends directly upon the value
and the accuracy of the $\epsilon$-dependent part of the radiative correction. 
Note that the radiative corrections for the data of \cite{andivahis} were made following Mo and Tsai 
\cite{tsai1,motsai,tsai2}, with the additional corrections introduced in Ref. \cite{walker}. 

More recently Maximon and Tjon \cite{maxtjon} have reconsidered the radiative 
correction calculation, and included additional terms with explicit emphasis of the hadronic effects. 
A similar
reexamination of the Mo-Tsai procedure was made by Vanderhaeghen {\it et al.} \cite{vdh00}
in the process of a detailed calculation of radiative corrections for virtual 
Compton (VCS).
Also recently Ent {\it et al.} \cite{ent} and Weissbach {\it et al.} \cite{weissbach} have
published improvements and  detailed studies of  the radiative correction
calculation technique for coincidence experiments (e,e'p). 

Most recently Bystritskiy {\it et al.} \cite{bystrit} have calculated the 
radiative corrections for elastic $ep$ scattering using the 
Drell-Yan electron structure function approach; no co-linearity 
approximation is made in such a calculation, but the proton vertex corrections
have not been included so far; the diagram with two hard photons has been
approximatively calculated using both nucleon and $\Delta$ intermediate states 
and was found to make a negligible contribution. 
The results of \cite{bystrit} suggest that hard bremsstrahlung 
may cause the difference between the Rosenbluth and polarization results.  
Usual bremsstrahlung calculations are for soft bremsstrahlung, 
where the emitted photon energy is kept only to linear order 
in denominators and entirely omitted in numerators.  
Soft bremsstrahlung multiplies all amplitudes by the same factor and 
does not, for a relevant example, change the slope on a Rosenbluth plot.  
If one makes no approximations in the photon energy, 
there can be different effects on different spin amplitudes.  
Thus the claim is that emitted photons that are energetic enough 
to affect the spin structure of the calculation but still small 
enough to escape detection, give rise to the difference between the two 
methods of measuring $G_{Ep}/G_{Mp}$.  
A contrasting numerical claim is that hard bremsstrahlung effects 
are noticeable and helpful in reconciling the Rosenbluth 
and polarization experiments, but are not decisive, 
see Ref.~\cite{afanasev05}.
These contrasting claims clearly need to be sorted out, 
but an independent reexamination is not available as of this writing.
%The results in \cite{bystrit}
%differ from the standard radiative corrections 
%\cite{tsai1,motsai,tsai2} applied in 
%\cite{andivahis} by ~-4 \% at $Q^2$=5 GeV$^2$ and $\epsilon$=1. 

The effect of these new radiative corrections is illustrated in Fig. \ref{fig:simplest}. The dashed-dot line 
is obtained from the  $Q^2$=5 GeV$^2$, uncorrected data from \cite{andivahis}, applying the radiative 
correction calculated with the code of \cite{vdh00}, with the same energy cuts as used in the original data. 
This correction is 2.5\% smaller than the one in \cite{andivahis} at $\epsilon$=1.  
The two-photon calculation result shown is obtained by removing the soft part of the two-photon contribution, 
and replacing it by the GPD based calculation of \cite{afanbrod}; the result is then refitted with a straight 
line (dotted line). The resulting value at $\epsilon$=1 is 4.5\% smaller than the original correction. 
The long dashed line represents the results of \cite{bystrit}, after correcting the 
experimental data points \cite{egle}, using the same energy 
cuts as in the data, and refitting with a straight line (dashed line); these results are almost identical to the
ones obtained with two-photon correction \cite{afanbrod}. 
The slope calculated from the fit to the JLab recoil polarization data, for $Q^2$=5 GeV$^2$, 
from Eq.~(\ref{eq:jlabfit}), is shown as the short dashed line. The value at $\epsilon$=1 is 6\% smaller than 
that of the 
original Rosenbluth data of Ref.\cite{andivahis}. In Fig. \ref{fig:simplest} all fits are drawn with a 
renormalized value of $\sigma_R$ at $\epsilon$=0, to emphasize the differences in slope, which 
determine $G_{Ep}$; based on the recoil polarization results the contribution of $G_{Ep}$ to the cross 
section at $Q^2$=5 GeV$^2$ is $\sim$1\%. 
\newline
\indent
All three corrections are 
different, and each one of them brings the Rosenbluth results 
closer to the recoil polarization results,
indicating that present uncertainties in the calculations of the 
radiative corrections of the cross section are at the level of several \%s. 
\newline
\indent
In~\cite{arrsick}, the effect of the Coulomb distortion of the 
incoming and outgoing electron waves on the 
extraction of the proton FFs was studied. Coulomb distortion 
corresponds to the exchange of one hard and one (or several) soft photons. 
It was found that it does yield an $\epsilon$ dependent correction to the 
elastic electron-proton cross sections. Although it reduces the cross
sections, its magnitude is too small to explain the discrepancy between 
Rosenbluth and polarization methods. It is however straightforward to 
calculate and should be included in the data analysis. 
\newline
\indent
Following the important discrepancy between the determinations of
$G_{Ep}/G_{Mp}$  using the polarization transfer and Rosenbluth
techniques, the role of two hard photon exchange effects, 
beyond those which have already 
been accounted for in the standard treatment of
radiative corrections has been studied. A general
study of two- (and multi)-photon exchange contributions to the
elastic electron-proton scattering observables was given 
in~\cite{guichon}.  
In that work, it was noted that the interference of the two-photon exchange
amplitude with the one-photon-exchange amplitude could
be comparable in size to the $(G_{Ep})^2$ term in the unpolarized cross
section at large $Q^2$.  In contrast, it was found that 
the two-photon exchange effects do not impact the
polarization-transfer extraction of $G_{Ep}/G_{pM}$ 
in an equally significant way.  
Thus a missing and un-factorisable part of the two-photon exchange 
amplitude at the level of a few percent may well explain 
the discrepancy between the two methods. 
\newline
\indent
Realistic calculations of
elastic electron-nucleon scattering beyond the Born approximation are
required in order to demonstrate in a quantitative way that 
$2 \gamma$ exchange effects are indeed able to resolve this discrepancy.  
\newline
\indent
Recently, several model calculations of the $2 \gamma$ exchange 
amplitude have been done. In ~\cite{blunden}, 
a calculation of the $2 \gamma$ exchange when the 
hadronic intermediate state is a nucleon was performed. 
It found that the $2 \gamma$ exchange correction with 
intermediate nucleon can partially resolve the discrepancy between the two 
experimental techniques. However, subsequently it was found 
in ~\cite{Kondratyuk:2005kk} 
that the effect is partly canceled when 
including the next hadronic intermediate state, the $\Delta(1232)$ resonance. 
The $2 \gamma$ exchange contribution to elastic 
$eN$ scattering has also been estimated 
at large momentum transfer~\cite{chen,afanbrod}, 
through the scattering off a parton in a proton by relating 
the process on the nucleon to the generalized parton distributions. 
This approach effectively sums all possible intermediate states corresponding to excitations of the
nucleon . Applying the two-photon exchange corrections to
the unpolarized data (see dotted curve in Fig.~\ref{fig:simplest}), 
yields a much flatter slope for the Rosenbluth plot, 
hence a much smaller value of $G_E$. 
The two-photon exchange corrections to the Rosenbluth process 
can therefore substantially reconcile the two
ways of measuring $G_E/G_M$ (compare dotted with thin solid curves in 
Fig.~\ref{fig:simplest}). 
\newline
\indent 
To push the precision frontier further in electron scattering, one needs 
a good understanding, of $2 \gamma$ exchange mechanisms, 
and of how they may or may not affect 
different observables. This justifies a systematic study of such 
$2 \gamma$ exchange effects, both theoretically and experimentally.  
Experimentally, the real part of the $2 \gamma$ exchange amplitude 
can be accessed through the difference between 
elastic electron and positron scattering off a nucleon. Such experiments 
are planned in the near future.  

To conclude, on the one hand the discussion above makes 
it clear that the radiative corrections, including two hard photon exchange, 
for the cross section data are not complete 
at this point in time. Therefore, the FFs 
$G_E^2$ and $G_M^2$ obtained using the 
Rosenbluth method above $Q^2$ of 2 GeV$^2$ are not correct. 
On the other hand, all the authors cited above agree that radiative 
corrections change the longitudinal and transverse polarization components, 
$P_t$ and $P_{\ell}$, in $\vec{e}+p\rightarrow{e}+\vec{p}$,
similarly, with the ratio $P_t/P_{\ell}$ affected only 
at the level of a few percent. The radiative 
corrections specifically calculated for the JLab polarization 
data by Afanasev {\it et al.}~\cite{afanasev} 
found that the corrections are $\sim$1\%,  whereas the hard two photon exchange 
effects are at the few percent level~\cite{chen, afanbrod}.  
Hence the polarization transfer method gives correct values for the FFs.

\section{Theoretical interpretation of nucleon electromagnetic form factors}
\label{sec4}

In this section we give an overview of the theoretical understanding 
of the nucleon e.m. FFs. 
These FFs encode the information on the structure of a  
strongly interacting many-body system of quarks and gluons, 
such as the nucleon. 
This field has a long history and many theoretical 
attempts have been made to understand the nucleon FFs. 
This reflects the fact that a direct calculation of nucleon FFs from the 
underlying theory, Quantum Chromodynamics (QCD), is complicated as it 
requires, in the few GeV momentum transfer region,  
non-perturbative methods. Hence, in practice it involves approximations 
which often have a limited range of applicability. 
Despite their approximations and limitations, some of these non-perturbative 
methods do reveal some insight in the nucleon structure. 
\newline
\indent
The earliest models to explain the global features of the nucleon FFs, such 
as its approximate dipole behavior, were vector meson dominance (VMD) models 
which are discussed in Sect.~\ref{theory_dispersion}. 
In this picture the photon couples to the nucleon through the exchange of 
vector mesons. Such VMD models are a special case of more general dispersion 
relation fits, 
which allow to relate time-like and space-like FFs, and which are
discussed subsequently. 
\newline
\indent
To understand the structure of the nucleon in terms of quark and gluon degrees
of freedom, constituent quark models have a long history. We discuss the 
intricacies in describing a bound system of relativistic constituent quarks 
and review the resulting predictions for FFs in Sect.~\ref{theory_models}. 
Despite some of their successes, models based on quarks alone do suffer from 
the evident shortcoming that they do not satisfy the global chiral symmetry 
of QCD when rotating left and right handed light quarks in flavor space. 
This chiral symmetry is broken spontaneously in nature, 
and the resulting Goldstone bosons are pions. Since they are the lightest 
hadrons, they dominate the low momentum transfer behavior of form factors, 
and manifest themselves in a pion cloud surrounding the nucleon. Such pion
cloud models will also be discussed in Sect.~\ref{theory_models}. 
\newline
\indent 
In Sect.~\ref{sec:rad}, we discuss the spatial 
information which can be obtained from the nucleon FFs, and discuss 
both radial densities and the issue of shape of the nucleon. 
\newline
\indent
Sect.~\ref{theory_eft} describes the chiral 
effective field theory of QCD and their predictions for nucleon FFs at low 
momentum transfers, where such perturbative expansions are applicable. 
\newline
\indent
In Sect.~\ref{sec:lattice}, we shall discuss the lattice QCD simulations, 
which have the potential to calculate nucleon FFs from first principles. 
This is a rapidly developing field and important progress has been made in the 
recent past. Nevertheless, the lattice calculations 
are at present still severely limited by available computing 
power and in practice are performed for quark masses sizably larger than their 
values in nature. We will discuss the issues in such calculations and compare 
recent results. It will also be discussed how the chiral effective field
theory can be useful in extrapolating present lattice QCD calculations to the 
physical pion mass.  
\newline
\indent
In Sect.~\ref{theory_gpd}, we discuss the quark structure of the nucleon and 
discuss generalized parton distributions (GPDs) of the nucleon. 
These GPDs are being accessed in hard exclusive reactions, which allow to 
remove in a controlled way a quark 
from the initial nucleon and implanting instead another quark in the final 
nucleon. The resulting GPDs can be interpreted as quark correlation functions 
and have the property that their first moments exactly coincide with the
nucleon FFs. We discuss the information which has been obtained on GPDs 
from fits of their first moments to the precise FF data set.   
\newline
\indent
Finally, in Sect.~\ref{theory_pqcd}, we discuss the nucleon FFs in the 
framework of perturbative QCD. These considerations are only valid 
at very small distances, where quarks nearly do not interact. In this limit, 
the nucleon FFs correspond to a hard photon which hits a valence quark 
in the nucleon, which then shares the momentum with the other (near collinear)
valence quarks through gluon exchange. 
We discuss the predictions made in this limit and confront them with the
experimental status for Dirac and Pauli FFs at large momentum transfers.

\subsection{Dispersion theory}
\label{theory_dispersion}

\subsubsection{Vector Meson Dominance (VMD)}

The starting point in understanding the interaction of a vector probe 
such as the photon with a hadronic system is provided by the observation 
that the lowest lying hadrons with vector quantum numbers 
are the vector mesons $\rho(770)$, $\omega(782)$ and $\phi(1020)$. 
In the process $e^+ e^- \to hadrons$, these vector mesons show up as 
prominent resonances at the corresponding values of the $e^+ e^-$  
squared $c.m.$ energy $q^2 > 0$. 
One therefore expects that in the elastic electron scattering process 
on the nucleon, $e N \to e N$, the nucleon electromagnetic FFs 
at low space-like momentum transfers, $q^2 < 0$, will be dominated by these 
lowest lying singularities from the time-like region. 
A large class of models for $F_1$ and $F_2$ are based on this vector 
meson dominance (VMD) hypothesis, as depicted in~\Figref{vmd}.
\begin{figure}[h]
\centerline{\epsfxsize=2.cm
\epsffile{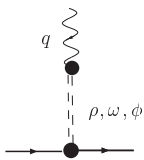}
}
\caption{\small The vector meson dominance picture for the coupling of the  
photon (with four-momentum $q$) to a nucleon. }
\figlab{vmd}
\end{figure}
\newline
\indent
Within such VMD models, the approximate dipole behavior of the nucleon 
e.m. FFs, see Eq.~(\ref{eq:dipole}), can be understood as being due to the 
contribution of two nearby vector meson poles which have opposite residua. 
Assume that one considers two vector meson pole contributions in \Figref{vmd} 
(with masses $m_{V1}$ and $m_{V2}$ 
and residua of equal magnitude and opposite sign $a$ and $-a$ respectively), 
one obtains~:
\begin{eqnarray}
F_{1,2} (q^2) &\sim& \frac{a}{q^2 - m_{V1}^2} + \frac{(-a)}{q^2 - m_{V2}^2} 
= \frac{a \, (m_{V1}^2 - m_{V2}^2)}{(q^2 - m_{V1}^2) (q^2 - m_{V2}^2)}.
\end{eqnarray}
\indent
An early VMD fit was performed by Iachello {\it et al.}~\cite{iach} and  
predicted a linear decrease of the proton $G_{Ep} / G_{Mp}$ ratio, which 
is in basic agreement with the result from the polarization transfer 
experiments. Such VMD models have been extended by Gari and 
Kr\"umpelmann~\cite{gari} to include the perturbative QCD (pQCD) 
scaling relations~\cite{brodlep}, which 
state that (see Sect.~\ref{theory_pqcd}) $F_1 \sim 1/Q^4$, and  
$F_2 / F_1 \sim 1/Q^2$.
\newline
\indent
In more recent years, extended VMD fits which provide 
a relatively good parameterization of  
all nucleon e.m. FFs have been obtained. An example is Lomon's 
fit~\cite{lomon}, using $\rho(770)$, $\omega(782)$, $\phi(1020)$, 
and $\rho^\prime(1450)$ mesons and containing 11 parameters. 
Another such recent parameterization 
by Bijker and Iachello~\cite{bijker} including 
$\rho(770)$, $\omega(782)$, and $\phi(1020)$ mesons only achieves a 
good fit by adding a phenomenological contribution attributed to a 
quarklike intrinsic $qqq$ structure (of $rms$ radius $\sim 0.34$~fm) 
besides the vector-meson exchange terms. The pQCD scaling relations are 
built into this fit which has 6 free parameters which are fit to the data. 
In contrast to the early fit of Ref.~\cite{iach}, the new fit of 
Ref.~\cite{bijker} gives a very good description of the neutron data, 
albeit at the expense of a slightly worse fit for the proton data. 
\newline 
\indent
It will be interesting to check the resulting VMD fits 
for the neutron FFs to larger $Q^2$. 
In this regard, an interesting ``prediction'' can be drawn when the FFs 
$F_2$ and $F_1$ obtained directly from double
polarization experiments are shown in the same graph for the 
proton and the neutron, as in Fig.~\ref{fig:f2andf1}. 
It is remarkable that both $F_1$ and $F_2$ tend toward the 
same value for proton and neutron, and may meet at a $Q^2$ value 
which will soon be accessible for the neutron. 
This conclusion is influenced by the VMD fits shown in the same figure, 
and rests on their extrapolation for the neutron to larger $Q^2$.  
Note that the VMD fits shown include all data for p and n, 
but selects the recoil polarization over the Rosenbluth results 
for $Q^2$ larger than 1 GeV$^2$.  
\begin{figure}[h]
\begin{center}
\epsfig{file=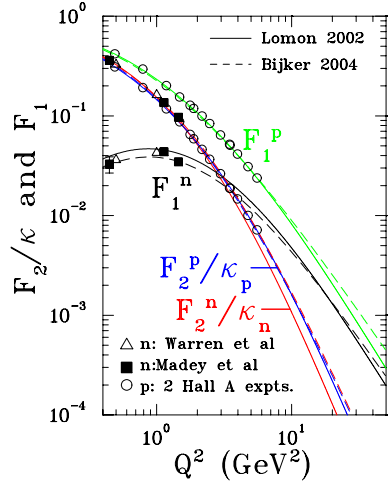,height=2.5in}
\caption{\small The FFs $F_1$ and $F_2$ for the proton and the neutron obtained
from double polarization experiments only.
The values of $F_2$ and $F_1$ were obtained from the experimental FF ratios
using fitted values to the data base for $G_{Mp}$ and $G_{Mn}$. 
For the proton data, the fit from \cite{kelly04} was used; 
for the neutron data in \cite{madey} the fit from \cite{kelly04}, and 
in \cite{warren} the fit from \cite{kubon} were used. 
The curves are the VMD fits of Lomon~\cite{lomon}
and of Bijker and Iachello~\cite{bijker}.}
\label{fig:f2andf1}
\end{center}
\end{figure}

\subsubsection{Dispersion analyses}

Despite the relatively good fits obtained by the VMD models, it was already 
pointed out in 1959 by Frazer and Fulco~\cite{frazer} that such an approach is 
at odds with general constraints from unitarity. 
Assuming an unsubtracted dispersion relation (DR), 
the nucleon e.m. FFs $F(q^2)$, 
where $F$ generically stands for any of the four FFs, can be obtained as~: 
\begin{eqnarray}
F(q^2) = \frac{1}{\pi} \int_{t_0}^{\infty} \, dq^{\prime \, 2} \, 
\frac{\mathrm{Im} F(q^{\prime \, 2})}{q^{\prime \, 2} - q^2}.  
\end{eqnarray}
The dispersion analyses are performed separately for nucleon isoscalar and 
isovector FFs.   
In the vector-isovector spectral function $\mathrm{Im} F(q^{\prime \, 2})$ 
one notices a large non-resonant contribution starting 
from $t_0 = 4 m_\pi^2$ and extending under the $\rho$-peak. Such a 
non-resonant contribution arises due to the two-pion continuum. 
For the isoscalar spectral function, the integral starts at 
$t_0 = 9 m_\pi^2$, corresponding to $3 \pi$ intermediate states. 
The two-pion continuum  
contribution was estimated by H\"ohler and collaborators~\cite{hohler} by 
using pion time-like FF data and $\pi \pi \to N \bar N$ amplitudes 
which were determined by extrapolating $\pi N$ partial waves to the time-like 
region~\cite{hohlerpiet}. 
\newline
\indent
H\"ohler's analysis has been updated by 
Mergell, Meissner, and Drechsel~\cite{mergell} in the mid-nineties and 
extended to include the nucleon time-like FF data~\cite{hmd96}. 
The inclusion of recent neutron FF data in such dispersion 
relation analysis has been performed in Ref.~\cite{hammer}. 
The resulting analysis describes the nucleon isovector FFs through the 
$2 \pi$ continuum (including the $\rho(770)$), and three additional 
vector isovector meson poles~: 
$\rho^\prime (1050)$,  
$\rho^{\prime \prime} (1465)$,  
$\rho^{\prime \prime \prime} (1700)$. The isoscalar FFs 
are described by four vector isoscalar meson poles~:
$\omega(770)$, $\phi(1020)$, $S^\prime(1650)$ and $S^{\prime \prime}(1680)$. 
In this approach, the masses of the mesons $\rho^\prime$,   
$\rho^{\prime \prime}$, $\rho^{\prime \prime \prime}$, 
$S^\prime, S^{\prime \prime}$ and the 14 residua (one for both the 
vector ($F_1$) and tensor ($F_2$) channels for each meson) are fitted and the 
pQCD scaling behavior is parameterized through three additional parameters. 
Note that for the isovector channel, the fitted masses for 
$\rho^{\prime \prime}$ and $\rho^{\prime \prime \prime}$ correspond to 
physical particles listed by the Particle Data Group (PDG), 
whereas enforcing the correct 
normalization of all FFs, the experimental value for the neutron charge 
radius, as well as the pQCD scaling behavior, requires the inclusion of 
an unphysical $\rho^\prime$ meson with mass 1050 MeV. The analysis of 
Hammer and Meissner~\cite{hammer} also finds that the residua for both 
isovector FFs $F_1^V$ and $F_2^V$ of  $\rho^{\prime \prime}$ and 
$\rho^{\prime \prime \prime}$ are relatively close in magnitude and 
opposite in sign, required by the approximate dipole behavior of the 
isovector FFs. For the isoscalar FFs $F_1^S$ and $F_2^S$, the fit 
also drives the residua of the nearby poles $S^\prime$ and $S^{\prime \prime}$ 
to values very close in magnitude and of opposite signs, required by the 
approximate dipole behavior of the isoscalar FFs. Using such an analysis, 
a good description of most FF data 
with the exception of the $G_{Ep} / G_{Mp}$ polarization data at 
$Q^2 > 3$~GeV$^2$, was obtained in \cite{hammer}.   
\begin{figure}[h]
\centerline{\epsfxsize=11.5cm
\epsffile{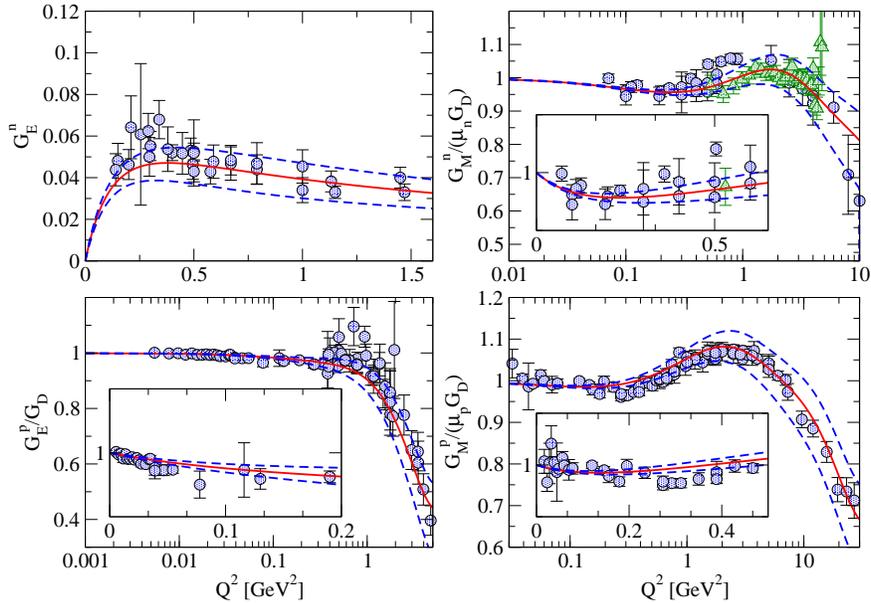}
}
\caption{\small Dispersion relation (15 parameter) 
fit for the four nucleon (space-like) e.m. FFs compared with the 
world data (circles) including 
the JLab/CLAS data for $G_{Mn}$ (triangles)~\cite{brooks}. 
The dashed curves indicate the $1 \sigma$ deviation from the fit, 
given by the solid curves.  
Figure from Ref.~\cite{Belushkin:2006qa}.}
\figlab{belushkin_dr}
\end{figure}

The dispersion relation analysis of nucleon e.m. FFs has been further 
improved by Belushkin {\it et al.} \cite{Belushkin:2006qa}. 
In addition to the $2 \pi$ continuum present in the isovector 
spectral functions of the previous DR  
analyses, also the $\rho \pi$ and $K \bar K$ continua were 
included as independent input in the isoscalar spectral functions. 
In Ref.~\cite{Belushkin:2006qa}, the $2 \pi$ continuum was reevaluated using 
the latest experimental data for the pion FFs in the time-like 
region~\cite{Belushkin:2005ds}. The $K \bar K$ continuum was obtained 
from an analytic continuation of $KN$ scattering data~\cite{HammerMusolf}. 
Following the work of Ref.~\cite{Meissner:1997qt}, 
the $\rho \pi$ continuum was approximated in the DR  
analysis by an effective pole term 
for a fictitious $\omega^\prime$ meson with mass~: 
$m_{\omega^\prime} = 1.12$~GeV. This approximate $\rho \pi$ continuum is 
found to yield an important negative contribution to $F_1^S$.  
The remaining contributions to the spectral functions are parameterized 
by vector meson poles from a fit to the FF data.  
The parameters in the fit were constrained to yield the correct 
normalization of the FFs at zero momentum transfers. The asymptotic 
constraints from pQCD were included in two different forms~: either as a 
superconvergence relation or by adding an explicit continuum term 
with the imposed pQCD behavior. A simultaneous fit to the world data 
for all four FFs in both the space-like and time-like 
regions was performed. \Figref{belushkin_dr} shows this fit 
for the nucleon space-like FFs where for 
 $G_{Ep}/G_{Mp}$ at larger $Q^2$ the JLab/Hall A 
polarization data \cite{jones,gayou2,punjabi05} have been used, 
and where for $G_{Mn}$ the preliminary JLab/CLAS data~\cite{brooks} 
have been included. In this fit, the pQCD limit was imposed through an 
explicit continuum term and the minimum number of poles in addition to the 
$\pi \pi$, $\rho \pi$ and $K \bar K$ continua were chosen to 
fit the data. In addition to the $\omega(782)$, the fit yields two more 
isoscalar poles ($m_{s1} \simeq 1.05$~GeV and  $m_{s2} \simeq 1.4$~GeV), 
and three additional isovector poles 
($m_{v1} \simeq 1.0$~GeV, $m_{v2} \simeq 1.6$~GeV, and 
$m_{v3} \simeq 1.8$~GeV ). The resulting 15 parameter fit shown 
in \Figref{belushkin_dr} has a total $\chi^2$/d.o.f. value of 2.2. 
\newline
\indent
It will be interesting to confront the most recent and sophisticated DR 
fit of~\cite{Belushkin:2006qa} with upcoming data for 
$G_{Ep}/G_{Mp}$ out to 8.5 GeV$^2$~\cite{E-04-108}. 
In all discussed VMD and DR fits starting with 
Gari-Kr\"umpelmann~\cite{gari} the asymptotic pQCD limit 
$F_2 / F_1 \sim 1/Q^2$ was built in, 
although the data do not support this limit at available momentum transfers, 
see Fig.~\ref{fig:scaling} 
in Sect.~\ref{theory_pqcd}. Besides, there is various 
theoretical work indicating that the pQCD prediction, in particular for 
$F_{2 p}$, might only set in at significantly larger values of $Q^2$, of 
the order of several tens of GeV$^2$. It might therefore be worthwhile  
to investigate how the DR analysis changes by removing the bias 
$F_2 / F_1 \sim 1/Q^2$ from the analysis, when fitting data in the range up to 
$Q^2 \sim 10$~GeV$^2$.

\subsection{Quark models versus pion-cloud models}
\label{theory_models}

\subsubsection{Constituent quark models}

In our quest to understand the structure of the nucleon in 
terms of the quark and gluon degrees of freedom which appear in the 
QCD Lagrangian, 
constituent quark models (CQMs) have a long history, which predates the 
establishment of the theory of strong interactions, QCD.
In a CQM, the nucleon appears as the ground state 
of a quantum-mechanical three-quark system in a confining potential.
In such a picture, the ground state baryons (composed of the 
light up ($u$), down ($d$) and strange ($s$) quark flavors) are 
described by $SU(6)$ spin-flavor wave functions (WFs), supplemented 
by an antisymmetric color WF.  
\newline
\indent
In the Isgur-Karl model~\cite{isgur}, 
the constituent quarks  
move in a harmonic oscillator type confining potential. 
For the ground state baryons, 
the three constituent quarks are in the $1s$ 
oscillator ground state, corresponding to the [56]-plet of $SU(6)$. 
In the Isgur-Karl model, the long-range confining potential is supplemented by 
an interquark force corresponding to one-gluon exchange. 
The one-gluon exchange leads to a color hyperfine interaction 
between quarks, which 
breaks the $SU(6)$ symmetry and leads to a mass splitting 
between $N(939)$ and $\Delta(1232)$, often referred to as the hyperfine 
splitting. 
It was found that it also predicts well the mass splittings between octet 
and decuplet baryons~\cite{derujula}. 
Furthermore, the color hyperfine interaction leads to a tensor force  
which produces a small $D$-state ($L = 2$) admixture in the 
$N$ (as well as $\Delta$) ground states~\cite{Koniuk:1979vy,Isgur:1981yz},  
corresponding to a $D$-state probability in the $N$ ground state 
around 0.2~\%.    
Even though such $D$-wave probability is small, 
it leads to a non-spherical charge distribution.  
For a static charge distribution, a measure of  
the non-sphericity (or deformation) is given by its quadrupole moment.  
Since the nucleon has spin 1/2, 
an intrinsic quadrupole moment of the nucleon 
cannot be directly measured because angular momentum conservation 
forbids a non-zero matrix element of a ($L = 2$) quadrupole operator 
between spin 1/2 states. 
However this quadrupole deformation may reveal itself in an 
electromagnetically induced transition 
from the spin 1/2 $N$ to the spin 3/2 $\Delta$ state. 
In this way, the tensor force between quarks gives rise to non-zero values 
for the electric quadrupole ($E2$) and Coulomb quadrupole ($C2$) 
transitions\footnote{  
The relation between the tensor force, $D$-wave admixture,
and the electromagnetic $N \to \Delta$ transition 
was already pointed out in the early paper of 
Glashow~\cite{Glashow79}. An up-to-date discussion of this field 
can be found in the review of~\cite{PVY06}.}. 
\newline
\indent
The non-relativistic CQM, despite its simplicity, 
is quite successful in predicting the spectrum of 
low-lying baryons, and gives a relatively good 
description of static properties such as the octet baryon magnetic moments.  
To calculate the FFs of a system of constituents with  
masses small compared with the confinement mass scale necessitates however 
a relativistic treatment even for low momentum transfers. For momentum 
transfers several times the nucleon mass squared a relativistic description  
becomes crucial. 
\newline
\indent
In contrast to the calculation of the spectrum, which uses eigenfunctions 
of a Poincar\'e invariant mass operator, 
a calculation of the nucleon electromagnetic FFs requires 
the relation between the rest frame spin and momenta 
(in the three-quark WF) and those in the moving frame.  
This requires 
an extension of eigenfunctions of the spin and mass operators, 
so as to transform consistently under 
the unitary representations of the Poincar\'e group.  
The way to implement relativity into a Hamiltonian formalism 
(describing e.g. a system of three interacting constituent quarks) 
has been laid 
out by Dirac~\cite{dirac}. There are three {\it forms of dynamics} 
(so called instant, point, and light-front forms) which 
differ in the choice of the kinematical subgroup of the Poincar\'e group. 
This is the subgroup of the Poincar\'e group whose commutator relations 
are not affected by the interactions between the constituents. 
The three (unitarily equivalent) forms therefore 
differ by which of the ten generators of the Poincar\'e group 
(four space-time translations, three spatial rotations, and three boosts) 
are kinematical (i.e. interaction free),  
and which are dynamical, i.e. depend on the interactions and 
necessarily have to be approximated in a practical calculation. 
\newline
\indent
In the {\it instant form}, the dynamical generators are the time component 
of the four-momentum and the three boost operators.  
Rotations do not contain interactions, which makes it 
easy to construct states of definite angular momentum in this form. 
\newline
\indent
In the {\it point form}, both boosts and rotations are kinematical. 
The point-form therefore has the important 
technical advantage that the angular momenta and Lorentz boosts are the same 
as in the free case. However all four components of the four-vector operator 
are dynamical in this form.  
\newline
\indent
In the {\it light-front form}, seven of the generators of the Poincar\'e 
group are kinematical (this corresponds to the symmetry group of a 
null plane), which is the maximum number possible. The 
remaining three dynamical generators which contain the interactions 
are one component of the four-momentum operator (the so-called light-cone 
Hamiltonian) and 2 transverse 
rotations. Light-front (as well as point form) calculations 
for relativistic CQMs are convenient as they allow to boost 
quark WFs independently 
of the details of the interaction. The drawback of the light-front 
calculations however is that because two generators 
of rotations are dynamical, the construction of states with good 
total angular momentum becomes interaction dependent. 
\newline 
 \indent
Any practical calculation in one of the three forms approximates the 
current operator. The common (so-called impulse) approximation 
is that the photon 
interacts with a single quark in the nucleon. 
\newline
\indent
The light-front form calculation of nucleon FFs has been pioneered by 
Berestetsky and Terentev~\cite{Berestetsky}, and more recently developed  
by Chung and Coester~\cite{chung}. In practice one starts from a 
rest frame nucleon WF for the three-quark state which ideally 
is fitted to the baryon spectrum. The nucleon WF 
in the light-front form (so-called light-front WF) is obtained by 
a Melosh rotation ~\cite{Melosh:1974cu} 
of each of the quark spinors, connecting the instant and light-front forms.  
When performing the front form calculation in a (Drell-Yan) 
frame where the photon light-cone 
momentum~\footnote{Defining light-cone components as 
$x^\pm = (x^0 \pm x^3)/\sqrt{2}$ and defining the null-plane by $x^+ = 0$.} 
component $q^+ = 0$, the space-like virtual photon 
only connects Fock components in the nucleon light-front WFs with 
the same number of constituents, i.e. matrix elements between $qqq$ and 
$qqqq \bar q$ states which would be present in an instant form calculation 
are zero in the light-front calculation. 
This property allows for a consistent calculation within 
the light-front formalism when truncating the Fock space to only the 
three-quark state. 
\newline
\indent
In~\cite{chung} a Gaussian WF in the quark internal 
(transverse) momentum variables was used. Although this model yields a 
surprisingly good agreement for the observed $G_{Ep}/G_{Mp}$ ratio, 
see \Figref{relcqm}, it yields nucleon FFs which 
drop too fast at larger $Q^2$ values when using constituent quark masses 
around 330 MeV.  
Schlumpf~\cite{schlumpf} allowed for high momentum components in the 
nucleon light-front WF by adopting 
a power law dependence in the 
quadratic quark internal momentum variables. The two parameters in Schlumpf's  
WF were fitted to magnetic moments and semi-leptonic decays of the 
baryon octet. The resulting e.m. FF calculations reproduce reasonably well 
the power behavior of the FF at larger $Q^2$. 
The WF of Schlumpf was also used 
by Frank, Jennings, and Miller~\cite{gamiller,miller02}. 
They found that such a light-front WF leads to a violation of 
hadron helicity conservation resulting in a  
$F_{2p}/F_{1p}$ ratio which drops less fast than 
$1/Q^2$~\cite{miller02}, in agreement with the $G_{Ep}/G_{Mp}$ polarization 
data.    
\begin{figure}[h]
\begin{center}
\includegraphics[width =5cm,angle=90]{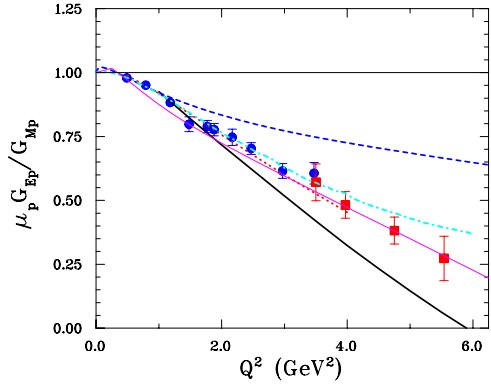}
\end{center}
\vspace{-0.25cm}
\caption{\small Comparison of relativistic CQM calculations 
with the data for $\mu_p G_{Ep} / G_{Mp}$. 
Dotted curve : front form calculation of Chung and Coester~\cite{chung} 
with point-like constituent quarks; 
thick solid curve : front form calculation of Frank {\it et al.}~\cite{gamiller}; 
dot-dashed curve : front form calculation of 
Cardarelli {\it et al.}~\cite{rome,cardarelli} with point-like constituent quarks; 
dashed curve : point form calculation of Boffi et 
al.~\cite{boffi} in the Goldstone boson exchange model with point-like 
constituent quarks; 
thin solid curve : covariant spectator model of 
Gross and Agbakpe~\cite{gross}. 
The data are from~\cite{punjabi05} (solid circles) and  
\cite{gayou2} (empty squares). 
}
\figlab{relcqm}
\end{figure}

The WFs in the calculations described above were however not 
constructed from a detailed fit to the baryon spectrum. 
Cardarelli {\it et al.}  
subsequently performed a more ``microscopic'' light-front 
calculation~\cite{rome,cardarelli} 
where the light-front WF was obtained from a rest frame WF 
which provided a fit to the spectrum. 
The rest frame WF was taken from 
the relativized Capstick-Isgur model~\cite{Capstick:1986bm}.  
Using this WF, 
the constituent quark momentum distribution in the nucleon 
was found to yield an important content 
of high-momentum components, which are generated by the short-range 
part of the quark-quark interaction, which is due to one-gluon exchange in 
the Capstick-Isgur model. These components are completely absent if one only 
considers the linear confinement potential in the model. 
\newline
\indent
In a CQM calculation,  
the effect of other degrees of freedom beyond three quarks are 
buried within the constituent quarks, which are considered as quasi-particles. 
In the absence of a microscopic calculation, 
such effects are parameterized in terms of constituent quark FFs. 
In~\cite{Petronzio:2003bw}, 
it was shown that the data for the proton unpolarized forward 
structure function at low momentum transfers exhibits a new scaling 
property and can be interpreted as quasi-elastic scattering off extended 
constituent quarks inside the proton described by a 
constituent quark FF. The resulting constituent size is 
around 0.2 - 0.3 fm. 
Using such effective constituent quark FF in 
the light-front form calculation of~\cite{cardarelli}, 
allows a good description of the individual nucleon FFs, see~\cite{pace}.
Note however that the experimental $G_{Ep} / G_{Mp}$ ratio 
can basically be reproduced using point-like constituent quarks, 
see \Figref{relcqm}. 
The suppression of the $G_{Ep}/G_{Mp}$ 
ratio with respect to the dipole-fit as predicted 
in the light-front form CQM calculation is 
attributed to relativistic effects generated by the Melosh rotations of the 
constituent quark spins. These Melosh rotations introduce kinematical 
$SU(6)$ breaking effects in addition to the dynamical 
$SU(6)$ breaking due to the (hyperfine) one-gluon exchange potential. 
\newline
\indent
A comparable amount of high-momentum components in the nucleon 
WF was obtained in the 
Goldstone-boson-exchange (GBE) 
quark model~\cite{Glozman:1997fs,Glozman:1997ag}.  
This model relies on constituent quarks and Goldstone 
bosons, which arise as effective degrees of freedom of low-energy QCD 
from the spontaneous breaking of the chiral symmetry. 
The resulting CQM assumes a linear confinement potential supplemented by 
a quark-quark interaction based on the exchange of pseudoscalar 
Goldstone bosons, which is the source of the hyperfine interaction. 
It was shown in~\cite{Glozman:1997fs,Glozman:1997ag}  that the GBE 
CQM yields a unified description of light- and strange-baryon spectra. 
The GBE CQM was used in~\cite{wagenbrunn,boffi} 
to calculate the nucleon e.m. FFs in the point-form. The neutron 
charge radius is well described in this model and is driven 
by the mixed-symmetry component in the neutron WF.  
In contrast to the light-front 
calculation~\cite{cardarelli,pace}, it was found that when performing 
a point-form calculation of the nucleon e.m. FFs at larger $Q^2$ 
within the impulse approximation, 
i.e. considering only single-quark currents, a surprisingly 
good overall description of the nucleon e.m. FFs can be obtained, using  
point-like constituent quarks only. 
When looking at details of Refs.~\cite{wagenbrunn,boffi}, 
the agreement is worse though for $G_{Mp}$ which 
is underpredicted at larger $Q^2$, and the ratio of $G_{Ep}/G_{Mp}$ is 
overpredicted at larger $Q^2$, see \Figref{relcqm}. 
Similar findings have also been obtained in the point-form calculation 
of~\cite{Wagenbrunn:2005wk} for the OGE CQM. 
The overall success of the point-form result using point-like constituent 
quarks was attributed in~\cite{wagenbrunn,boffi,Wagenbrunn:2005wk} 
to the major role played by relativity.  
Such a finding is remarkable in view of the expected finite 
size of the constituent quarks, as discussed above.  
\newline
\indent
An explanation for the above finding for the nucleon 
e.m. FFs in the point form, 
using the single-quark current approximation, 
has been suggested by Coester and Riska~\cite{Coester:2003rw}. 
When the spatial extent of the three-quark WF 
is scaled (unitarily) to zero, both  
instant- and front-forms yield FFs independent of the momentum 
transfer. Therefore, to reproduce 
the experimental fall-off of the nucleon e.m. FFs  
at large momentum transfers requires the introduction of constituent quark 
FFs. In contrast, when the WF in point form is scaled unitarily to 
zero (so-called point limit), a non-trivial scaling limit is obtained for 
the FFs, depending on the shape of the WF. 
At high values of momentum transfer, the scaled FFs decrease with an inverse 
power of the momentum transfer. The power is determined 
by the current operator and is independent of the WF. 
An explicit comparative calculation of the baryon e.m. FFs 
between the three different forms was performed in~\cite{Julia-Diaz:2003gq} 
using a simple algebraic form 
for the three-quark WF, depending on two parameters. 
It was verified that a qualitative description of the nucleon FF data 
demands a spatially extended WF in the 
instant- and front-form descriptions, in contrast to 
the point-form description which demands a much more compact WF.  
\newline
\indent
A manifestly covariant CQM calculation within the Bethe-Salpeter formalism 
and using an instanton-induced interaction between quarks has been performed 
by Merten {\it et al.}~\cite{Merten:2002nz}. Although this model 
reproduces the baryon spectrum, it can only qualitatively 
account for the $Q^2$ dependence of the nucleon e.m. FFs. 
\newline
\indent
Another covariant CQM calculation was performed by 
Gross and Agbakpe~\cite{gross}, using a covariant spectator model. 
Assuming a simple pure $S$-wave form for the nucleon three-quark wave 
function, evaluating the current matrix element in 
a relativistic impulse approximation, and assuming constituent quark FFs 
including a phenomenological term which parameterizes the pion cloud, 
an eleven parameter description of the nucleon FF data was obtained, 
see \Figref{relcqm}. 
\newline
\indent
As a next step for CQMs, it would clearly be very worthwhile to investigate 
the approximations in the current operator within each form.
The quality of the commonly made impulse approximation 
may differ between the different forms. 
Within the context of a toy model calculation in 
Refs.~\cite{Desplanques:2003nk,Desplanques:2005vu}, 
it has e.g. been shown that the neglect of two-body currents in the point form 
does affect the FFs in a more drastic way than their neglect in the instant 
or light-front forms. 
\newline
\indent
The importance of two-body currents has also been 
shown in the work of De Sanctis {\it et al.}~\cite{Desanctis}. 
In that work, a calculation 
within the hypercentral CQM was performed of the (two-body) quark pair 
contribution to the e.m. current resulting from the 
one-gluon exchange interaction between the quarks. This pair current 
contribution was found to lead to a sizeable reduction of $G_{Ep}$ 
compared with $G_{Mp}$.

\subsubsection{Pion cloud models}

Despite their relative success in describing the spectrum and structure of 
low-lying baryons, models based on constituent quarks alone 
suffer from evident shortcomings as they do not satisfy all symmetry 
properties of the QCD Lagrangian. In nature, the up and down (current) quarks 
are nearly massless. In the exact massless limit, the QCD Lagrangian is 
invariant under $SU(2)_L \times SU(2)_R$ rotations of left ($L$) and 
right ($R$) handed quarks in flavor space. This {\it chiral symmetry} 
is spontaneously broken in nature leading to the appearance of massless 
Goldstone modes. For two flavors, there are three Goldstone bosons ---
pions, which acquire a mass due to the explicit breaking of chiral 
symmetry by the current quark masses. 
\newline
\indent
Since pions are the lightest hadrons, they 
dominate the long-distance 
behavior of hadron WFs and yield characteristic 
signatures in the low-momentum transfer behavior of hadronic 
FFs. 
Therefore, a natural way to qualitatively improve 
on the above-mentioned CQMs is to include the pionic degrees of freedom 
~\cite{Manohar:1983md}.  
\newline
\indent
An early quark model with chiral symmetry 
is the chiral (or, cloudy) bag model. This model  
improves the early MIT bag model by introducing 
an elementary, perturbative pion which couples to quarks in the bag 
in such a way that chiral symmetry is restored~\cite{Thomas:1982kv}. 
Within the cloudy bag model, Lu {\it et al.}~\cite{lu} 
performed a calculation of the nucleon e.m. FFs 
improving upon previous calculations by 
applying a correction for the center-of-mass motion of the bag. 
This calculation also implemented Lorentz covariance in an approximate way 
by using a prescription for the Lorentz contraction of the 
internal structure of the nucleon. Using a bag radius $R \simeq 1$~fm, 
this model provides a good description of the nucleon e.m. FFs in the range 
$Q^2 < 1$~GeV$^2$.  
\newline
\indent
To extend such a calculation to larger $Q^2$, Miller performed a 
light-front cloudy bag model calculation~\cite{miller02b}. 
Starting from a model in terms of constituent quarks~\cite{miller02}, 
described by the light-front WF of Schlumpf, 
the effects of the pion cloud were 
calculated through one-loop diagrams, including relativistic 
$\pi N N$ vertex FFs. The model gives a relatively good gobal 
account of the data both at low $Q^2$ and larger $Q^2$, 
though tends to show too much structure around the dipole form 
for the magnetic FFs at low $Q^2$. 
\newline
\indent
The cloudy bag model is one chiral quark model which 
treats the effect of pions perturbatively. Other quark models 
which calculated nucleon e.m. FFs using perturbative pions can be found 
e.g. in the early works of~\cite{Oset:1984tv,Jena:1992qx}, as well as in 
the already discussed works of~\cite{Glozman:1997fs,Glozman:1997ag}.  
Recently, the above chiral quark models where pions are included 
perturbatively have been improved in~\cite{faessler}. 
This work extends a previous work of ~\cite{Lyubovitskij:2001nm} by 
dynamically dressing bare constituent quarks by mesons to fourth order 
within a manifestly Lorentz covariant formalism. 
Once the nucleon and $\Lambda$ hyperon magnetic moments are fitted, 
other e.m. properties, such as the
nucleon e.m. FFs at low momentum transfers, follow as a prediction. 
It was found in~\cite{faessler} 
that the meson cloud is able to nicely describe 
the FF data in the momentum transfer region up to about 0.5 GeV$^2$.  
To extend the calculations to larger $Q^2$, a phenomenological approach 
has been adopted in \cite{faessler} by introducing 
bare constituent quark FFs which were parameterized 
in terms of 10 parameters. Such parameterization makes it plausible to 
simultaneously explain the underlying dipole structure in the nucleon e.m. FF
as well as the meson cloud contribution at low $Q^2$ which results from 
the underlying chiral dynamics. In a later paper~\cite{Faessler:2006ky}, a 
model calculation for the bare constituent quark FFs has been performed and 
applied to the e.m. properties of the $N \to \Delta$ transition. 
\newline
\indent
When pion effects dominate nucleon structure, their effects have to be 
treated non-perturbatively. A non-perturbative 
approach which has both quark and pion degrees of freedom 
and interpolates between a CQM and 
the Skyrme model (where the nucleon appears as a soliton solution of an 
effective nonlinear pion field theory) is 
the chiral quark soliton model ($\chi$QSM). 
As for the Skyrme model, 
the $\chi$QSM is based on a $1/N_c$ expansion  
(with $N_c$ the number of colors in QCD).  
Its effective chiral action has been
derived from the instanton model of the QCD vacuum \cite{Dia86}, which
provides a natural mechanism of chiral symmetry breaking 
and enables one to generate dynamically the constituent 
quark mass.  
Although in reality the number of colors $N_c$ is equal to three, 
the extreme limit of large $N_c$ is 
known to yield useful insights. At
large $N_c$ the nucleon is heavy and can 
be viewed as $N_c$ ``valence" quarks bound by a self-consistent pion
field (the ``soliton")~\cite{Dia88}.
A successful description of static properties of baryons, 
such as mass splittings, axial constants, magnetic moments, 
FFs, has been achieved (typically at the 30 \% level or better, 
see~\cite{Chr96} for a review of early results). 
After reproducing masses and decay constants in the mesonic sector, 
the only free parameter left to be fixed in the baryonic sector
is the constituent quark mass. 
When taking rotational ($1/N_c$) corrections into account, 
this model achieved a qualitative good description of the nucleon e.m. 
FFs in the range $Q^2 < 1$~GeV$^2$, using a constituent 
quark mass around $420$~MeV~\cite{Christov:1995hr}.
The chiral soliton model naturally accounts for the decrease of the 
$G_{Ep} / G_{Mp}$ ratio with increasing $Q^2$. This can be understood 
from the hedgehog structure in soliton models which couples spatial rotations 
with isorotations. In the soliton rest frame, the isovector electric FF 
$G_E^V$ therefore measures the rotational inertia density $\rho^V(r)$, in 
contrast to the isoscalar electric FF $G_E^S$ which measures the 
isoscalar baryon density $\rho^S(r)$. For a rigid rotor, 
the inertia density is obtained from the baryon density as 
$\rho^V = r^2/r_B^2 \, \rho^S$, with $r_B$ a free parameter characterizing 
the spatial extent. Assuming a Gaussian density for $\rho^S(r)$, 
this yields~\cite{holzwarth}~:
\begin{eqnarray}
\frac{\mu_p G_{Ep}(Q^2)}{G_{Mp}(Q^2)} = 
1 - \frac{1}{18} Q^2 r_B^2. 
\label{eq:gegmsol}
\end{eqnarray} 
With the choice $r_B^2 \approx (0.3 \, \mathrm{fm})^2$, one can obtain an 
excellent fit of the polarization data for $G_{Ep}/G_{Mp}$. 
Although in the chiral soliton model calculation 
the baryon density is not exactly Gaussian, 
and the rigid rotor calculation does not hold exactly, these relations can be 
considered as approximate relations~\cite{holzwarth}. 
\newline
\indent
Holzwarth~\cite{holzwarth} extended the chiral soliton model by including the 
$\rho$ ($\omega$) meson propagators for the isovector (isoscalar) channels 
respectively. Furthermore, to extend the range in $Q^2$ of the predictions, 
he uses a relativistic prescription to boost the soliton rest frame densities 
to the Breit frame. Such prescription is also used to extract 
radial charge and magnetization rest frame densities 
from experimental FFs, as will be discussed in Sect.~\ref{sec:rad}. 
Using 4 fit parameters (one effective boost mass and three free parameters 
to fix the couplings of $\rho$ and $\omega$ mesons), 
the model was found to provide a good account of 
the detailed structure of the nucleon e.m. FFs in the low $Q^2$ region. 
In particular, for $G_{Ep}/G_{Mp}$ it 
predicts a decreasing ratio in good agreement with the data. 
At larger $Q^2$, the boost prescription gives a reasonably good account of the 
data (except for $G_{Mn}$) and predicts a zero in $G_{Ep}$ around 
10~GeV$^2$. Due to the uncertainty introduced from 
the particular choice for the boost prescription, the 
high $Q^2$ behavior (for $Q^2$ larger than about $4 M^2$) 
of the e.m. FFs is however not a profound prediction of the 
low-energy effective model.

\subsection{Radial distributions and shape of the nucleon}
\label{sec:rad}

\begin{figure}[h]
\centerline{\epsfxsize=10cm
\epsffile{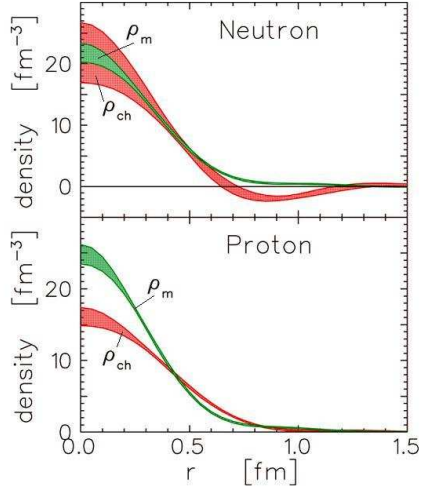}
}
\vspace{-.5cm}
\caption{\small Radial distributions of charge $\rho_{ch}$ 
and magnetization $\rho_m$ in the proton 
and neutron from~\cite{kelly02}, 
obtained from the Fourier Bessel transforms Eq.~(\ref{eq:fb2}) 
using the relativistic transformation of Eqs.~(\ref{eq:geft},\ref{eq:gmft}) 
with $n_E = n_M = 2$.  Note that the neutron charge distribution 
has been multiplied by a factor of 6 to emphasize the similarity in shape 
of charge and magnetization densities.}
\label{fig:densitynandp}
\end{figure}

As discussed in Sect.~\ref{Breitframe}, 
in the Breit frame the nucleon charge operator 
depends only on the electric FF $G_E$, whereas the e.m. three-current operator 
depends only on the magnetic FF $G_M$. This suggest to interpret the 
Fourier transforms of $G_E$ ($G_M$) as the nucleon charge (magnetization) 
densities. This identification is only appropriate for a 
non-relativistic (static) system however, 
as in general there is a variation of the Breit frame with $Q^2$. 
For the nucleon, where FF data have been obtained for $Q^2$ values 
much larger than $M^2$, one needs to take the effect of relativity 
into account. 
Recently Kelly \cite{kelly02} has used a relativistic prescription 
to relate the Sachs FFs to the nucleon charge and magnetization densities, 
accounting for the Lorentz contraction of the densities in the 
Breit frame relative to the rest frame. 
\newline
\indent
One starts from the spherical 
nucleon charge $\rho_{ch}(r)$ and magnetization $\rho_m(r)$  
densities in the nucleon rest frame. 
These densities are normalized so as to yield the total charge 
for $\rho_{ch}$, or one for $\rho_m$ 
(the magnetic moment is taken out of the density) as~:
\begin{eqnarray}
\int_0^\infty dr r^2 \, \rho_{ch}(r) = Z , \quad \quad \quad
\int_0^\infty dr r^2 \, \rho_{m}(r) = 1 , 
\end{eqnarray}  
where $Z = 0,1$ is the nucleon charge. 
From these intrinsic (rest frame) densities, one can construct 
intrinsic FFs $\tilde \rho(k)$ which are related 
through a Fourier-Bessel transform as~:
\begin{eqnarray}
\tilde \rho(k) \equiv  \int_{0}^\infty dr \, r^2 \, j_0(k r) 
\rho(r), \quad \quad \quad 
\rho(r) = \frac{2}{\pi} \int_{0}^\infty dk \, k^2 \, j_0(k r) 
\tilde \rho(k),
\label{eq:fb2}
\end{eqnarray}
with $k \equiv |\vec q|$ is the wave vector in the nucleon rest frame. 
For a non-relativistic system, the intrinsic FFs are obtained from the 
Sachs FFs using $k \to Q$ as~: $\tilde \rho_{ch}(k) \to G_E(Q^2)$, 
and $\mu_N \tilde \rho_{m}(k) \to G_M(Q^2)$. 
\newline
\indent
To properly relate the intrinsic FFs evaluated in the 
rest frame to the Breit frame, where the nucleon moves with velocity  
$v = \sqrt{\tau / (1 + \tau)}$ relative to the rest frame, 
involves a Lorentz boost with~:
$\gamma^2 = (1 - v^2)^{-1} = 1 + \tau$. 
This Lorentz boost leads to a contraction of the nucleon densities as 
seen in the Breit frame. Consequently, in the Fourier-transforms, this amounts 
to replace in the intrinsic FF arguments~:
\begin{eqnarray}
k^2 \to Q^2 / ({1 + \tau}).
\label{eq:kqsqr}
\end{eqnarray}
To relate intrinsic FFs $\tilde \rho(k)$ 
with the Sachs FFs $G(Q^2)$ is not 
unambiguous however because the boost operator for a composite system 
depends on the interactions among its constituents. 
There exist different prescription in the literature which can be 
written in the form~:
\begin{eqnarray}
\tilde \rho_{ch}(k) &=& \gamma^{2 \, n_E} \, 
G_E(Q^2) = (1 + \tau)^{n_E} \, G_E(Q^2), 
\label{eq:geft} \\
\mu_N \tilde \rho_m(k) &=& \gamma^{2 \, n_M} \, 
G_M(Q^2) = (1 + \tau)^{n_M} \, G_M(Q^2),
\label{eq:gmft}
\end{eqnarray}
where $k$ and $Q^2$ are related as in Eq.~(\ref{eq:kqsqr}). 
For $Q^2 \to \infty$, the boost maps $G(Q^2 \to \infty)$ to 
$\tilde \rho(2 M)$. One sees that there is a limiting wave vector 
$k_{max} = 2 M$ determined by the nucleon Compton wavelength. 
In the rest frame, no information can be obtained on distance scales 
smaller than the Compton wavelength due to relativistic position fluctuations 
(known as the {\it Zitterbewegung}). 
To account for an asymptotic $1/Q^4$ FF behavior, 
Mitra and Kumari~\cite{mitra} proposed    
the choice $n_E = n_M = 2$. 
Kelly~\cite{kelly02} followed this choice when extracting the 
rest frame densities from the measured nucleon e.m. FFs.  
\newline
\indent
In his analysis, Kelly furthermore 
minimized the model dependence of the fitted densities 
by using an expansion in a complete 
set of radial basis functions.
For $Q^2 >$ 1~GeV$^2$ the $G_{Ep}$ analysis used recoil polarization 
data from JLab \cite{jones,gayou2}
rather than the Rosenbluth separation data. 
Fig.~\ref{fig:densitynandp} compares 
the fitted charge and magnetization densities for neutron and proton. 
The uncertainty bands include both statistical and 
incompleteness errors. The excess of negative charge near $r~\sim0.8-1.0$ fm 
is a characteristics of the $\pi^-$-meson cloud in the neutron. 
The proton charge density is significantly broader than the magnetization 
density, 
a direct consequence of $G_{Ep}$ being softer than $G_{Mp}$ in $Q^2$-space. 

To investigate the pion cloud as revealed through the neutron electric 
charge distribution further, Friedrich and Walcher~\cite{walcher} have 
performed a phenomenological analysis of all four nucleon e.m. FFs. 
They performed a two-component fit of the four FFs starting from 
a smooth part (parameterized by a sum of two dipoles) and by adding 
on top of it a Gaussian ``bump'' structure.  
The choice of such a two-component form was triggered 
by the behavior of $G_{En}$ 
at small $Q^2$, and by the observation of the noticeable 
oscillations of the other three e.m. FFs around the dipole form. 
Their parameterization allows for 6 fit parameters for each FF, which 
provide an excellent fit to the FFs. 
When subtracting from the data two dipoles with suitably chosen parameters, 
the remaining part displays a bump structure 
as shown in \Figref{friedwalcher}. 
Friedrich and Walcher made the striking observation that all four 
FFs display such a bump structure around $Q^2 \approx 0.25$~GeV$^2$. 
They intepret this structure as a signature of the pion cloud. 
Upon Fourier transforming, the corresponding Breit-frame densities, 
corresponding to the ``bump'' structure in the FFs, were found to extend 
as far out as 2~fm. It is interesting to compare this with the findings of 
the dispersion theory, in which the longest range part of the pion cloud 
contribution to the nucleon e.m. FFs is given by the $2 \pi$ continuum. 
These $2 \pi$ continuum contributions were found to be much more confined 
in coordinate space~\cite{Belushkin:2005ds}. In order to get a bump  
structure in $G_{En}$ in the DR theory requires to introduce additional 
strength in the spectral functions below 1 GeV. 
New high precision data for $G_{Ep}/G_{Mp}$ 
from the BLAST experiment at MIT-Bates~\cite{crawford}, shown in 
Fig.~\ref{fig:gepgdpol},  
confirms the dip structure around $Q^2 \approx 0.3$~GeV$^2$. 
It will also be interesting 
to compare upcoming data of BLAST for $G_{En}$ in the same range with a 
parameterization as in \Figref{friedwalcher}~\footnote{For preliminary data 
from BLAST for $G_{En}$, see Ref.~\cite{Ziskin}.}.  
\begin{figure}[h]
\vspace{-.5cm}
\centerline{ \epsfxsize=19cm \epsfysize=16cm
\epsffile{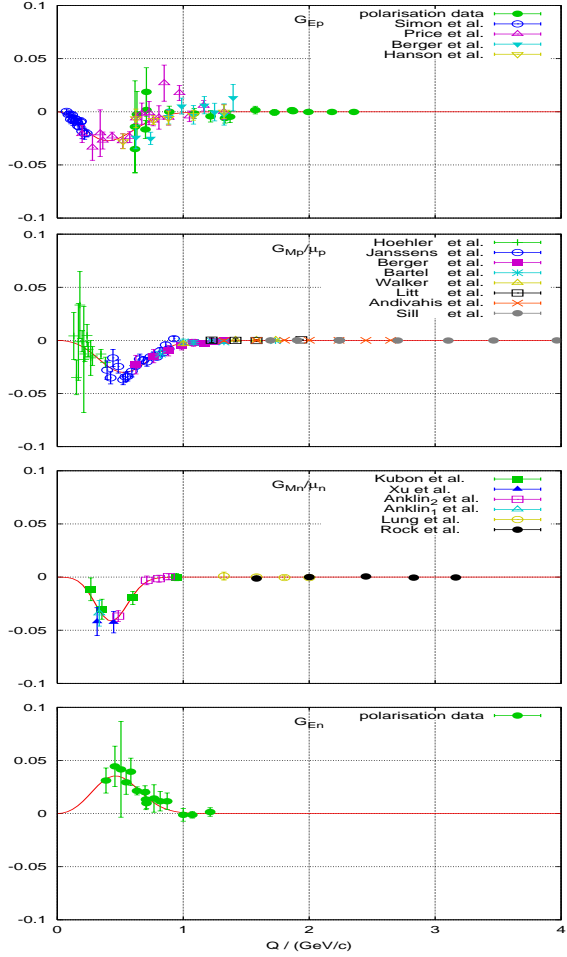}
}
\vspace{-3cm}
\caption{\small  
Phenomenological two-component fit of the nucleon e.m. FFs 
according to Friedrich and Walcher~\cite{walcher}. The 6-parameter fit 
for each FF consists of a smooth part, described by a sum of two dipoles, 
and a Gaussian bump part. The latter is displayed in the figure.
}
\figlab{friedwalcher}
\end{figure}

Miller~\cite{miller03} has defined spin-dependent quark densities as 
matrix elements of density operators in proton states of definite
spin-polarization. Within a constituent quark picture, 
the spin-dependent density operator for 
a quark in the proton to be found at position ${\bf r}$ 
and with spin-direction ${\bf \hat n}$ is given by~:
\begin{eqnarray}
\hat \rho({\bf r}, {\bf \hat n}) = 
\sum_i \frac{e_i}{e} \delta({\bf r} - {\bf r}_i) 
\frac{1}{2} ( 1 + {\bf \sigma}_i \cdot {\bf \hat n} ),
\end{eqnarray}
where the sum runs over the three constituent quarks $i$ with 
fractional charge $e_i/e$. 
Relative to the spin-direction ${\bf \hat s}$ of the proton, 
Miller then studied the 
distribution of quarks for different quark spin orientations ${\bf \hat n}$. 
The so defined densities may become non-spherical as shown in 
Refs.~\cite{miller03,miller06}.  
Averaging over quark spin ${\bf \hat n}$ or over nucleon spin ${\bf \hat s}$  
yields a spherical distribution~\cite{gross}.

\subsection{Chiral perturbation theory}
\label{theory_eft}

At low momentum transfers $Q^2$, 
the nucleon e.m. FFs can also be studied within
chiral perturbation theory ($\chi$PT) expansions based on chiral Lagrangians 
with pion and nucleon fields. 
In $\chi$PT, the short-distance physics is 
parameterized in terms of low-energy-constants (LECs) which ideally 
can be determined by matching to QCD but in practice are fitted 
to experiment or are estimated using resonance saturation. 
In the calculation of the nucleon e.m. 
FFs, the LECs can be fitted to the nucleon charge radii and the anomalous
magnetic moments. Once they are fixed, the $Q^2$ dependence of the FFs  
follows as a prediction.
\newline
\indent 
To calculate the nucleon e.m. FFs in $\chi$EFT 
involves a simultaneous expansion in soft scales~: 
$Q^2$ and $m_\pi$, which are  
understood to be small relative to the chiral symmetry breaking 
scale $\Lambda_{\chi SB}~\sim$~1~GeV. 
Several expansion schemes (also called 
power-counting schemes) have been developed in the literature. 
They all yield the same non-analytic dependencies (e.g. terms proportional to 
$m_\pi$, $m_\pi^3$, $ m_\pi^2 \ln m_\pi^2$, ...) 
but differ in analytic terms 
(e.g. terms proportional to $m_\pi^2$, $m_\pi^4$, ...). 
\newline
\indent
Because the first nucleon excitation, the $\Delta(1232)$ resonance, 
has an excitation energy of only about 
$\vDe \equiv M_\Delta - M \simeq 300$~MeV, 
the $\Delta$ resonance is often included as an explicit degree of 
freedom in the theory. The resulting chiral effective theory ($\chi$EFT) 
includes pion, nucleon, and $\Delta$ fields. 
When including the $\Delta$ as an explicit degree of freedom in the chiral 
Lagrangian, the counting scheme has to specify how the expansion parameter 
$\varepsilon \equiv m_\pi / \Lambda_{\chi SB}$ is counted relative to 
$\delta \equiv \vDe / \Lambda_{\chi SB}$. 
In the small scale expansion (SSE)~\cite{HHK97}, 
also called $\varepsilon$-expansion, 
the pion mass and the $M_\Delta - M_N$ mass 
difference are counted on the same footing,
i.e. $\varepsilon \sim \delta$. 
The recently developed $\delta$-expansion scheme, 
see~\cite{PVY06} for a review and applications, counts 
the pion mass as $\varepsilon \sim \delta^2$, 
which is the closest integer power 
relation between these parameters in the real world. 
\newline
\indent
Early calculations of the nucleon e.m. FFs in the SSE at order $\varepsilon^3$ 
have been performed in Ref.~\cite{Bernard:1998gv}. Because such an approach 
is based on a heavy baryon expansion it is limited to $Q^2$ values much below 
0.2~GeV$^2$. 
\newline
\indent
Subsequently, several calculations of the nucleon e.m. FFs 
have been performed in manifestly Lorentz invariant $\chi$PT. 
Kubis and Meissner~\cite{kubis01} performed a calculation to fourth order 
in relativistic baryon $\chi$PT, employing the 
infrared regularization (IR) scheme. 
They showed that the convergence of the chiral expansion is improved 
as compared to a heavy baryon $\chi$PT results. 
Schindler {\it et al.}~\cite{schindler} also performed a 
manifestly Lorentz invariant calculation to fourth order, employing the 
extended on-mass-shell (EOMS) renormalization scheme. 
Both groups found that when including 
pion and nucleon degrees of freedom alone, one is not able to describe the 
nucleon e.m. FFs over a significant range of $Q^2$. In both calculations, 
the proton electric FF would cross zero for $Q^2$ values as low as
0.4~GeV$^2$. Both calculations also confirm that a 
realistic description of the nucleon e.m. FFs is only obtained once the vector 
mesons are included as explicit degrees of freedom in the chiral Lagrangian.  
The vector meson loop diagrams were found to play only a minor role, 
the dominant contribution coming from the pole diagrams, 
confirming the findings of VMD models and dispersion theory. 
\newline
\indent
The corresponding results for the nucleon e.m. FFs in both the 
IR and EOMS schemes are shown in \Figref{chpt_results}. 
The covariant  baryon $\chi$PT results 
including vector mesons of Ref.~\cite{kubis01} are shown 
in \Figref{chpt_results} at both third and fourth order. 
The electric FFs of proton and neutron require fixing one LEC for
each, corresponding to the charge radii. 
One sees from \Figref{chpt_results} that the resulting 
fourth order results, including vector mesons, give a 
reasonably good description of the 
$Q^2$ dependence of the data up to $Q^2$ around 0.4~GeV$^2$. 
For the magnetic FFs, at third order the two LECs are fixed from the 
corresponding proton and neutron magnetic moments, 
whereas at fourth order two more LECs are fixed from the magnetic radii.  
Also for the magnetic FFs, a good description is only obtained 
once the vector mesons are included. 
The results of Schindler {\it et al.}~\cite{schindler}  
in a covariant $\chi$PT calculation and using a consistent power counting 
scheme which includes to fourth order both vector meson pole and loop 
contributions are also shown in \Figref{chpt_results}. 
Again the explicit vector meson contributions 
play a major role at the higher end of this momentum transfer range 
in order to obtain 
a reasonable description of the data, as can be 
seen from \Figref{chpt_results} (compare dotted and dashed-dotted curves).
\begin{figure}[h]
\epsfysize=5.4cm \epsffile{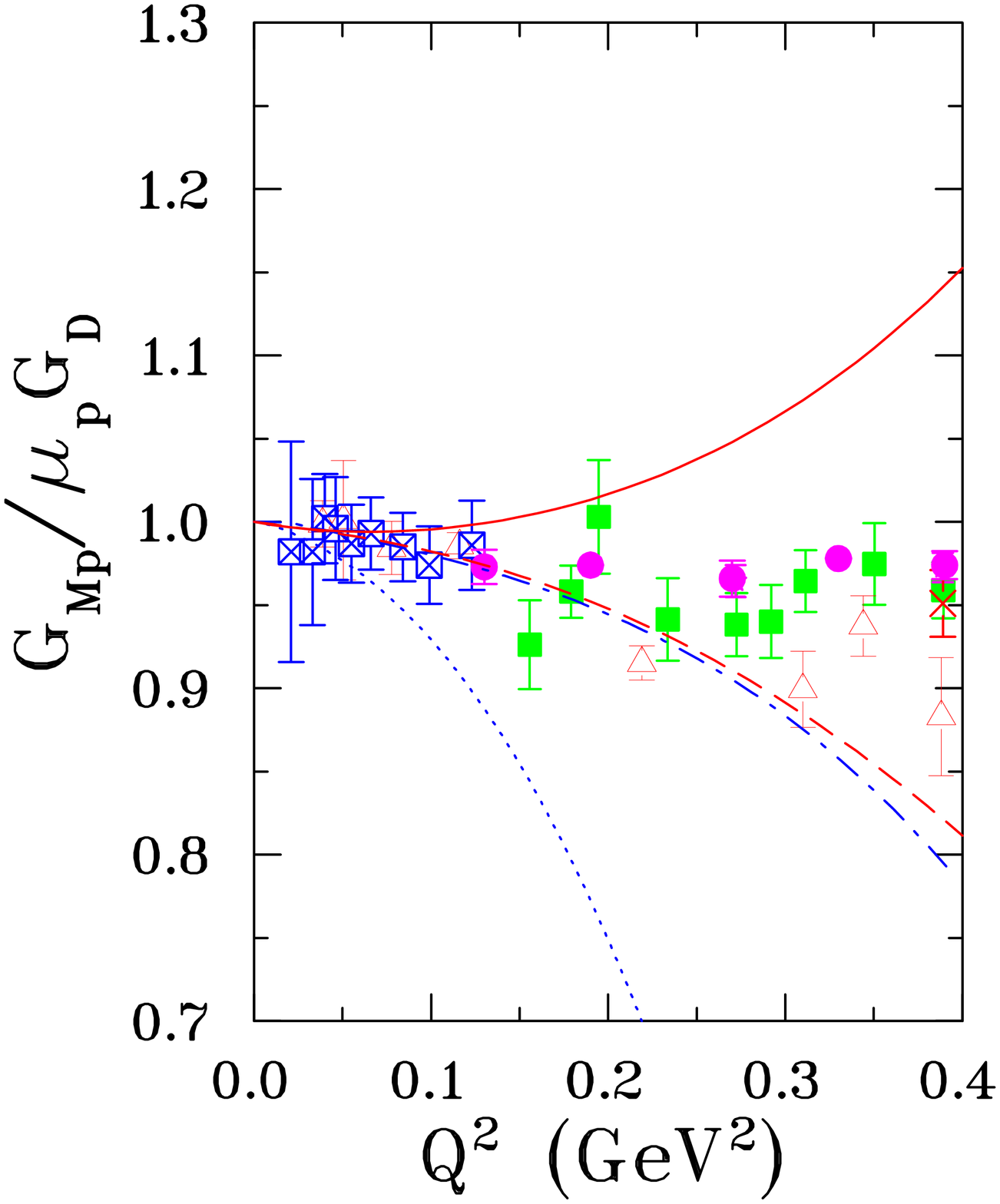} 
\epsfysize=5.4cm \epsffile{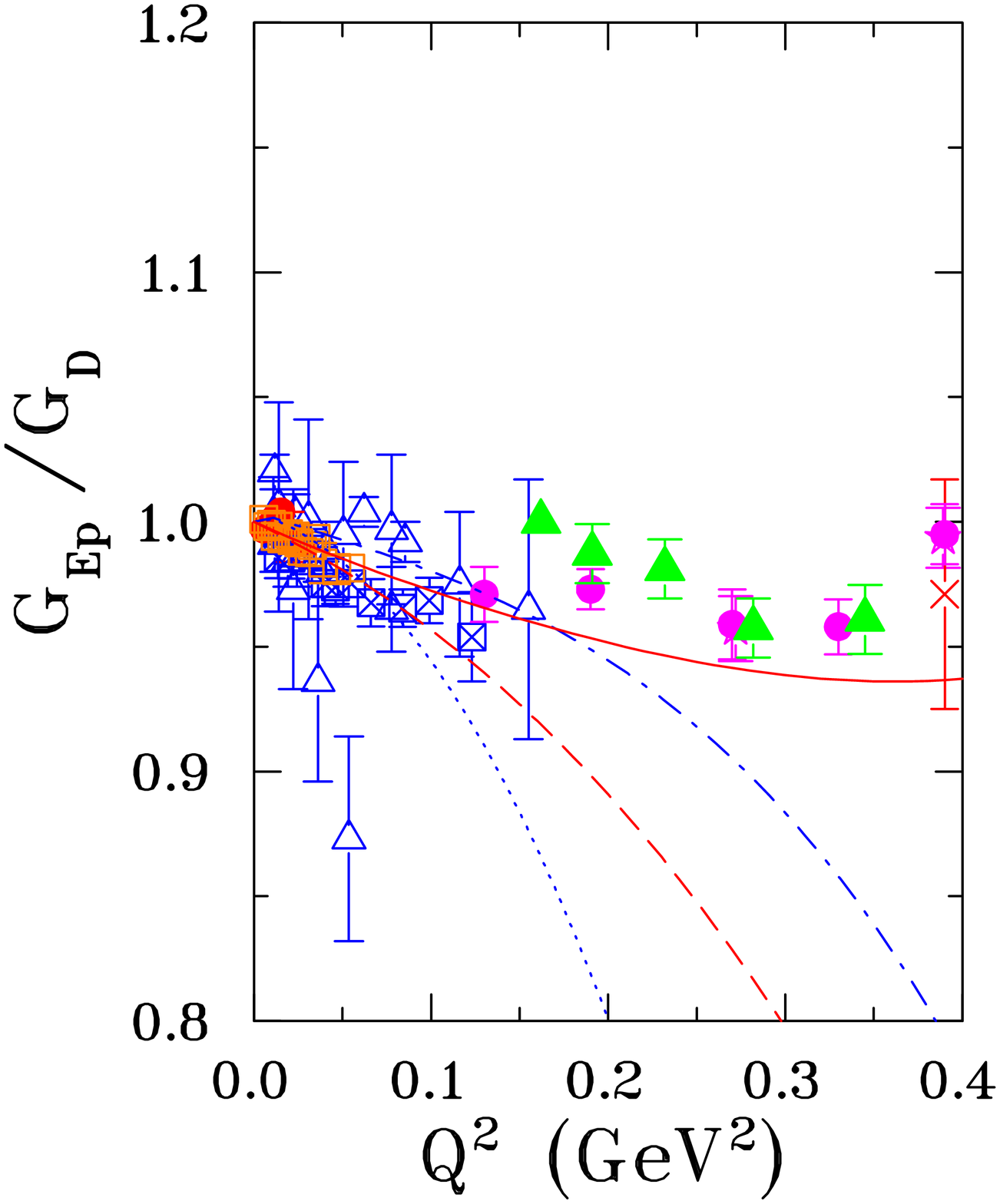} 
\epsfysize=5.4cm \epsffile{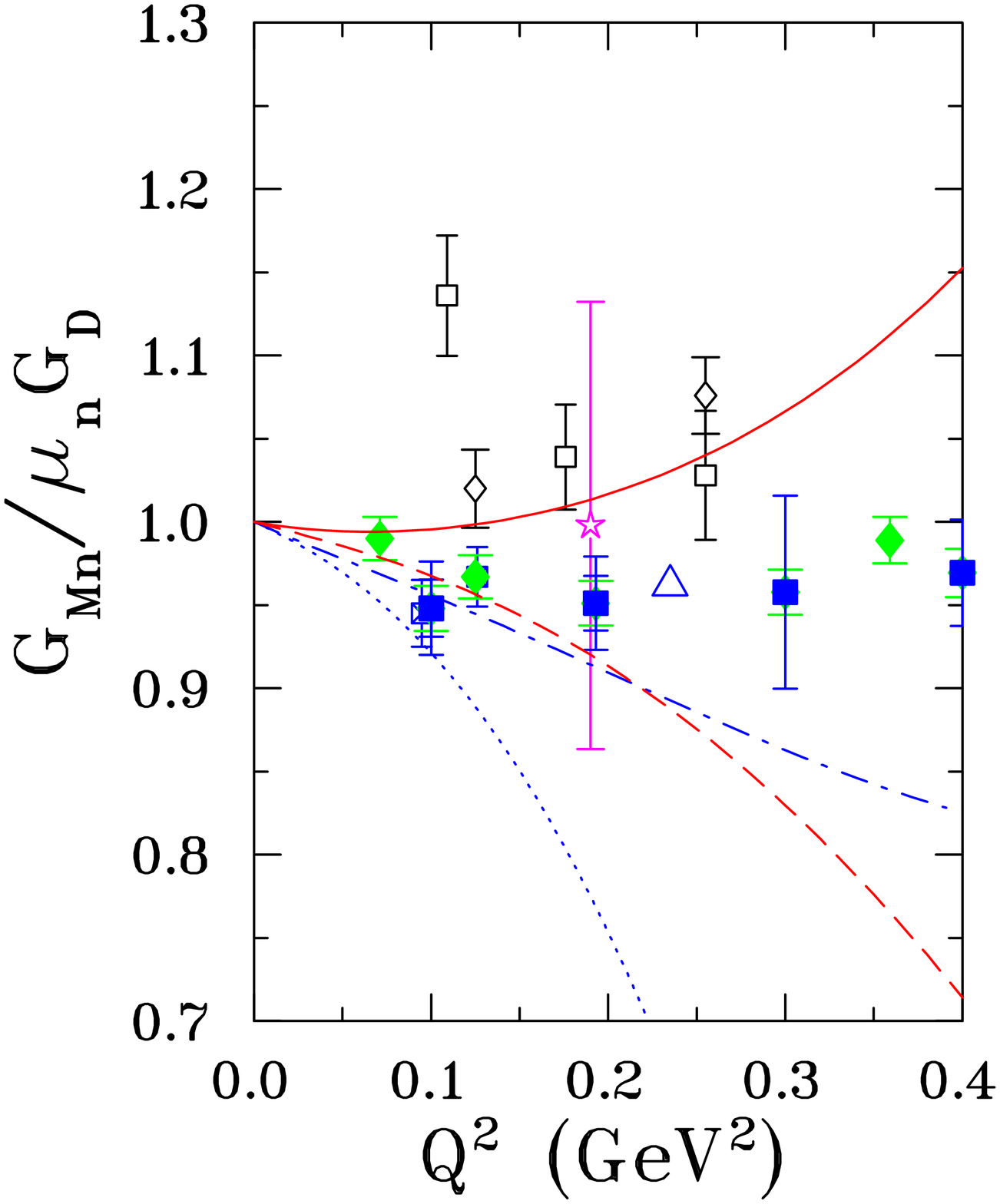} 
\epsfysize=5.4cm \epsffile{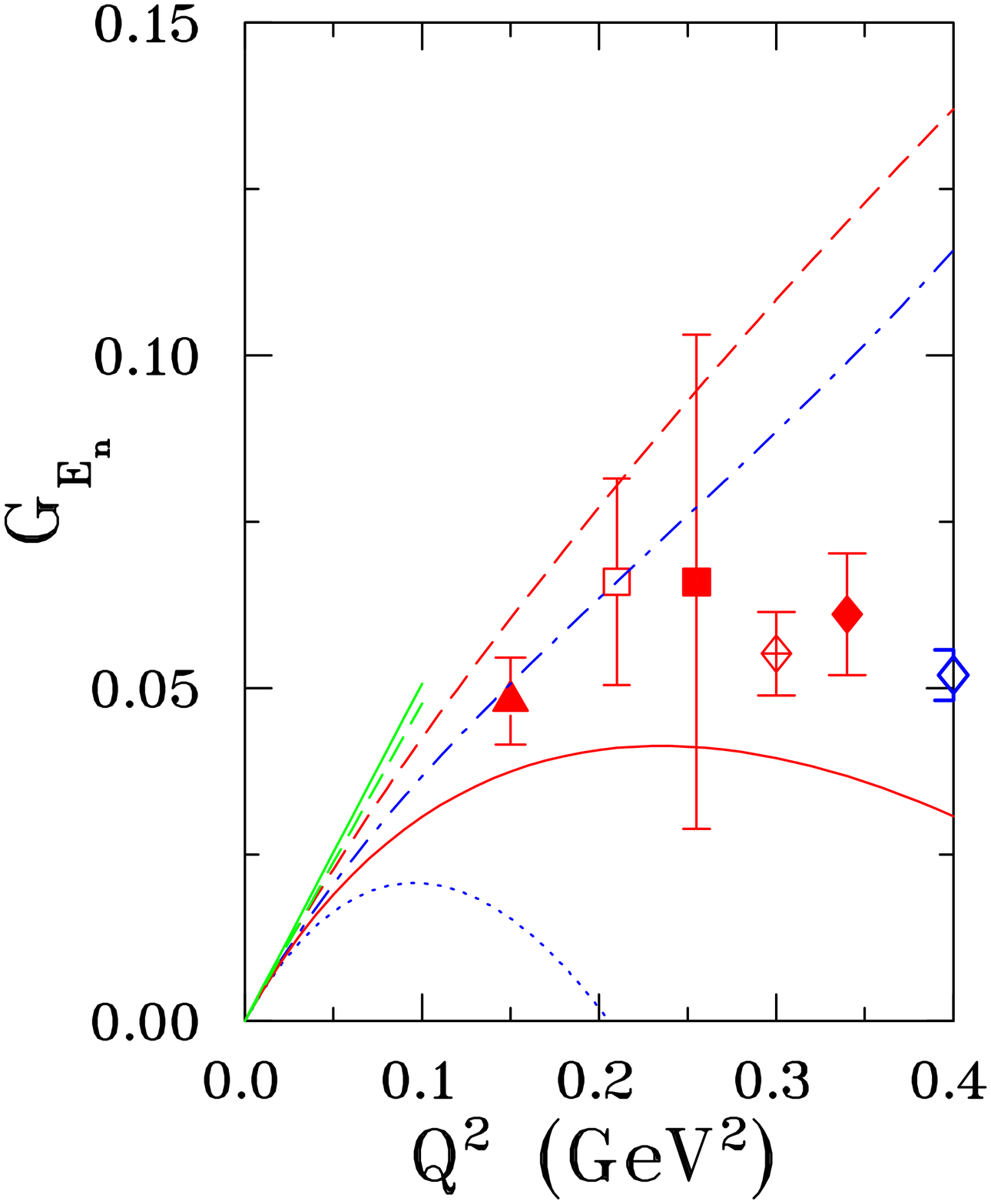} 
\caption{\small The nucleon e.m. FFs 
in the relativistic baryon $\chi$PT of~\cite{kubis01} (IR scheme) 
and ~\cite{schindler} (EOMS scheme).  
The results of ~\cite{kubis01} including vector mesons are shown 
to third (dashed curves) and fourth (solid curves) orders.  
The results of ~\cite{schindler} to fourth order are displayed both 
without vector mesons (dotted curves) and 
when including vector mesons (dashed-dotted curves). 
For references to the data : 
$G_{Mp}$ (see Fig.~\ref{fig:gmpgd}); 
$G_{Ep}$ (see Fig.~\ref{fig:gepgd}, green triangles 
are data of ~\cite{crawford});  
$G_{Mn}$ (see Fig.~\ref{fig:gmn_comp});  
$G_{En}$ (see Fig.~\ref{fig:gen_pol}, and constraint from rms radius 
is given by green slopes~\cite{Kopecky:1997rw}). 
}
\figlab{chpt_results}
\end{figure}

\subsection{Lattice QCD and chiral extrapolation}
\label{sec:lattice}

\subsubsection{Lattice simulations}

Lattice QCD calculations of nucleon structure quantities have matured 
considerably in the recent past. They provide an {\it ab initio} calculation 
of quantities such as the nucleon e.m. FFs 
from the underlying theory of QCD. 
\newline
\indent
Lattice QCD is a discretized version of QCD 
formulated in terms of path integrals on a space-time lattice~\cite{wilson} 
with only parameters the bare quark masses and the coupling constant. 
One recovers the continuum theory by 
extrapolating results obtained at finite lattice spacing $a$ to $a=0$.
In order to perform the continuum extrapolation a separate calculation at
several  values of $a$ is required. 
As lattice calculations necessarily are performed for 
a finite lattice size, one must 
keep the size of the box large enough to 
fit the hadrons inside the box. This requires to 
increase the number of sites as one decreases $a$. 
On the other hand, to keep finite volume effects small
one must have a box that is much larger than the Compton wavelength
of the pion. Present lattice QCD calculations 
take $Lm_\pi\stackrel{>}{\sim} 5$ where $L$ is the spatial length 
of the box and $m_\pi$ the pion mass.
As the computational costs of such calculations 
increase like $m_\pi^{-9}$, one uses quark mass values 
for the $u$ and $d$ quarks which are larger than in the real world.  
This enables the inversion of the fermionic matrix, 
which is needed for the calculation of hadronic matrix elements, with
currently available  resources.
\newline
\indent
State-of-the-art lattice calculations for nucleon structure studies  
use $a\stackrel{<}{\sim}0.1$~fm and
$L\sim 3$~fm and reach pion mass values down to about 350~MeV.
To connect those results with the physical world requires an extrapolation 
down to the physical quark masses 
(note that $m_q$ is proportional to $m_\pi^2$ for small quark mass values). 
This so-called chiral extrapolation will be discussed further on. 
It is only very recently that pion mass values 
below $350$~MeV~\cite{MILC,tmQCD} have been reached. 
This continuous effort is important to eliminate
one source of systematic error associated with the extrapolation to 
the light quark masses. 
\newline
\indent
The bare coupling constant and quark masses are
tuned as $a$ changes to leave physical quantities unchanged.
In a typical lattice calculation one starts
  by choosing the bare coupling constant  $g$,  which fixes the lattice
spacing, and the bare masses for the u-, d- and s-quarks. One
then computes a physical quantity such as the mass of the pion and the nucleon
in lattice units as a function of the  quark mass. The
pion mass is used to fix the u- and d-  quark masses (assumed degenerate)
 and the mass of the kaon or $\phi$ to fix the strange quark mass whereas the
lattice spacing is determined by extrapolating the results,
for instance, for the nucleon mass to the physical pion mass. 
Any other physical quantity in the light quark sector then follows.
\begin{figure}[h]
\begin{center}
{\includegraphics[height=3.5cm,angle=90]{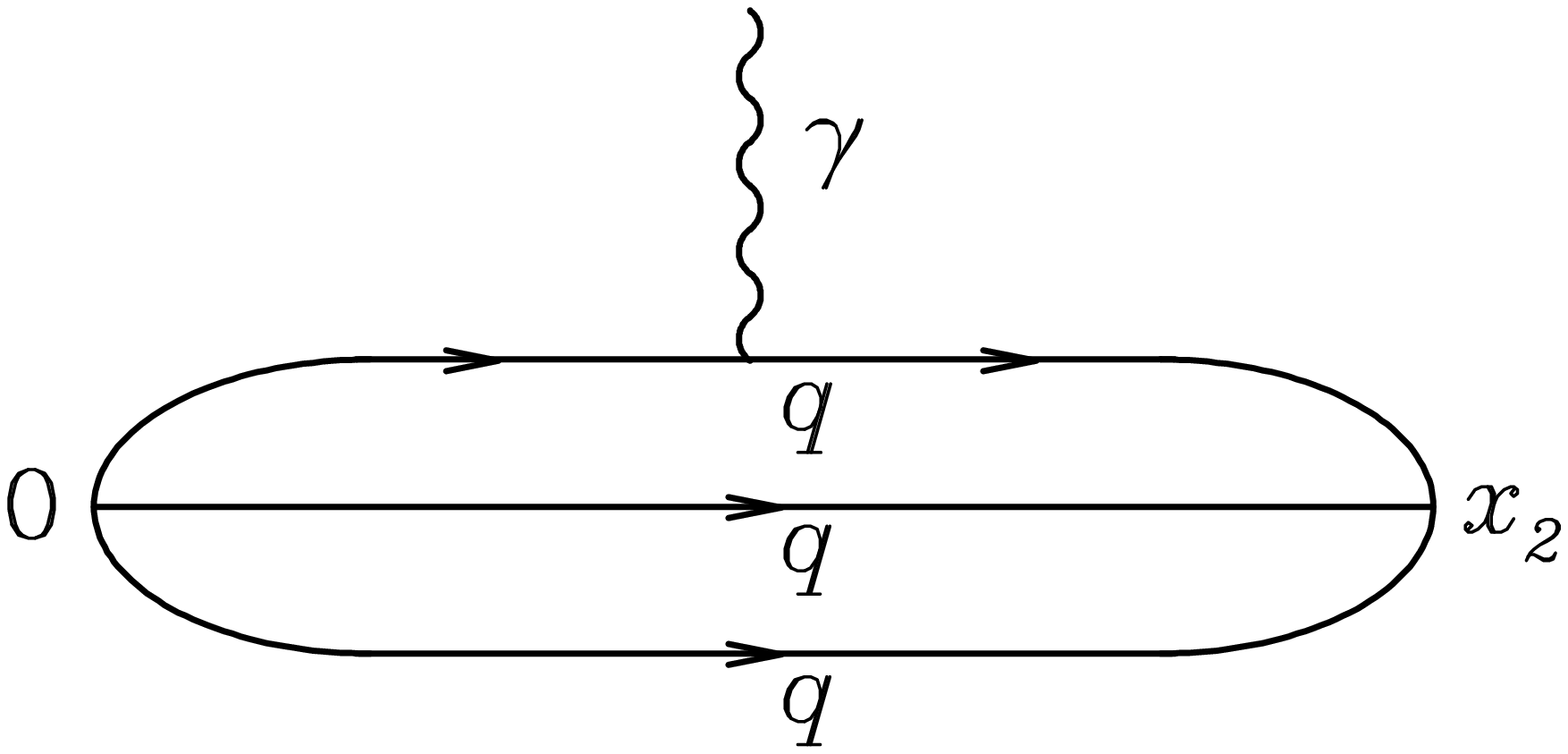} \hspace{2cm}
 \includegraphics[height=3.5cm,angle=90]{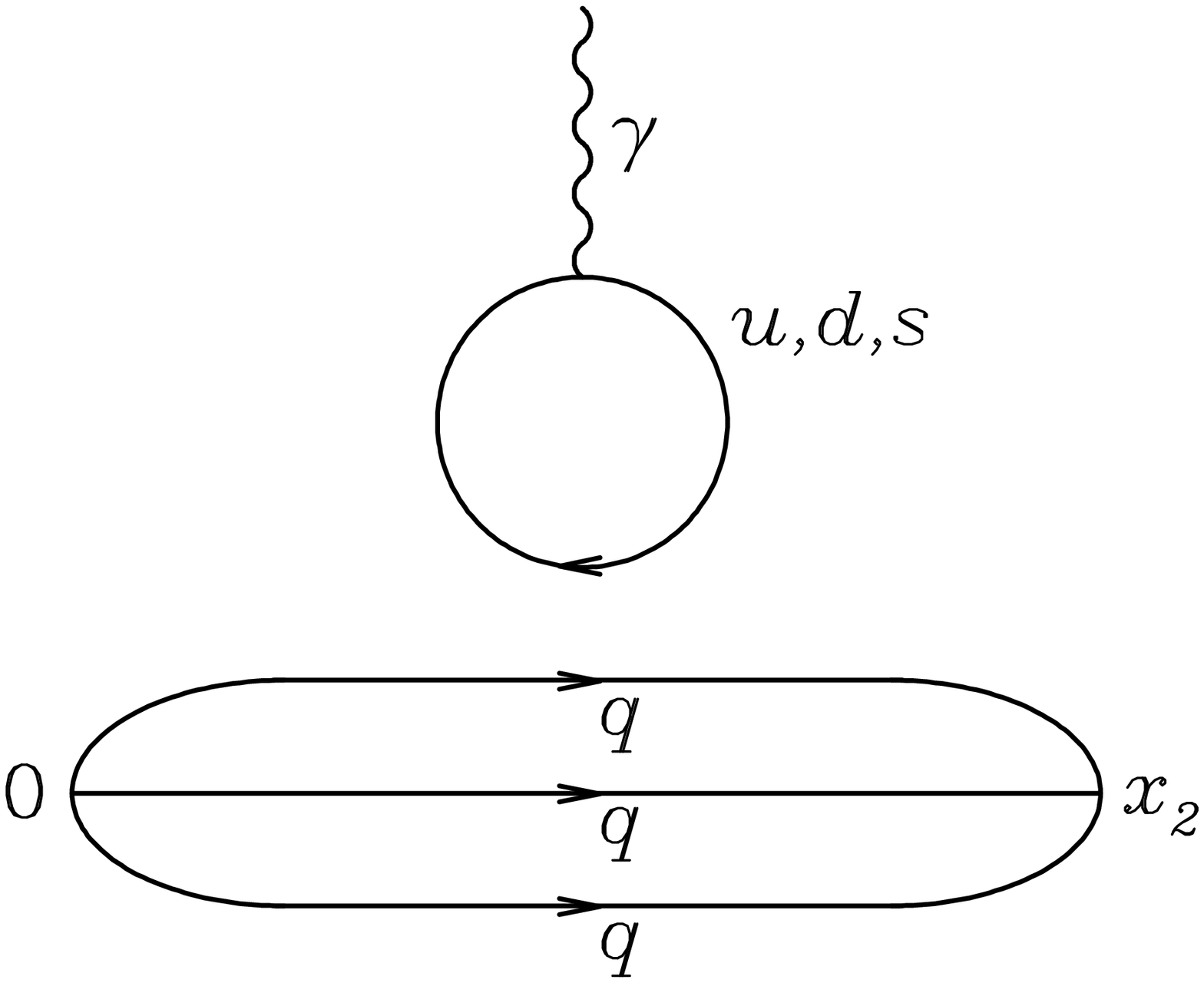}}
\end{center}
\caption{\small Diagrams illustrating the two topologically different 
contributions when calculating nucleon e.m. FFs in lattice QCD. 
Left (right) panels show the connected (disconnected) diagrams. 
Figure from~\cite{Boinepalli:2006xd}.  
}
\figlab{disconn}
\end{figure}

In the following, we will discuss lattice calculations for the 
(space-like) nucleon e.m. FFs. These calculations require 
the evaluation of three-point functions, which 
involve two topologically different contributions as illustrated 
in \Figref{disconn}. 
In the connected diagram contribution (left panel of \Figref{disconn}),  
the photon couples to one of the quarks connected 
to either the initial or final nucleon. 
The quark lines in \Figref{disconn} are 
understood to be dressed with an arbitrary number of gluons 
exchanged between the quarks. If the fluctuations of such gluons into 
$q \bar q$ pairs are neglected, 
one speaks of the quenched approximation. The full QCD
(unquenched) results include as well these sea-quark loop
insertions into the gluon lines.   
The disconnected diagram (right panel of \Figref{disconn}) 
involves a coupling to a 
$q \bar q$ loop, which then interacts with the nucleon through gluon
exchange. The disconnected diagram, which requires a numerically 
more intensive calculation, is at present neglected in most lattice studies. 
When taking the difference between proton and neutron e.m. FFs, i.e. 
for the isovector combination of nucleon e.m. FFs, 
the disconnected contribution drops out. 
Therefore, all following calculations in which the disconnected 
diagram is neglected are applicable only to the isovector e.m. FFs. To 
directly calculate the proton and neutron e.m. FFs, involves the  
evaluation of the disconnected contribution, which awaits the next generation 
of dynamical-fermion lattice QCD simulations.
\newline
\indent
The calculation of the connected diagram contribution to the nucleon e.m. FFs  
involves the computation of a sequential propagator, which 
can be done in two different ways. 
In an early pioneering 
work~\footnote{The first lattice QCD calculations for the pion e.m. FF  
were performed by Wilcox and Woloshyn~\cite{Wilcox:1985iy}, 
whereas the first attempt 
at a lattice QCD calculation for the proton electric FF was reported by  
Martinelli and Sachrajda~\cite{Martinelli:1988rr}.} 
of Leinweber, Woloshyn and Draper~\cite{Leinweber:1990dv}, 
this was done in the so-called 
fixed current approach, which requires the current to have a fixed direction 
and to carry a fixed momentum.  
This method allows one to use different initial or final states without 
requiring further inversions, which is the time-consuming part of the 
calculation. For a recent calculation of charge radii and magnetic moments of 
the whole baryon octet using this method, see~\cite{Boinepalli:2006xd}. 
The drawback of this method is that a new calculation 
is required for each momentum transfer. 
\newline
\indent
More recently, a second method has been used by different groups to evaluate 
nucleon e.m. FFs, in which one fixes the initial and final states 
to have the quantum numbers of the nucleon. In this so-called fixed sink 
method, the current can couple to a quark line 
at any intermediate time slice, see \Figref{disconn} (left panel), 
carrying any possible value of the lattice momentum, which makes it the 
method of choice for a detailed study of the momentum transfer dependence 
of the nucleon e.m. FFs.  
\newline
\indent
The Nicosia-MIT group~\cite{Alexandrou:2006ru,Alexandrou:2006pr} 
has performed a high-statistics calculation of nucleon isovector e.m. FFs 
in the fixed sink method, both in the quenched approximation and 
in full QCD, using two dynamical Wilson fermions,   
and for one value of the lattice spacing $a$, around 0.09 (0.08)~fm 
for the quenched (unquenched) results.  
The finite box size of length $L$  
imposes a smallest available non-zero momentum transfer, which 
for the quenched calculation is 
around $Q^2 \simeq 0.17$~GeV$^2$. The largest $Q^2$ value accessible 
is around $Q^2 \simeq 2$~GeV$^2$. Beyond such value, the Fourier transforms 
needed to evaluate the two- and three-point functions become noise dominated. 
Furthermore, the quenched calculation was performed for pion masses in the 
range $m_\pi \simeq 410 - 560$~MeV. 
The unquenched calculation 
was performed in the range of $m_\pi \simeq 380 - 690$~MeV.   
\begin{figure}[h]
\hspace{2.cm} \epsfxsize=6.1cm  \epsffile{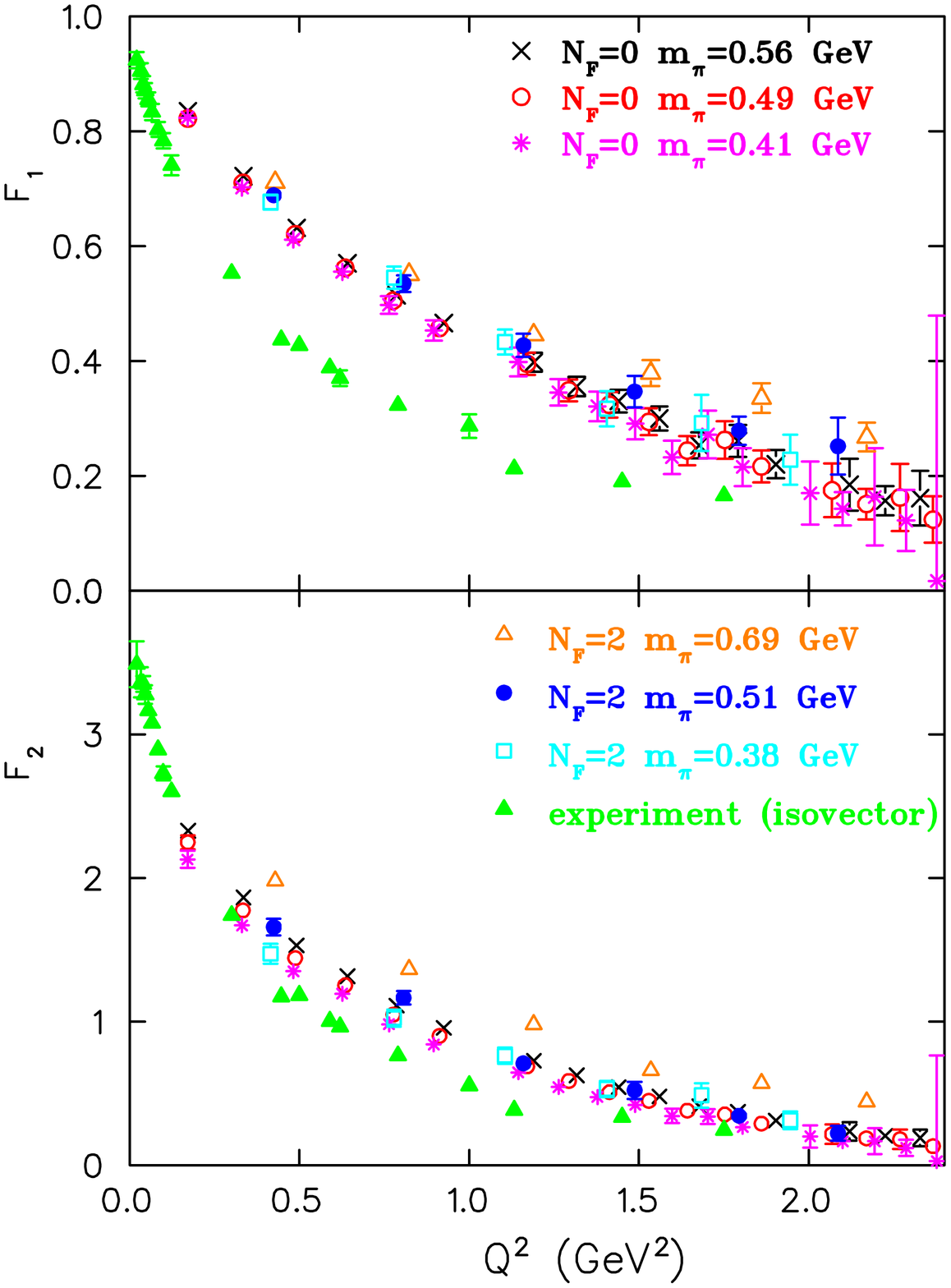} 
\hspace{2.5cm}  \epsfxsize=6.1cm \epsffile{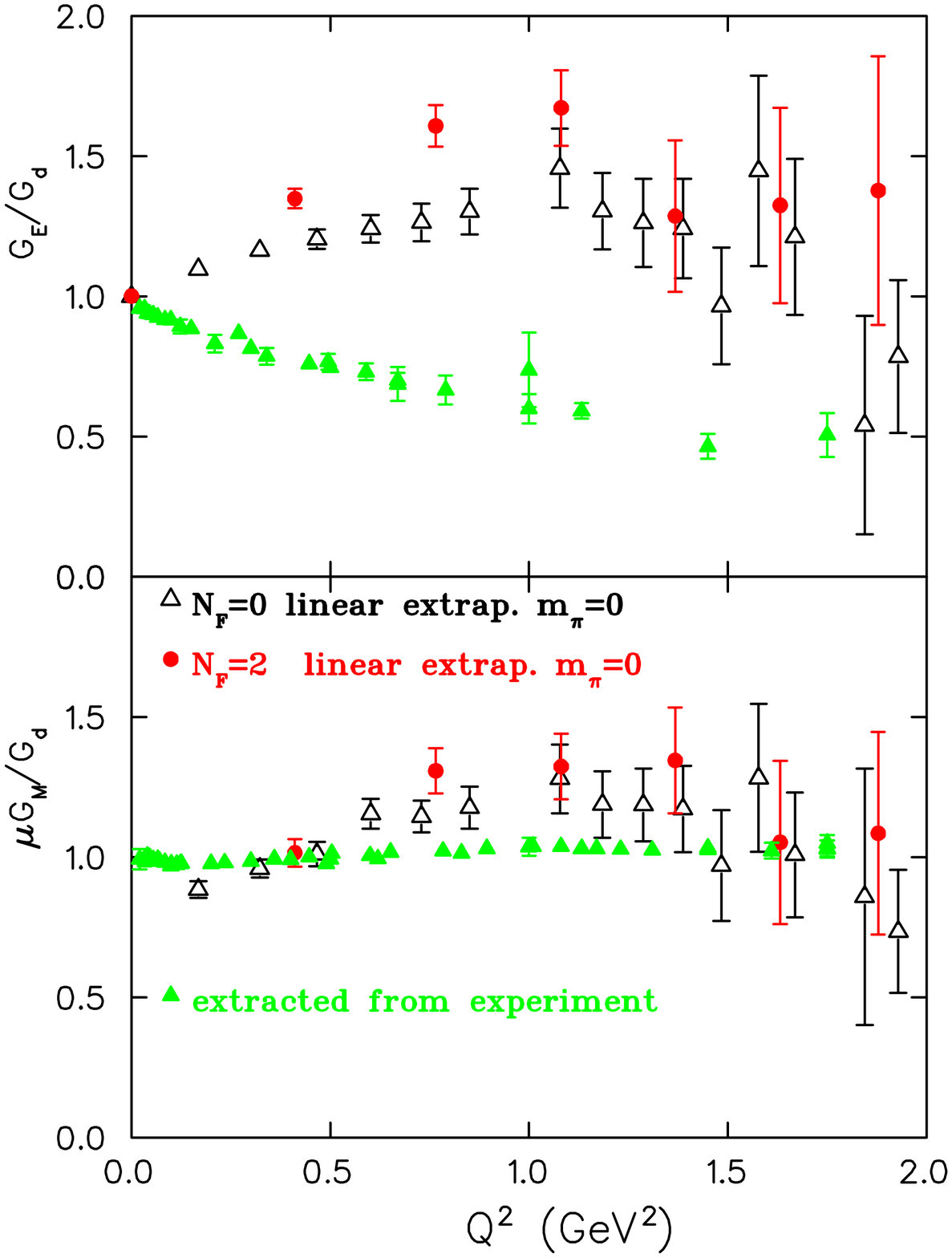} 
\caption{\small 
Lattice QCD results (from the Nicosia-MIT group~\cite{Alexandrou:2006ru})  
for the isovector FFs $F_1^V$ (upper left panel) 
and $F_2^V$ (lower left panel) as a function of $Q^2$. 
Both the quenched results ($N_F=0$) and 
unquenched lattice results with two dynamical Wilson fermions 
($N_F=2$) are shown for three different pion mass values.  
The right panels show the results for 
$G_E^V$ (upper right panel) and 
$G_M^V$ (lower right panel), divided by the standard dipole FF, 
as a function of  $Q^2$ in the chiral limit 
(using a linear extrapolation in $m_\pi^2$). 
The filled triangles show the 
experimental results for the isovector FFs
extracted by interpolating the experimental data for the
proton and neutron e.m. FFs.  
Figure from~\cite{Alexandrou:2006ru}. 
}
\figlab{f1f2lattice}
\end{figure}
\indent
The lattice QCD results of~\cite{Alexandrou:2006ru} 
for the nucleon Dirac and Pauli isovector FFs  
are shown in \Figref{f1f2lattice}. 
One observes that both the quenched and unquenched results for 
$F_1^V$ show only a very weak quark mass dependence in the 
range $m_\pi \simeq 400 - 700$~MeV. When comparing with experiment, 
one sees that both the quenched and unquenched lattice results 
of~\cite{Alexandrou:2006ru} largely overestimate the data for 
$F_1^V$. 
For $F_2^V$, one observes a stronger quark mass dependence, 
bringing the lattice results closer to experiment when decreasing $m_\pi$. 

The two main uncertainties in this calculation are the continuum 
extrapolation (i.e. finite $a$ effects) and whether one is close enough to 
the chiral limit (i.e. extrapolation in quark mass or $m_\pi$).  
To check the latter, and to extrapolate the lattice results down to the 
physical pion mass value in order to directly compare with experiment, 
the Nicosia-MIT group uses a linear fit in $m_\pi^2$ 
(corresponding to a linear fit in the quark mass). Such a linear fit,  
which is supported by the lattice results in the range $m_\pi \simeq 400 -
700$~MeV, is also shown in \Figref{f1f2lattice}.
The thus extrapolated lattice results for $F_2^V$ and $G_M^V$ are in 
agreement with experiment for $Q^2$ larger than about $0.3$~GeV$^2$. 
At smaller $Q^2$, an agreement can also be expected as 
one can calculate in this range the $m_\pi$ dependence using $\chi$PT. 
We will discuss in the following that the pion loops 
lead to non-analytic behaviors in the  
quark mass, yielding e.g. a more rapid (than linear in $m_\pi^2$) variation 
of the isovector magnetic moment as one approaches the chiral limit. 
However, the linearly in $m_\pi^2$ extrapolated results for 
$F_1^V$ still show strong disagreement with the data. This translates 
into an electric FF $G_E^V$ which drops less fast than the dipole FF $G_D$, 
whereas the data tell us that $G_E^V$ drops faster than the dipole. 
It is puzzling that this strong disagreement 
is seen at larger values of $Q^2$, where 
effects of pion loops are already suppressed, making it unlikely that the 
chiral extrapolation alone can explain this discrepancy. 
As both quenched and unquenched calculations 
of~\cite{Alexandrou:2006ru} were only performed at one value of $a$, 
it would be very worthwhile, in order to shed light on this puzzle, 
to repeat such calculations for different values of $a$  
and check the continuum extrapolation.   
\newline
\indent
Unquenched lattice calculations using two mass degenerate flavors of 
dynamical Wilson fermions have also been reported by the 
QCDSF Coll.~\cite{Gockeler:2006uu}. These results improve on previous 
calculations by the QCDSF Coll.~\cite{Gockeler:2003ay} which were performed 
using Wilson fermions in the quenched approximation.  
The $Q^2$ dependence of the lattice results for the nucleon isovector FFs 
was parameterized (as a first approximation) in terms of a dipole behavior. 
For $F_1^V$, the unquenched QCDSF lattice results find that  
the corresponding dipole mass becomes smaller with decreasing pion mass 
values. However at the 
smallest available pion mass values of around 340~MeV, 
the dipole mass reaches a value around 1.3~GeV, which still lies 
significantly above the experimental value of 0.843~GeV. 
This yields an isovector FF $F_1^V$ which has a too flat 
$Q^2$ dependence, confirming the 
puzzling finding of~\cite{Alexandrou:2006ru}, which was  
also obtained for two dynamical Wilson fermions.   
\newline
\indent
Recently, unquenched lattice QCD calculations for the nucleon e.m. FFs have 
also been performed by the LHPC Coll.~\cite{Edwards:2006qx} 
based on the Asqtad improved action, using different fermions for valence 
and sea quarks. 
This hybrid action uses for the valence quarks domain wall fermions, 
preserving chiral symmetry on the lattice.   
For the sea quarks, the configurations generated by 
the MILC Coll.~\cite{MILC} are used, 
with two degenerate light and one strange staggered quarks, 
allowing for economical calculations. 
The discretization errors in this action are of order $a^2$, in comparison 
with order $a$ for the above discussed Wilson action. 
Although this action has generated quite a number of 
encouraging results when applied to nucleon structure studies, 
such as e.g. moments of unpolarized, helicity, and transversity distributions, 
see~\cite{Edwards:2006qx} for a recent overview, some controversy 
remains around the fourth root of the fermion 
determinant~\cite{Sharpe:2006re}. 
\newline
\indent
In \Figref{fig:f1f2_lhpc}, we show the unquenched lattice QCD 
results from the LHPC Coll. for the nucleon e.m. FFs, 
performed for one lattice spacing of 
$a \simeq 0.125$~fm, and for pion mass values in the 
range $m_\pi = 360 - 775$~MeV. It is seen that in contrast to the above 
discussed Wilson results, this action yields a 
noticeable dependence on $m_\pi$ for the Dirac isovector FF $F_1^V$ at larger 
values of $Q^2$. The $Q^2$ dependence of $F_1^V$ at the 
smallest $m_\pi$ value of around 360~MeV is found 
to be in qualitative agreement with the data. 
One also sees from \Figref{fig:f1f2_lhpc} that the isovector ratio 
$F_2^V / F_1^V$ approaches the experimental result when decreasing $m_\pi$. 
So far this is the only lattice calculation which yields a 
qualitative consistent picture for both $F_1^V$ and $F_2^V$. 
Evidently, it will be very worthwhile to corroborate the results at the
lowest pion masses and improve their statistics in future calculations. 
If confirmed by higher statistics results, it remains to be understood 
why different actions may yield significantly different results, 
in particular for $F_1^V$. Unquenched calculations at a couple of different 
lattice spacings using different actions 
would be very helpful in this respect. 
\begin{figure}
\hspace{1cm}
\epsfxsize=7cm \epsffile{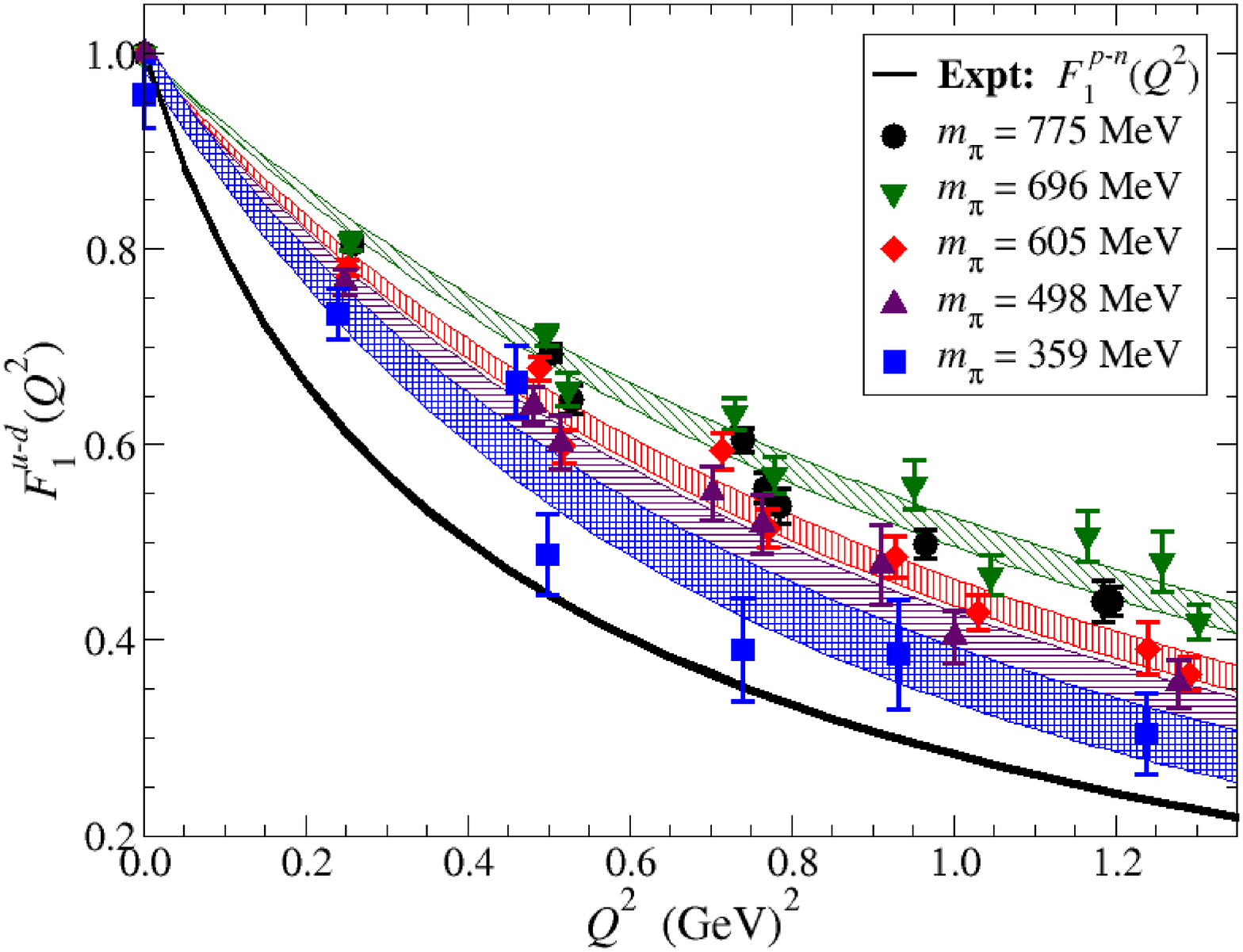} 
\hspace{1.25cm}
\epsfxsize=8.3cm \epsffile{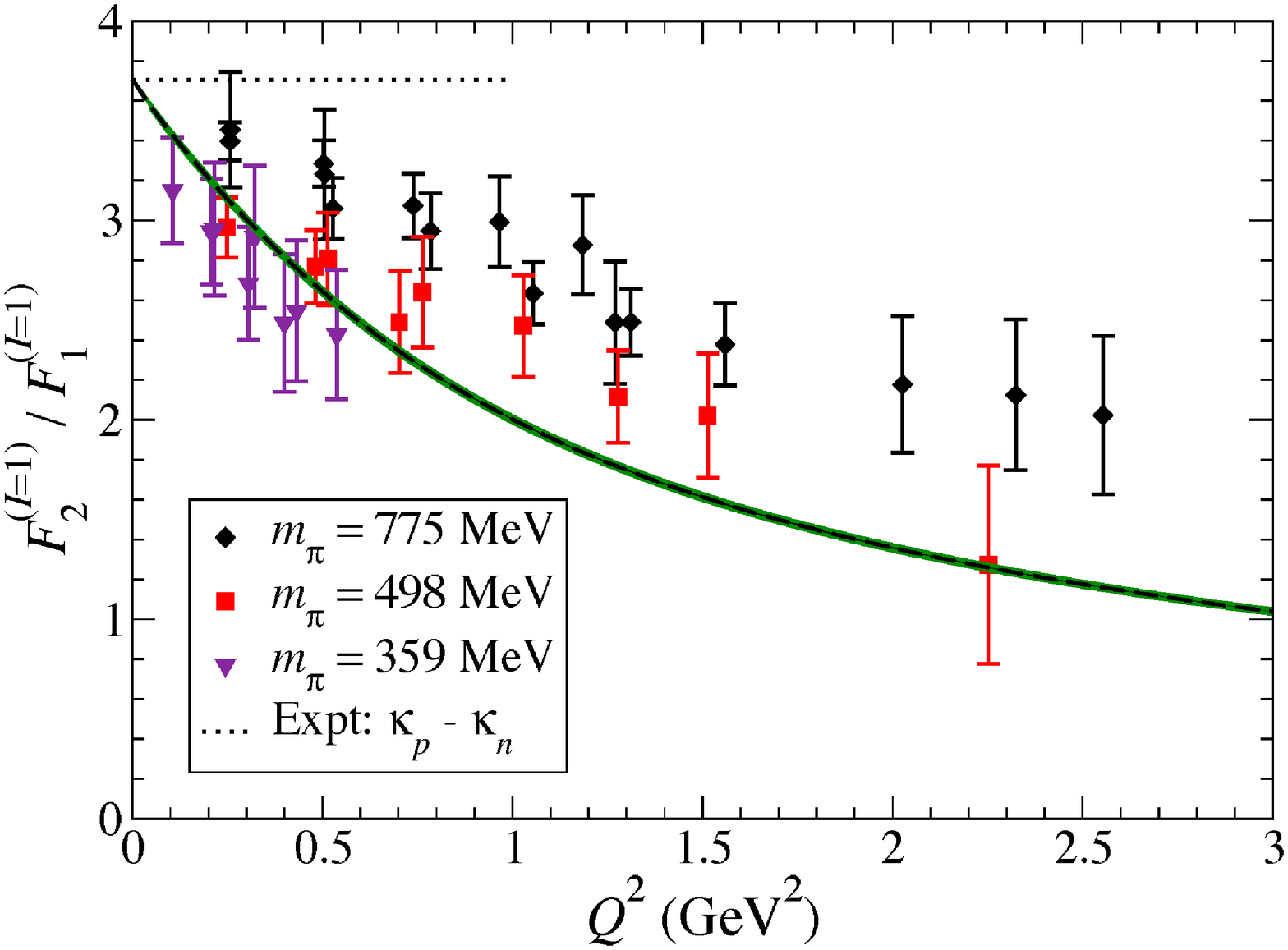} 
\caption{\small 
Lattice QCD results (from the LHPC Coll.~\cite{Edwards:2006qx})   
for the nucleon isovector FFs $F_1^V$ (left panel) and for the 
ratio $F_2^V / F_1^V$ (right panel). The unquenched results 
using  a hybrid action of domain wall valence quarks and 
2+1 flavor staggered sea quarks are shown for   
different values of $m_\pi$ and are compared with experiment 
(solid curve, using the parametrization of~\cite{kelly04}).   
Figure from~\cite{Edwards:2006qx}. 
}
\figlab{fig:f1f2_lhpc}
\end{figure}

\subsubsection{Chiral extrapolations}

Present lattice calculations are possible for larger than physical  
quark masses, and therefore necessitate an extrapolation 
procedure in order to make contact with experiment. 
The extrapolation in the quark mass $m_q$ is not straightforward,
because the non-analytic dependencies, such as $\sqrt{m_q}$
and $\ln m_q$, become important as one approaches the
small physical value of $m_q$. Therefore naive extrapolations 
often fail, while spectacular non-analytic effects
are found in a number of different quantities, 
such as nucleon magnetic moments and charge radii, 
see {\it e.g.} Refs.~\cite{Leinweber:2001ui,Hemmert:2002uh}. 
The $\chi$EFT, discussed in the Sect.~\ref{theory_eft}, provides 
a framework to compute these non-analytic dependencies, 
for small quark masses.  

As an example, in the SSE ($\varepsilon$-expansion) 
to order $\varepsilon^3$, the $\gamma^\ast NN$ 
vertex has been calculated in~\cite{Gockeler:2003ay} 
through pion one-loop diagrams. 
Due to the pion loops, 
the isovector Dirac radius $\langle r^2 \rangle_1^V$
acquires non-analytic dependencies in the quark mass. 
Its leading dependence in the pion mass is given 
by~\cite{Gasser:1987rb,Leinweber:1992hj}~:
\begin{eqnarray}
\langle r^2 \rangle_1^V \equiv 
\langle r^2 \rangle_{1 p} - \langle r^2 \rangle_{1 n} 
= a_0(\mu) 
-\frac{1+5g_A^2}{(4 \pi f_\pi)^2}\ln\left(\frac{m^2_\pi}{\mu^2}\right)
+ {\mathcal O}(m_\pi^2) \,+\, \pi \Delta \; \mathrm{loops},
\label{eq:rms-extrap}
\end{eqnarray}
where the logarithmic term in $m_\pi$ is the leading non-analytic (LNA) 
dependence originating from the $\pi N$ loop diagrams, and depends 
only on the nucleon axial coupling 
$g_A = 1.2695$ and the pion decay constant $f_\pi = 92.4$~MeV.
Furthermore in Eq.~(\ref{eq:rms-extrap}), 
$\mu$ is the renormalization scale, and $a_0(\mu)$ is a LEC (evaluated at 
scale $\mu$). 
Note that, in contrast to most chiral extrapolations 
which contain finite terms of
the form $m_\pi^2\ln m_\pi^2$, the isovector radius diverges like
$\log m_\pi^2$, rendering the variation of the radius quite
substantial near the physical pion mass.  
\newline
\indent
Analogously, the $\pi N$ loop diagrams give rise to non-analytic terms in the 
quark-mass expansion of the nucleon isovector magnetic moment. Its leading 
dependence in the pion mass is given by~\cite{Gasser:1987rb}~:
\begin{eqnarray}
\kappa^V \equiv \kappa_p - \kappa_n 
= \kappa_0^V - \frac{4 \, g_A^2 \, M}{(4 \pi f_\pi)^2} \, m_\pi 
+ {\mathcal O}(m_\pi^2) \,+\, \pi \Delta \; \mathrm{loops},
\label{eq:kappa-extrap} 
\end{eqnarray}
where the LEC $\kappa_0^V$ corresponds to the isovector anomalous magnetic 
moment in the chiral limit. 
\newline
\indent
The LNA behavior of the isovector Pauli radius 
$\langle r^2 \rangle_2^V$ due to $\pi N$ loops 
shows a $1/m_\pi$ divergence in the chiral 
limit~\cite{Gasser:1987rb}~:
\begin{eqnarray}
\langle r^2 \rangle_2^V \equiv 
\langle r^2 \rangle_{2 p} - \langle r^2 \rangle_{2 n} |_{LNA} 
= \frac{1}{\kappa^V} \, \frac{g_A^2 \, M}{8 \pi f_\pi^2 \, m_\pi}.
\end{eqnarray} 
\indent
First attempts have been made to compare the lattice results for the 
pion mass dependence of the isovector magnetic 
moment, and the isovector Dirac and Pauli squared radii  
with the $\chi$PT results of~\cite{Gockeler:2006uu} in the SSE 
to order $\varepsilon^3$. 
For $\kappa^V$, 
allowing for one extra higher order parameter beyond $\varepsilon^3$, 
a four parameter $\chi$PT fit was performed in~\cite{Gockeler:2003ay} 
to the lattice data at relatively large $m_\pi$ values.  
It was found that for both $\kappa^V$ and the isovector Pauli radius,  
$\langle r^2 \rangle_2^V$, the quenched and unquenched results 
from the Nicosia-MIT group\cite{Alexandrou:2006ru} and the unquenched 
results from the QCDSF Coll.\cite{Gockeler:2006uu}, at the 
smallest available values of $m_\pi$ indeed 
seem to follow the strong rise predicted by the $\chi$PT fit. 
On the other hand, the lattice results of~\cite{Alexandrou:2006ru} 
and \cite{Gockeler:2006uu} for the isovector Dirac radius, 
$\langle r^2 \rangle_1^V$, 
show no strong indication of the logarithmic $\ln m_\pi$ 
divergence. The SSE results of \cite{Gockeler:2003ay}  
are not able to account for the lattice results for $\langle r^2 \rangle_1^V$. 
\newline
\indent
One may of course wonder if any agreement or disagreement with $\chi$PT 
for $m_\pi$ values as large as 0.5 - 1~GeV is very meaningful. Surely at such
large $m_\pi$ values, higher order contributions not accounted for in e.g. the 
$\varepsilon^3$ calculations are  
important~\footnote{See e.g. Ref.~\cite{McGovern:2006fm}, where it was shown 
that the surprisingly good agreement of fourth-order $\chi$PT  
when extrapolating lattice data for the nucleon mass out to  
large pion mass values (in the range 0.5 - 1~GeV) is spoiled 
once the fifth-order terms (due to 2-loop $\pi N$ diagrams) are included. }. 
A conservative strategy is to restrict $\chi$PT to its limited range of 
applicability and await lattice results for $m_\pi$ value below 300~MeV where
the effect of higher order terms is still relatively small. 
Alternatively, one may choose to build upon $\chi$EFT 
and extend its range of applicability - leaving the domain of power 
counting - by resumming higher order 
effects using additional physics principles.  
\newline
\indent
One such strategy has been adopted in 
Refs.~\cite{Pascalutsa:2004ga,Holstein:2005db} by using analyticity 
to resum higher order (analytic) terms in $m_\pi^2$ 
to the nucleon magnetic moments. 
By requiring the anomalous magnetic moments to satisfy (a generalization of) 
the Gerasimov-Drell-Hearn sum rule~\cite{gerasimov,drellhearn}, 
a relativistic one-loop $\pi N$ 
calculation has the correct chiral behavior at the small values 
of $m_\pi$, and yields a convergent $1/m_\pi^2$ behavior 
at larger values of $m_\pi$. 
It was found to yield a much smoother $m_\pi$ behavior 
for the magnetic moment than a truncated $\chi$PT calculation, 
while encompassing the correct behavior for small $m_\pi$ values.  
\newline
\indent
Another strategy has been pursued by the Adelaide group by modifying 
the one-loop $\chi$PT results and taking into account the finite size of the 
nucleon through a finite range regularization procedure. This method was 
found successful when applying it to the calculation of the $m_\pi$ 
dependence of nucleon and $\Delta(1232)$ masses,  
see e.g.~\cite{Leinweber:1999ig,Leinweber:2003dg}. 
For $\langle r^2 \rangle_1^V$, one may try in this spirit a modification 
of the $\chi$PT formula of Eq.~(\ref{eq:rms-extrap}) as~\cite{Dunne:2001ip}~:
\begin{eqnarray}
\langle r^2 \rangle_1^V  
= a_0
-\frac{1+5g_A^2}{(4 \pi f_\pi)^2}\ln \left(\frac{m^2_\pi}{m_\pi^2 + 
\Lambda^2}\right),
\label{eq:rms-extrap2}
\end{eqnarray}
where $\Lambda$ is a phenomenological cut-off which reflects the finite size 
of the nucleon. Such a fit for the isovector Dirac radius 
is shown in \Figref{diracrms_lhpc} and compared with the most recent 
unquenched lattice results using the hybrid action 
(domain wall valence quarks on top of a 2+1 flavor staggered sea) 
of the LHPC Coll. 
One firstly sees, that these lattice results do show appreciable $m_\pi^2$ 
variation over the pion mass range $m_\pi = 360 - 775$~MeV and provide a 
first clear hint of the logarithmic $m_\pi$ divergence. 
As the pion mass approaches the physical value, the calculated nucleon 
size increases and approaches the correct value. 
Using the simple extrapolation formula of Eq.~(\ref{eq:rms-extrap2}),  
which has the $\ln m_\pi$ divergence at low $m_\pi$ values built in, one 
obtains a consistent description of the $m_\pi$ dependence 
of the lattice results using $\Lambda \sim 500$~MeV.
\begin{figure}
\centerline{  \epsfxsize=7.5cm
  \epsffile{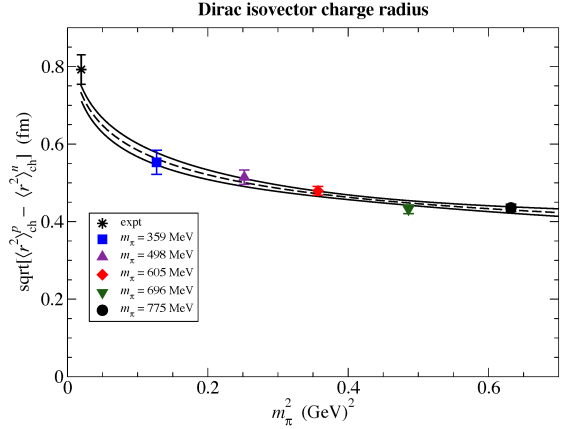} 
}
\caption{\small 
Chiral extrapolation of the nucleon isovector Dirac radius 
$\langle r^2 \rangle_1^V$.   
The unquenched results are from the LHPC Coll.~\cite{Edwards:2006qx}. 
The experimental value is shown by the star. The curves are fits using the 
chiral extrapolation formula Eq.~(\ref{eq:rms-extrap2}) 
of~\cite{Dunne:2001ip}. Figure from~\cite{Edwards:2006qx}. 
}
\figlab{diracrms_lhpc}
\end{figure}

Finally we would like to emphasize that 
presently there is no systematic framework for 
extrapolating lattice QCD results for FFs 
at values of $Q^2$ larger than about 0.3~GeV$^2$, i.e.  
beyond the region where a $\chi$PT expansion is expected to be applicable. 
The development of such a framework remains a challenge for future work. 
Even when lattice results become available 
for $m_\pi$ values below 300~MeV, at larger $Q^2$,   
one is confronted with the problem of performing a chiral extrapolation 
(in the small scale $m_\pi$) in the presence of a large scale $Q^2$.  
A first attempt in this direction 
has been performed in~\cite{Matevosyan05}, 
within the context of a light-front cloudy bag model.

\subsection{Generalized parton distributions (GPDs)}
\label{theory_gpd}

So far we have discussed the $N \to N$ transition 
as revealed with the help of the electromagnetic probe. By measuring 
the response of the hadron to a virtual photon, one measures the matrix 
element of a  well-defined quark-gluon operator (in this case the vector 
operator $\bar q \gamma^\mu q$) over the hadronic state. This matrix element  
can be parametrized in terms of the nucleon e.m. FFs, 
revealing the quark-gluon structure of the nucleon. 
We are however not limited in nature to probes such as photons 
(or $W$, $Z$ bosons for the axial transition). The phenomenon of asymptotic 
freedom of QCD, meaning that at short distances the interactions between 
quarks and gluons become weak, provides us with more sophisticated 
QCD operators to explore the structure of hadrons. Such operators can 
be accessed by selecting a small size configuration of quarks and gluons, 
provided by a hard reaction, such as deep inelastic scattering (DIS), or 
hard exclusive reactions such as deeply virtual Compton scattering (DVCS).  
We will be mostly interested here in DVCS reactions which are of the type 
$\gamma^*(q_h) + N(p) \to \gamma(q^\prime) + N(p^\prime)$, where the 
virtual photon momentum $q_h$ is the hard scale. 
The common important feature of such hard reactions is the possibility
to separate clearly the perturbative and nonperturbative stages of
the interactions~: this is the so-called factorization property. 
\newline
\indent
The all-order factorization theorem for the DVCS process on the 
nucleon has been proven in~\cite{Ji98a,Col99,rady}.
Qualitatively one can say that the hard reactions allow
one to perform a ``microsurgery'' of a nucleon by removing in a
controlled way a quark of one flavor and spin and implanting
instead another quark in the final nucleon. 
It is illustrated in \Figref{n_dvcs} 
for the case of the DVCS process. 
The non-perturbative stage of such hard exclusive 
electroproduction processes is described by 
universal objects, so-called generalized parton distributions 
(GPDs)~\cite{Muller:1998fv,ji,Radyushkin:1996nd}, 
see~\cite{Ji:1998pc,Goeke:2001tz,Diehl:2003ny,Belitsky:2005qn,Ji:2004gf} 
for reviews and references. 
\newline
\indent
The nucleon structure information entering the nucleon DVCS process, 
can be parametrized at leading twist-2 level, in
terms of four quark chirality conserving GPDs. 
The GPDs depend on three variables: the quark longitudinal 
momentum fractions $x$ and $\xi$, and the momentum transfer 
$Q^2 = - q^2$ to the nucleon.
The light-cone momentum fraction $x$ is defined by $k^+ = x P^+$,
where $k$ is the quark loop momentum and
$P$ is the average nucleon momentum 
$P = (p + p^{\ \prime})/2$, where $p (p^{\ \prime})$
are the initial (final) nucleon four-momenta respectively,  
see \Figref{n_dvcs}.
The skewedness variable $\xi$ is
defined by $q^+ = - 2 \xi \,P^+$, where 
$q = p^{\ \prime} - p$ is the
overall momentum transfer in the process, and where
$2 \xi \rightarrow x_B/(1 - x_B/2)$ in the Bjorken limit: 
$x_B = Q_h^2/(2 p \cdot q_h)$ is the usual Bjorken scaling variable, 
with $Q_h^2 = -q_h^2 > 0$ the virtuality of the hard photon.
\newline
\indent
The DVCS process 
corresponds to the kinematics $Q_h^2 \gg Q^2, M^2$, 
so that at twist-2 level, terms proportional to $Q^2 / Q_h^2$ 
or $M^2 / Q_h^2$ are neglected in the amplitude.  
In a frame where the virtual photon momentum \( q_h^{\mu } \) and the average
nucleon momentum \(  P^{\mu } \) are collinear
along the \( z \)-axis and in opposite directions, one can parameterize
the non-perturbative object entering the nucleon DVCS process as 
(following Ji~\cite{ji})\footnote{In all non-local
expressions we always assume the gauge link:
P$\exp(ig\int dx^\mu A_\mu)$, ensuring the color gauge
invariance.}:
\begin{eqnarray}
&& \frac{1}{2\pi} \, \int dy^{-}e^{ix  P^{+}y^{-}}
\left. \langle N(p^\prime)|\bar{\psi } (-y/2) \; \gamma \cdot n \; \psi (y/2)
| N(p) \rangle \right|_{y^{+}=\vec{y}_{\perp }=0} \nonumber \\
&&=\; H^{q}(x,\xi ,Q^2)\; \bar{N}(p^{'}) \; \gamma \cdot n \; N(p)
\,+\, E^{q}(x,\xi ,Q^2)\; \bar{N}(p^{'}) \; i\sigma^{\mu \nu} 
\frac{n_\mu \, q_\nu}{2 M} \; N(p) ,
\label{eq:qsplitting}
\end{eqnarray}
where \( \psi  \) is the quark field
of flavor $q$, \( N \) the nucleon spinor, and $n^\mu$ is a 
light-cone vector along the negative $z$-direction. 
The {\it lhs} of Eq.~(\ref{eq:qsplitting}) can be interpreted as a Fourier
integral along the light-cone distance $y^-$ of a quark-quark
correlation function, representing the process where
a quark is taken out of the
initial nucleon (having momentum $p$) at the space-time point $y/2$, and
is put back in the final nucleon (having momentum $p^{\ \prime}$) 
at the space-time
point $-y/2$. This process takes place at equal light-cone time ($y^+
= 0$) and at zero transverse separation ($\vec y_\perp = 0$) between
the quarks. The resulting one-dimensional Fourier integral along the
light-cone distance $y^-$ is with respect to the quark light-cone
momentum $x  P^+$.
The {\it rhs} of Eq.~(\ref{eq:qsplitting}) parametrizes this
non-perturbative object in terms of the GPDs $H^q$ and $E^q$ 
for a quark of flavor $q$.  
The quark vector operator ($\gamma \cdot n$) 
corresponds at the nucleon side to a vector transition 
(parametrized by the function $H^q$) and
a tensor transition (parametrized by the function $E^q$). 
Analogously, there are two GPDs corresponding to 
a quark axial vector operator ($\gamma \cdot n \gamma_5$), which are 
commonly denoted by the polarized GPDs $\tilde H^q$ and $\tilde E^q$. 
\begin{figure}[t]
\centerline{  \epsfxsize=6cm
  \epsffile{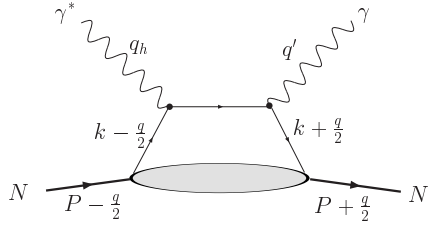} 
}
\caption{\small 
The ``handbag'' diagram for the nucleon DVCS process. 
Provided the virtuality of the initial photon (with momentum $q_h$) 
is sufficiently large, the 
QCD factorization theorem allows to express the 
total amplitude as the convolution  
of a Compton process at the quark level and a non-perturbative 
amplitude parameterized in terms of generalized parton distributions 
(lower blob). The diagram with the photon lines crossed is also understood.    
}
\figlab{n_dvcs}
\end{figure}
\newline
\indent
The variable $x$ in the GPDs runs from $-1$ to 1.
Therefore, the momentum fractions of the
active quarks ($x + \xi$) for the initial quark and ($x - \xi$) for the final 
quark can either be positive or negative. Since positive
(negative) momentum fractions correspond to quarks (antiquarks), it
has been noted in \cite{Radyushkin:1996nd} that in this way, one can
identify two regions for the GPDs:
when $x > \xi$ both partons represent quarks, whereas for
$x < - \xi$ both partons represent antiquarks. In these regions,
the GPDs are the generalizations of the usual parton distributions from
DIS. Actually, in the forward direction, the GPD $H$ 
reduces to the quark (anti-quark) density distribution $q(x)$ 
($\bar q(x)$) obtained from DIS: 
$H^q(x,0,0) = q(x)$, for $x > 0$;  
$H^q(x,0,0) = - \bar q(-x)$, for $x < 0$. 
The GPD $E$ 
is not measurable through DIS because the associated tensor 
in Eq.~(\ref{eq:qsplitting}) vanishes in the forward limit ($q \to 0$).
Therefore, $E$ is a new leading twist function, which
is accessible by measuring hard exclusive electroproduction reactions, such 
as DVCS.
\newline
\indent
Besides coinciding with the quark distributions at vanishing momentum
transfer, the GPDs have interesting links with other
nucleon structure quantities. The first moments of the GPDs are related to
the elastic FFs
of the nucleon through model independent sum rules.
By integrating Eq.~(\ref{eq:qsplitting}) over \( x \), one
obtains for any $\xi$
the following relations for a particular quark flavor \cite{ji} :
\begin{eqnarray}
\int_{-1}^{+1}dx\, H^{q}(x,\xi ,Q^2)\,=\, F_{1}^{q}(Q^2)\, ,
\hspace{2.cm}
\int _{-1}^{+1}dx\, E^{q}(x,\xi ,Q^2)\,=\, F_{2}^{q}(Q^2)\, ,
\label{eq:ffsumrulehe}
\end{eqnarray}
where $F_1^q$ ($F_2^q$) represents the elastic Dirac (Pauli) FFs for the
 quark flavor $q$ in the nucleon. These quark FFs are expressed, 
using $SU(2)$ isospin, 
as flavor combinations of the proton and neutron elastic FFs as:
\begin{eqnarray}
F_{1}^u \,=\, 2\,F_{1 p}\,+\,F_{1 n}\,+\,F_{1}^{s}\; , 
\hspace{2.5cm}
F_{1}^d \,=\, 2\,F_{1 n}\,+\,F_{1 p}\,+\,F_{1}^{s}\; ,
\label{eq:vecff}
\end{eqnarray}
where $F_{1}^{s}$ 
is the strangeness FF of the nucleon (which is neglected in 
the calculations discussed below).
Relations similar to Eq.~(\ref{eq:vecff}) hold for the
Pauli FFs \( F_{2}^{q} \).
At $Q^2 = 0$, the normalizations of the Dirac FFs are given by:
$F_{1}^{u}(0) = 2$ ($F_{1}^{d}(0) = 1$) 
so as to yield the normalization of 2 (1) for the
$u$ ($d$)-quark distributions in the proton. 
The normalizations of the Pauli FF at $Q^2 = 0$ are 
given by $F_{2}^{q}(0) = \kappa^q$ (for $q = u, d$), where 
$\kappa^u, \kappa^d$ can be expressed 
in terms of the proton ($\kappa_p$) 
and neutron ($\kappa_n$) anomalous magnetic moments as:
\begin{eqnarray}
\kappa^u \,\equiv \, 2 \kappa_p + \kappa_n = +1.673, 
\quad \quad \quad \quad 
\kappa^d \,\equiv \, \kappa_p + 2 \kappa_n = -2.033. 
\label{eq:kappaud}
\end{eqnarray}
\indent
The above sum rules allow us to make a prediction for the 
nucleon e.m. FFs provided we have a model for the nucleon GPDs. 
Note that the sum rules of Eq.~(\ref{eq:ffsumrulehe}) only involve valence 
quark GPDs, since the sea-quark and anti-quark contributions 
cancel each other in the sum rules. 
Since the results of the integration 
in Eq.~(\ref{eq:ffsumrulehe}) do not depend on the skewness 
$\xi$~\footnote{This is the simplest example
of a so-called polynomiality condition when calculating moments of GPDs.}, one 
can choose $\xi = 0$ in these sum rules. We therefore only discuss 
the GPDs $H$ and $E$ at $\xi = 0$ in the following. 
\newline
\indent 
In~\cite{diehl,guidal}, parameterizations of GPDs were developed 
which have a Regge behavior at small $x$ and $Q^2$, and which were 
modified to larger $Q^2$ behavior so as to lead to the observed power 
behavior of the FFs~\cite{Burkardt:2002hr,Burkardt:2004bv}. 
A modified Regge parameterization for $H$ and $E$ was proposed in 
\cite{guidal}~:
\begin{eqnarray}
H^q (x,0,Q^2) = q_v (x)\,  x^{\alpha^\prime \, (1 - x) \, Q^2}, 
\quad \quad \quad 
E^q (x, 0, Q^2) = \frac{\kappa^q}{N^q} 
\, (1 - x)^{\eta^q} \, q_v(x) \, 
{{x^{\alpha' \, (1 - x) \, Q^2}}} \, , 
\label{eq:gpd_r2}
\end{eqnarray}
depending on 3 parameters.  
The Regge slope $\alpha^\prime$ is determined from the Dirac radius, 
and two parameters $\eta^u$ and $\eta^d$, entering the GPD $E$, ensure that 
the $x\sim 1 $ limit of $E^q$ 
has extra powers of $1-x$ compared to that  of $H^q$. 
This results in a proton helicity flip FF $F_2$ which has a faster 
power fall-off at large $Q^2$ than $F_1$, as observed experimentally.   
Furthermore, in Eq.~(\ref{eq:gpd_r2}), 
the normalization factors $N^u$ and $N^d$ are given by~:  
\begin{eqnarray}
N^u \,=\,  \int _{0}^{1}dx \; (1 - x)^{\eta^u} \, u_v(x) \, ,
\quad \quad \quad \quad 
N^d \,=\,  \int _{0}^{1}dx \; (1 - x)^{\eta^d} \, d_v(x) \, ,
\label{eq:nd}  
\end{eqnarray} 
and guarantee the normalization condition for the GPD $E^q$. 
\newline
\indent
Diehl {\it et al.}~\cite{diehl} chose a more 
general functional form for $E^q$ at the expense of more free parameters. 
In the following, we discuss the `minimal' model with 3 parameters, 
and refer the interested reader to~\cite{diehl} for a study of more 
general functional forms. 
The 3 free parameters in the resulting modified Regge ansatz 
are to be determined from a fit to the FF data. 
\begin{figure}[h]
\vspace{-1.2cm}
\hspace{1cm}
\epsfxsize=8.5cm
  \epsffile{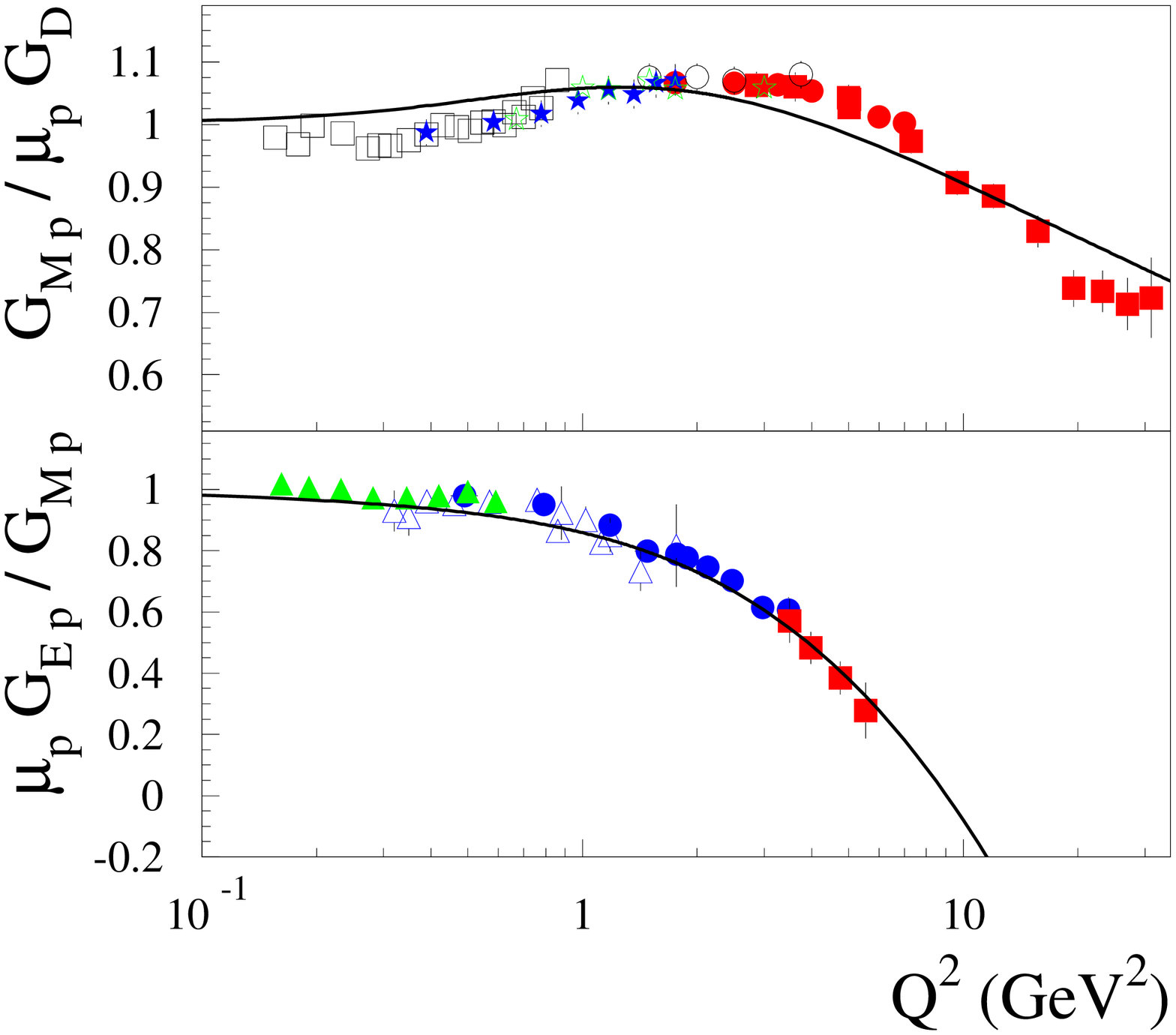}
\epsfxsize=8.5cm
  \epsffile{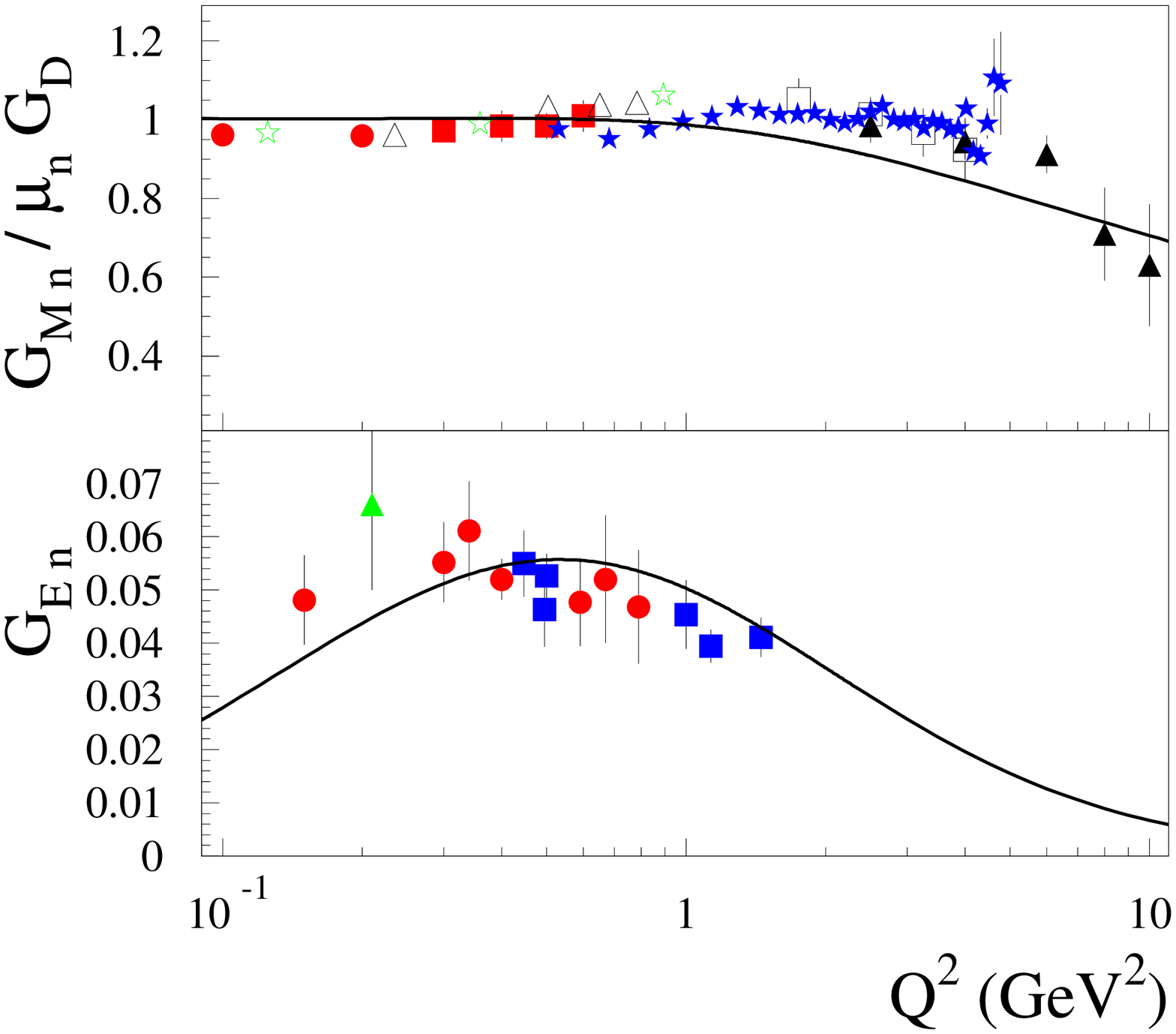}
\caption[]{\small GPD calculation of 
$G_{Mp}$ and $G_{Mn}$ relative to the dipole $G_D$ (upper panels), 
ratio of $G_{Ep}/G_{Mp}$ (lower left panel), 
and $G_{En}$ (lower right panel), according to Ref.~\cite{guidal}. 
The curves are a 3 parameter modified Regge parameterization : 
$\alpha^\prime = 1.105$\, GeV$^{-2}$, $\eta^u$ = 1.713 and $\eta^d$ = 0.566. 
Data for $G_{M p}$ are from   
\cite{janssens} (open squares), \cite{litt} (open circles),
\cite{berger} (blue solid stars), \cite{bartel} (green open stars), 
\cite{andivahis} (red solid circles), \cite{sill} 
(red solid squares), 
according to the recent re-analysis of Ref.~\cite{brash}. 
Data for the ratio $G_{E p} / G_{M p}$ are from 
\cite{gayou1} (blue open triangles), 
\cite{gayou2} (red solid squares),
\cite{punjabi05} (blue solid circles), 
and \cite{crawford} (green solid triangles). 
The data for $G_{M n}$ are from 
\cite{xu00} (red solid circles), 
\cite{xu02} (red solid squares),    
\cite{anklin2} (open triangles), \cite{kubon} (green open stars), 
\cite{lung} (open squares), \cite{rock} (solid triangles), 
and \cite{brooks} (blue solid stars).
The data for $G_{E n}$ are from 
double polarization experiments at 
MAMI~\cite{herberg,ostrick,becker,rohe,glazier} (red solid circles), 
NIKHEF~\cite{passchier} (green solid triangle),  
and JLab~\cite{zhu,madey,warren} (blue solid squares). 
}
\label{fig:gegmpn}
\end{figure}

In Fig.~\ref{fig:gegmpn}, the proton and neutron Sachs electric 
and magnetic FFs are shown. 
One observes that the 3-parameter 
modified Regge model gives a rather good overall 
description of the available FF data for both proton and neutron 
in the whole $Q^2$ range,   
using as value for the Regge trajectory 
$\alpha^\prime $ = 1.105 \,GeV$^{-2}$, 
and the following values for the coefficients 
governing the $x \to 1$ behavior of the $E$-type GPDs:  
$\eta^u$ = 1.713 and $\eta^d$ = 0.566.  
Note that a value $\eta^q = 2$ 
corresponds to a $ 1/Q^2$ asymptotic behavior of the ratio
$F_2^q / F_1^q$ at large $Q^2$. 
The modified Regge GPD parameterization allows 
to accurately describe the decreasing ratio of $G_{E p} / G_{M p}$ 
with increasing $Q^2$, and 
also leads to a zero for $G_{E p}$ at a 
momentum transfer of $Q^2 \simeq 8$~GeV$^2$, which will be within the range 
covered by an upcoming JLab experiment~\cite{E-04-108}.

\subsection{Perturbative QCD (pQCD)}
\label{theory_pqcd}

\begin{figure}[h]
\centerline{  \epsfxsize=6cm%
  \epsffile{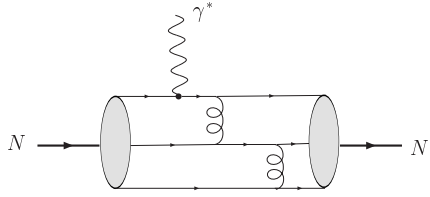} 
}
\caption{\small Perturbative QCD picture for the nucleon 
e.m. FFs. The highly virtual photon resolves 
the leading three-quark Fock states of the nucleon, described by 
a distribution amplitude. The large momentum is transferred between the quarks 
through two successive gluon exchanges (only one of several possible 
lowest-order diagrams is shown). }
\figlab{ff_pqcd}
\end{figure}
\begin{figure}[h]
\leftline{\hspace{1cm} \epsfysize=8.8cm \epsffile{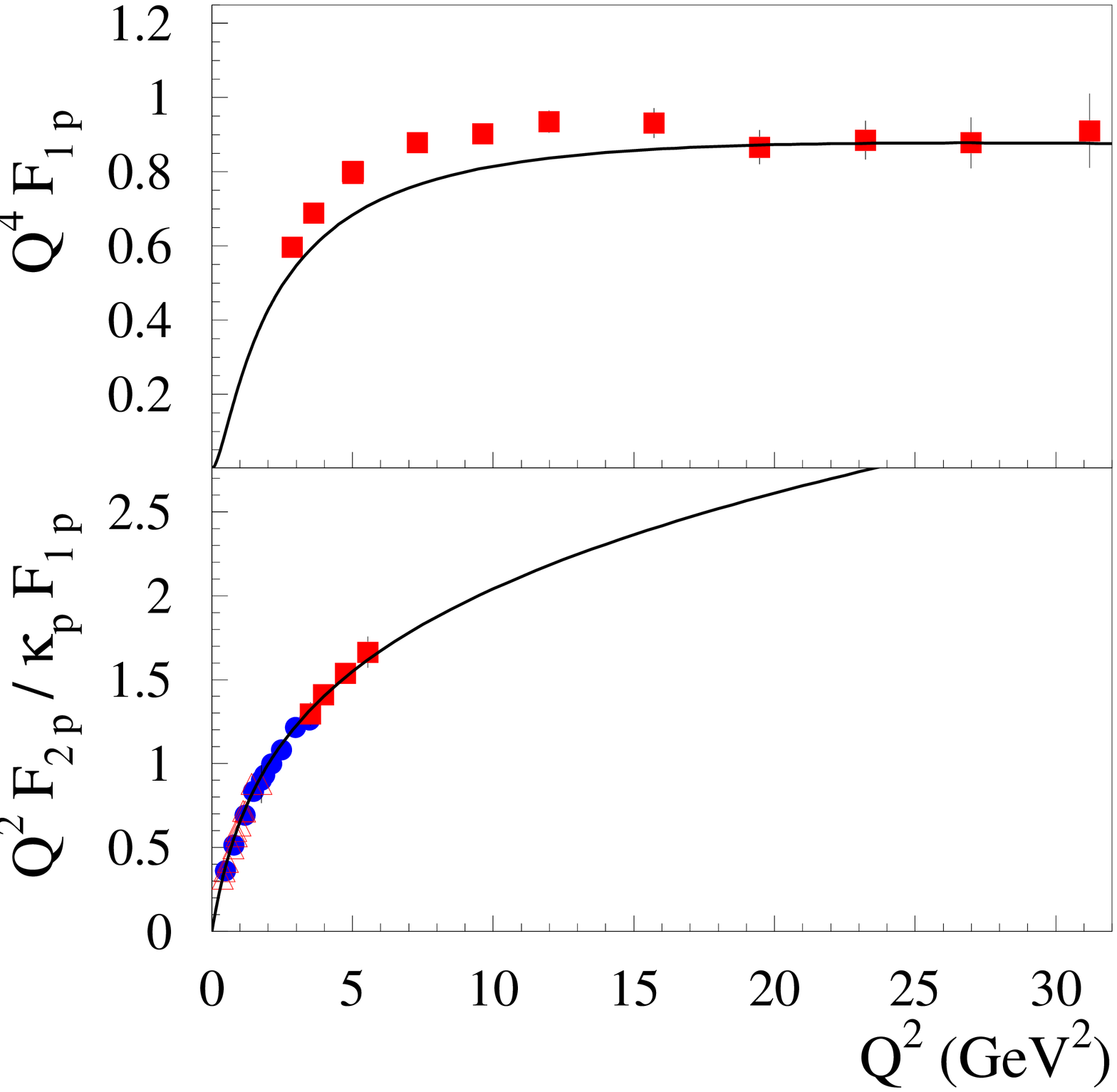}} 
\vspace{-7.4cm}
\centerline{\hspace{9cm} \epsfysize=6.95cm \epsffile{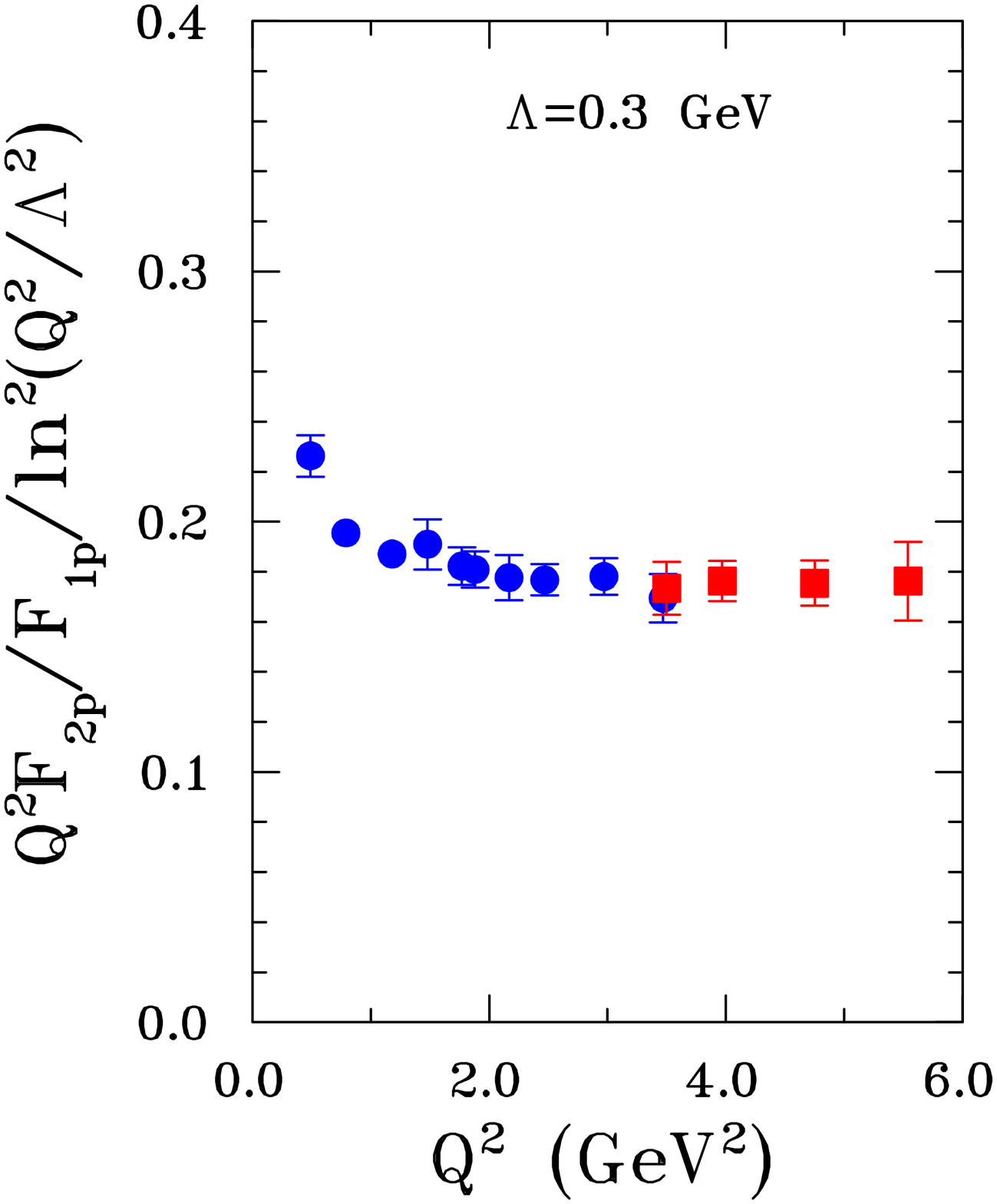}}
\caption{\small Test of the scaling behavior of the proton FFs.  
Upper left panel : proton Dirac FF multiplied by $Q^4$.   
Lower left panel : ratio of Pauli to Dirac proton FFs 
multiplied by $Q^2$. 
Right panel : test of the modified scaling prediction for 
$F_{2p}/F_{1p}$ of \cite{Belitsky:2002kj}. 
The data for $F_{1 p}$ are from \cite{sill} (solid squares).  
Data for the ratio $F_{2 p} / F_{1 p}$ on both panels are 
from \cite{punjabi05} (blue solid circles), 
\cite{gayou1} (red open triangles), 
and \cite{gayou2} (red solid squares). 
The curves on the left panels represent the calculation based on the  
three parameter modified Regge GPD parametrization of~\cite{guidal}.}
\label{fig:scaling}
\end{figure}

\indent
The nucleon e.m. FFs 
provide a famous test for perturbative QCD. 
Brodsky and Farrar derived scaling rules for dominant helicity amplitudes 
which are expected to be valid at sufficiently high momentum 
transfers $Q^2$~\cite{brodsky}. A photon of sufficient high 
virtuality will see a nucleon consisting of three massless quarks moving 
collinear with the nucleon. 
When measuring an elastic nucleon FF, 
the final state consists again of 
three massless collinear quarks. In order for this (unlikely process) to 
happen, the large momentum of the virtual photon has to be transferred 
among the three quarks through two hard gluon exchanges as illustrated in 
\Figref{ff_pqcd}. This hard scattering mechanism is generated by 
valence quark configurations with small transverse size and finite 
light-cone momentum fractions of the total hadron momentum carried by each 
valence quark. The hard amplitude can be written in a factorized 
form~\cite{Chernyak:1977as,Chernyak:1977fk,Efremov:1979qk,brodlep}, 
as a product of a perturbatively calculable hard scattering amplitude and 
two distribution amplitudes (DAs) 
describing how the large longitudinal momentum 
of the initial and final nucleons is shared between their constituents. 
Because each gluon in such hard scattering process carries a virtuality 
proportional to $Q^2$, this leads to the pQCD prediction that the helicity 
conserving nucleon Dirac FF $F_1$ should fall as $1/Q^4$ 
(modulo $\ln Q^2$ factors) at sufficiently high $Q^2$. 
Processes such as in 
\Figref{ff_pqcd}, where the interactions among the quarks proceed 
via gluon or photon exchange, both of which are vector interactions, 
conserve the quark helicity in the limit when the quark masses or off-shell 
effects can be neglected. 
In contrast to the helicity conserving FF $F_1$, the nucleon 
Pauli FF $F_2$ involves a helicity flip between the initial and 
final nucleons. Hence it requires one helicity flip at the quark level, which 
is suppressed at large $Q^2$. Therefore, for collinear quarks, i.e. 
moving in a light-cone WF state with 
orbital angular momentum projection 
$l_z = 0$ (along the direction of the fast moving hadron), 
the asymptotic prediction for $F_2$ leads to a 
$1/Q^6$ fall-off at high $Q^2$. 
\newline
\indent
We can test how well the above pQCD scaling predictions for the 
nucleon e.m. FFs 
are satisfied at currently available momentum transfers, see~\Figref{scaling}. 
One firstly sees that $F_{1p}$, which has been measured up to about 
30 GeV$^2$, displays an approximate $1/Q^4$ scaling above 10 GeV$^2$. 
For the proton ratio $F_{2 p}/F_{1 p}$, the data up to 5.6 GeV$^2$ show 
no sign of a $1/Q^2$ behavior as predicted by pQCD. 
Instead, the data show that the ratio $F_{2 p}/F_{1 p}$ falls less 
fast than $1/Q^2$ with increasing $Q^2$.
Belitsky, Ji, and Yuan \cite{Belitsky:2002kj} investigated the 
assumption of quarks moving collinearly with the proton, 
underlying the pQCD prediction. 
It has been shown in~\cite{Belitsky:2002kj} that by 
including components in the nucleon light-cone WFs with quark 
orbital angular momentum projection $l_z = 1$, one obtains 
the behavior $F_2/F_1 \to \ln^2 (Q^2 / \Lambda^2)/ Q^2$ at large $Q^2$, 
with $\Lambda$ a non-perturbative mass 
scale~\footnote{In~\cite{jain,Brodsky:2003pw}, 
it has also been discussed that inclusion
  of quark orbital angular momentum yields a ratio $F_{2p}/F_{1p}$ which drops
  less fast than $1/Q^2$ with increasing $Q^2$.  
}. 
Choosing $\Lambda$ around 
$0.3$~GeV, Ref.~\cite{Belitsky:2002kj} noticed that the 
data for $F_{2 p}/F_{1 p}$ support such double-logarithmic enhancement, as 
can be seen from \Figref{scaling} (right panel). 
The arguments of~\cite{Belitsky:2002kj} 
still rely on pQCD and it remains to be seen by 
forthcoming data at higher $Q^2$ if this prediction already starts in the 
few GeV$^2$ region. 
\newline
\indent
Although at high enough $Q^2$, the pQCD scaling predictions should set in, 
the available data for the nucleon e.m.  
FFs show that one is still far away from this regime. 
Nesterenko and Radyushkin~\cite{Nesterenko:1983ef} argued that the 
above described hard scattering mechanism is suppressed at accessible 
momentum transfers relative to the Feynman mechanism~\cite{feynman}, 
also called soft mechanism. The soft mechanism involves only one active 
quark, and the FF is obtained as an overlap of initial and final 
hadron WFs. 
The hard scattering mechanism on the other hand, 
involving three active quarks, requires the exchange of two gluons, each of 
which brings in a suppression factor $\alpha_s / \pi \sim 0.1$. One therefore 
expects the hard scattering mechanism for $F_{1 p}$  
to be numerically suppressed by a factor 1/100 compared to the soft term, see 
also~\cite{bolz,kroll}. 
Even though the soft mechanism is suppressed asymptotically by a power of 
$1/Q^2$ relative to the hard scattering mechanism, 
it may well dominate at accessible values of $Q^2$.
In~\cite{Nesterenko:1983ef}, the soft contribution 
to the nucleon e.m. FFs was 
estimated using a model based on local quark-hadron duality, and  
was found to yield an approximate $1/Q^4$ behavior 
in the range $Q^2 \sim 10 - 20$~GeV$^2$, 
in qualitative agreement with the $F_{1p}$ data. 
\newline
\indent
In a more recent work, the soft contribution  
was evaluated within the light-cone sum rule (LCSR) approach~\cite{braun06}. 
Using asymptotic DAs for the nucleon, 
the LCSR approach yields values of $G_{Mp}$ and 
$G_{Mn}$ which are within 20 \% compatible with the data in the range 
$Q^2 \sim 1 - 10$~GeV$^2$. The electric FFs however were found to be much
more difficult to describe, with $G_{En}$ overestimated, and 
$G_{Ep}/G_{Mp}$ near constant when using an asymptotic nucleon DA. 
Only when including twist-3 and twist-4 nucleon DAs within a simple model, 
is a qualitative description of the electric
proton and neutron FFs obtained. Such higher twist components hint at the
importance of quark angular momentum components in the nucleon WF. 
\newline
\indent
In Sect.~\ref{theory_gpd}, we have shown that the nucleon 
e.m. FFs can be obtained from model independent 
GPD sum rules. These GPDs, represented by the lower blob in 
\Figref{n_dvcs}, are non-perturbative objects which include 
higher Fock components in the nucleon WFs. One 
can use a GPD parametrization to provide an estimate of the soft 
contributions, and expects this non-perturbative approach 
to be relevant in the low and intermediate $Q^2$ region for the FFs.  
This is shown in \Figref{scaling} 
(solid curves) from which one sees that the GPD Regge parametrization  
discussed above is able to explain 
at the same time an approximate $1/Q^4$ 
behavior for $F_{1 p}$ and a behavior for $F_{2 p}/F_{1 p}$ 
which falls less steep than $1/Q^2$.  
Forthcoming experiments at the Jefferson Lab 12 GeV facility will extend the 
data for $F_{2 p}/F_{1 p}$ to $Q^2$ values around 
13 GeV$^2$. Such measurements will allow to quantify in detail the higher 
Fock components in the nucleon WF 
(which are all included in the nucleon GPD) versus 
the simple three-quark Fock component, and to map out the 
transition to the perturbative QCD regime.

\section{Conclusions and outlook}
\label{conclude}

The increasingly common use of the double-polarization technique to 
measure the nucleon FFs, in the last 15 years, has resulted 
in a dramatic improvement 
of the quality of all four nucleon e.m. FFs, 
$G_{Ep}$, $G_{Mp}$, $G_{En}$ and
$G_{Mn}$. It has also completely changed our understanding of the proton 
structure, having resulted in a distinctly different $Q^2$- dependence for
$G_{Ep}$ and $G_{Mp}$,
contradicting the prevailing wisdom of the 1990's based on
cross section measurements and the Rosenbluth separation method, namely that 
$G_{Ep}$ and $G_{Mp}$ obey a ``scaling'' relation
$\mu G_{Ep}\sim G_{Mp}$. A direct consequence  
of the faster decrease of $G_{Ep}$ revealed by
the JLab polarization experiments was the disappearance of the early 
scaling $F_2/F_1 \sim 1/Q^2$ predicted by perturbative QCD. 
\newline
\indent
The main origin of this abrupt change in results 
is now understood in simple terms.
The faster decrease of $G_{Ep}$ reduces its contribution to the cross section
significantly below the natural ratio prevailing at small $Q^2$, namely
$G_{Ep}^2/G_{Mp}^2\sim 1/\mu_{p}^2$. At the highest $Q^2$ for which 
we now have polarization data, 5.6 GeV$^2$, the contribution from the 
electric FF to the cross section is less than 1\%. 
It has been realized in recent years, that to extract $G_{Ep}$ from 
Rosenbluth separations at larger $Q^2$ requires a much better 
quantitative understanding of several of the radiative corrections contributions,
including in particular the one due two hard photon exchange. As discussed in section
\ref{RadCorr}, there are currently differences of order several \%s between the 
results of various radiative correction calculations. The two hard photon correction
by itself might explain the whole discrepancy between Rosenbluth and recoil polarization 
results, but it does not affect recoil polarization results measurably, because these 
are measurements of ratios of FFs and both FFs are, in first order, modified similarly.
Until the origin of the difference between cross section- 
and polarization results is understood in full quantitative detail, 
it is safe to take the polarization results as the closest to the real, 
Born approximation, proton FFs. 
\newline
\indent 
The use of the polarization technique has also resulted in a 
constant progress in the 
measurement of $G_{En}$, which is intrinsically more difficult to obtain 
because of the
smallness of this form factor, due to the overall zero charge of the neutron. 
Recent times have seen the maximum $Q^2$ for which we have polarization FFs  
grow to 1.5 GeV$^2$, with new data 
obtained and under analysis up to 3.4 GeV$^2$, and several experiments 
planned or proposed to significantly higher $Q^2$ values. 
Important progress has been made for $G_{Mn}$ too, 
with new data with much improved error bars up 
to 4.8 GeV$^2$. 
\newline
\indent
A basic understanding of the nucleon e.m. FFs
can be obtained within the VMD picture, in which the 
photon couples to the nucleon through the exchange of vector
mesons. Dispersion analyses build on this picture by including, besides 
vector mesons, also non-resonant contributions in the coupling of the photon 
with the nucleon. 
The state-of-the-art dispersion analyses, which include the $2 \pi$ 
continuum in the isovector spectral function, 
and the $K \bar K$ and $\rho \pi$ continua as input in the isoscalar spectral
function, are found to yield 
a reasonably good overall description of the data 
for all four nucleon e.m. FFs using a 15 parameter fit.  
\newline
\indent
The effect of pionic degrees of freedom in the nucleon e.m. FFs can be
systematically calculated within chiral effective field theory. 
The latest relativistic $\chi$EFT calculations found  
that calculations based on pions and nucleons alone are not able
to explain the nucleon e.m. FFs. 
Only upon inclusion of explicit vector meson degrees of freedom, these 
calculations were found to describe the FFs in the range $Q^2 \lesssim
0.4$~GeV$^2$.  
\newline
\indent
Calculations of the nucleon e.m. FFs within constituent quark models have
highlighted the role of relativity when trying to arrive at a microscopic 
description of nucleon FFs based on quark degrees of freedom in the few 
GeV$^2$ region. Although a complete calculation is independent of the 
specific choice of relativistic form chosen to describe the dynamics, present 
approximations destroy this independence.   
\newline
\indent
We have also reviewed the recent progress made by lattice QCD calculations of 
the nucleon e.m. FFs. In present lattice simulations, 
disconnected diagrams, which are numerically more intensive, have not 
yet been evaluated. Therefore the current lattice calculations are for 
the isovector combination of nucleon FFs where this 
disconnected diagram contribution drops out. 
They are performed at pion mass values above 
about 350 MeV and have to be extrapolated to the physical pion mass to allow
for a comparison with experiment.  
The calculations using dynamical Wilson quarks 
are able to provide a reasonably good 
description of the isovector Pauli FF $F_2^V$. At the lower $Q^2$, the 
non-analytic terms in the chiral extrapolation were found to be important to 
arrive at a description of the isovector magnetic moment and Pauli radius. 
The present dynamical Wilson calculations largely overestimate the isovector 
Dirac FF $F_1^V$ however. In contrast, the results using a hybrid 
action, consisting of domain wall valence quarks and staggered sea quarks, 
are compatible for $F_2^V$, but differ from the Wilson results for 
$F_1^V$. The $F_1^V$ results using the hybrid action are  
in qualitative better agreement with the data, and provide a first clear hint 
of the logarithmic $m_\pi$ divergence in the isovector Dirac radius. 
\newline
\indent
The quark structure of the $N \to N$ electromagnetic transition can be 
accessed in hard scattering processes such as deeply virtual Compton
scattering. The non-perturbative information in this process can be 
parameterized in terms of GPDs, and the nucleon
e.m. FFs can be obtained as first moment sum rules 
in the quark longitudinal momentum fraction of these GPDs. 
In particular, the GPDs contain the information on 
the quark distribution of the 
nucleon anomalous magnetic moment which can not be accessed 
from inclusive deep inelastic scattering experiments. 
We discussed a GPD parameterization, and found that 
the first moment of such a 3-parameter modified Regge parameterization 
yields a good overall description of the nucleon e.m. FF data 
over the whole available $Q^2$ range. 
The modified Regge GPD parameterization predicts that, 
at moderately large $Q^2$ values, 
$F_{1p}$ follows an approximate $1/Q^4$ scaling, 
whereas $F_{2p}/F_{1p}$ drops less fast than the $1/Q^2$ pQCD behavior, 
in agreement with the polarization data. 
It furthermore predicts that $G_{Ep}$ reaches a 
zero around $Q^2 \sim 8$~GeV$^2$, which is in the reach of an upcoming 
experiment.  
\newline
\indent
We like to end this review by spelling out a few open issues and challenges 
in this field~:
\begin{enumerate}
\item {\it Quantifying the two-photon exchange processes both 
experimentally and theoretically} 

In order to use electron scattering as a precision tool, it is clearly 
worthwhile to arrive at a quantitative understanding of two-photon exchange 
processes. This calls for detailed experimental studies, and 
several new experiments are already planned.  
Differences between elastic $e^-$ and $e^+$ scattering  
directly access the real part of the two-photon exchange amplitude. 
The predicted small effect of two-photon processes on the polarization data 
can be checked by measuring
the $\epsilon$ dependence in polarization transfer experiments. 
These upcoming experiments also call for further refinements  
on the theoretical side. 

\item {\it Dispersion analyses}

In the present dispersion analysis for the nucleon e.m. FFs, the pQCD limit 
$F_2 / F_1 \sim 1/Q^2$ was built in as a constraint, although the data do not 
support this limit. It might therefore be worthwhile to investigate how the 
dispersion fits change by removing this bias from the analysis 
at the largest available $Q^2$ values. 

\item {\it Relativistic constituent quark model calculations}

The quality of the commonly introduced impulse approximation 
when describing nucleon FFs in relativistic CQMs 
may differ between different forms.  As a next step for CQMs, it 
would clearly be worthwhile to investigate the approximations made in 
the current operator within each form,  
and quantify e.g. the effect of explicit two-body currents. 

\item {\it Lattice calculations and chiral extrapolations}
\begin{itemize}
\item[(a)]
One would clearly like to understand the present disagreement between the 
unquenched lattice predictions when using different actions 
(Wilson action vs. hybrid action). 
Understanding the structure of the FFs at low $Q^2$, 
such as the Dirac charge radius, depends crucially on the effect of pion loops,
which yield strong non-analytic dependence (in particular a log $m_\pi$
singularity). 
At the larger $Q^2$ values, between 0.5 and 2~GeV$^2$, the effects of
pion loops are expected to be reduced however. 
One may therefore suspect that the differences  
between different actions at larger $Q^2$ values 
are due to discretization errors, which have only partly been studied. 
Unquenched calculations at a couple of different lattice spacings 
using both the Wilson and hybrid actions would be very helpful to shed a
further light on this issue. 
\item[(b)]
In order to provide predictions for proton and neutron e.m. FFs
separately, the calculation of the disconnected diagrams awaits the next 
generation of dynamical fermion lattice QCD simulations.
\item[(c)]
A further challenge for the lattice calculations 
is a fully consistent treatment of both valence and sea quarks which 
respect chiral symmetry on the lattice.  
\item[(d)]
As future lattice calculations for pion mass values 
around and below 300~MeV become
available for FFs in the range $Q^2 \stackrel{>}{\sim} 0.5$~GeV$^2$, 
i.e. beyond the range where present $\chi$PT calculations are applicable, one 
is confronted with a two-scale problem. A challenge is to theoretically 
study the extrapolation (in the small scale $m_\pi$) 
in the presence of a (moderately) large scale $Q^2$. 
\end{itemize}

\item {\it Precision measurements in the low $Q^2$ regime}

Precision measurements of the nucleon e.m. FFs in the $Q^2$ range below 
0.5~GeV$^2$, may bring the effects of the pion could sharper into focus. 
In this respect, new measurements for $G_{En}$ are needed 
to better quantify the conjectured ``bump'' 
structure in $G_{En}$ around $Q^2 \sim 0.3$~GeV$^2$.  

\item {\it Extending the FF measurements to larger $Q^2$}

The anticipated upgrade of JLab to 12 GeV beam energy, 
offers promises of measurements of 
all four FFs up to or larger than $\sim$10 GeV$^2$. It is unlikely 
that we will see indications of a clear departure from the soft physics 
dominance to the hard collision regime of pQCD. However these data will 
constrain the parameterizations of GPDs and yield information on the
spatial distribution of partons which carry a 
large momentum fraction of the nucleon momentum, i.e. partons with $x \sim 1$.

\end{enumerate}

The recent unexpected results in the nucleon e.m. FFs 
using double-polarization high-precision experiments, 
have challenged our theoretical understanding of the structure of the nucleon. 
They have triggered several new theoretical developments, which were reviewed
in this work. As a result of the unexpected findings, 
several further experiments are planned, which will 
bring the quark-gluon structure of the most common constituent 
of visible matter in the universe into sharper focus.

\section*{\center{Acknowledgments}}

We like to thank 
C. Alexandrou, T. Averett, D. Day,
J. Friedrich, 
K. Orginos,  
V. Pascalutsa,
B. Pasquini,
and T. Walcher,
for useful discussions and correspondence during the course of this work and 
E.J. Brash, K. de Jager, and B. Mecking 
for their careful reading. 
C.F.~P. acknowledges support from a NSF grant, 
PHY-0456645. 
The work of V.~P. is supported by the Physics Division, 
DOE, grant DE-FG02-89ER40525.
Furthermore, the work of M.~V. is supported in part by DOE grant
DE-FG02-04ER41302 and contract DE-AC05-06OR23177 under
which Jefferson Science Associates operates the Jefferson Laboratory.

\end{document}